\documentclass[preprint,nopreprintline,3p,12pt]{elsarticle}
\pdfoutput=1
\usepackage{graphicx}
\usepackage{hyperref}
\usepackage{amssymb}
\usepackage{gensymb}
\usepackage{amsmath}
\usepackage{enumitem}
\usepackage{caption}

\newcommand{\hyph}{\text{--}}

\begin{document}

\begin{frontmatter}

\title{Measurement of the $e^+e^-\to\pi^+\pi^-$ cross section from threshold to 1.2 GeV with the CMD-3 detector}

\author[adr1,adr2]{F.V.~Ignatov\fnref{tnot}}
\author[adr1,adr2]{R.R.~Akhmetshin}
\author[adr1,adr2]{A.N.~Amirkhanov}
\author[adr1,adr2]{A.V.~Anisenkov}
\author[adr1,adr2]{V.M.~Aulchenko}
\author[adr1]{N.S.~Bashtovoy}
\author[adr1,adr2]{D.E.~Berkaev}
\author[adr1,adr2]{A.E.~Bondar}
\author[adr1]{A.V.~Bragin}
\author[adr1,adr2]{\fbox{S.I.~Eidelman}}
\author[adr1,adr2]{D.A.~Epifanov}
\author[adr1,adr2,adr3]{L.B.~Epshteyn}
\author[adr1,adr2]{A.L.~Erofeev}
\author[adr1,adr2]{G.V.~Fedotovich}
\author[adr1,adr3]{A.O.~Gorkovenko}
\author[adr6]{F.J. Grancagnolo}
\author[adr1,adr2]{A.A.~Grebenuk}
\author[adr1,adr2]{S.S.~Gribanov}
\author[adr1,adr2,adr3]{D.N.~Grigoriev}
\author[adr1,adr2]{V.L.~Ivanov}
\author[adr1]{S.V.~Karpov}
\author[adr1]{A.S.~Kasaev}
\author[adr1,adr2]{V.F.~Kazanin}
\author[adr1]{\fbox{B.I.~Khazin}}
\author[adr1]{A.N.~Kirpotin}
\author[adr1,adr2]{I.A.~Koop}
\author[adr1,adr2]{A.A.~Korobov}
\author[adr1,adr3]{A.N.~Kozyrev}
\author[adr1,adr2]{E.A.~Kozyrev}
\author[adr1,adr2]{P.P.~Krokovny}
\author[adr1]{A.E.~Kuzmenko}
\author[adr1,adr2]{A.S.~Kuzmin}
\author[adr1,adr2]{I.B.~Logashenko}
\author[adr1,adr2]{P.A.~Lukin}
\author[adr1]{A.P.~Lysenko}
\author[adr1,adr2]{K.Yu.~Mikhailov}
\author[adr1,adr2]{I.V.~Obraztsov}
\author[adr1]{\fbox{V.S.~Okhapkin}}
\author[adr1]{A.V.~Otboev}
\author[adr1,adr2]{E.A.~Perevedentsev}
\author[adr1]{Yu.N.~Pestov}
\author[adr1,adr2]{A.S.~Popov}
\author[adr1,adr2]{\fbox{G.P.~Razuvaev}}
\author[adr1,adr2]{Yu.A.~Rogovsky}
\author[adr1]{A.A.~Ruban}
\author[adr1]{\fbox{N.M.~Ryskulov}}
\author[adr1,adr2]{A.E.~Ryzhenenkov}
\author[adr1,adr2]{A.V.~Semenov}
\author[adr1]{A.I.~Senchenko}
\author[adr1]{P.Yu.~Shatunov}
\author[adr1]{Yu.M.~Shatunov}
\author[adr1,adr2]{V.E.~Shebalin}
\author[adr1,adr2]{D.N.~Shemyakin}
\author[adr1,adr2]{B.A.~Shwartz}
\author[adr1,adr2]{D.B.~Shwartz}
\author[adr4]{A.L.~Sibidanov}
\author[adr1,adr2]{E.P.~Solodov}
\author[adr1,adr2]{A.A.~Talyshev}
\author[adr1]{M.V.~Timoshenko}
\author[adr1]{V.M.~Titov}
\author[adr1,adr2]{S.S.~Tolmachev}
\author[adr1]{A.I.~Vorobiov}
\author[adr1]{I.M.~Zemlyansky}
\author[adr1]{D.S.~Zhadan}
\author[adr1]{Yu.M.~Zharinov}
\author[adr1]{A.S.~Zubakin}
\author[adr1,adr2]{Yu.V.~Yudin}

\address[adr1]{Budker Institute of Nuclear Physics, SB RAS, 
Novosibirsk, 630090, Russia}
\address[adr2]{Novosibirsk State University, Novosibirsk, 630090, Russia}
\address[adr3]{Novosibirsk State Technical University, 
Novosibirsk, 630092, Russia}
\address[adr4]{University of Victoria, Victoria, British Columbia, Canada V8W 3P6}
\address[adr6]{Instituto Nazionale di Fisica Nucleare, Sezione di Lecce, Lecce, Italy}
\fntext[tnot]{Corresponding author: F.V.Ignatov@inp.nsk.su}



\begin{abstract}
  The cross section of the process $e^+e^-\to\pi^+\pi^-$ has been
  measured in the center of mass energy range from 0.32 to 1.2 GeV
  with the CMD-3 detector at the electron-positron collider VEPP-2000.
  The measurement is based on a full dataset collected below 1 GeV
  during three data taking seasons, corresponding to an integrated
  luminosity of about 62~pb$^{-1}$. In the dominant $\rho$-resonance region, 
  a systematic uncertainty of 0.7\% has been reached. 
  At energies around $\phi$-resonance the $\pi^+\pi^-$ production cross section was measured 
  for the first time with high beam energy resolution. 
  The forward-backward charge asymmetry in the $\pi^+\pi^-$ production has 
  also been measured.  It shows a strong deviation from the
  theoretical prediction based on the conventional scalar quantum
  electrodynamics framework, 
  and it is in good agreement with the generalized vector-meson-dominance and dispersive-based predictions.
  The impact of the presented results on the evaluation of the hadronic contribution to
  the anomalous magnetic moment of muon is discussed.  
\end{abstract}

\end{frontmatter}


\section{Introduction}
The $e^+e^-\to\pi^+\pi^-$ channel provides the dominant contribution to the
production of hadrons from $e^+e^-$ annihilation at the energy range below $\sqrt{s}<1$~GeV.
The total hadron production cross section normalized to the two muon production cross section,
$R(s)$, is one of the fundamental quantities in high energy physics 
as it reflects a number of quark flavors opened for the production at the particular 
$s$ and a number of colors in quantum chromodynamics (QCD): 
\begin{equation}
R(s)=\frac{\sigma^0(e^+e^- \to \gamma^* \to
  hadrons)}{\sigma^0(e^+e^- \to \gamma^* \to \mu^+\mu^-)}
\sim  N_c\sum_{q}e_q^2(1+\delta_{QCD}(s)),\nonumber
\end{equation}
where $N_c=3$ is a number of QCD colors, $e_q$ is electric charge of $q$th quark, 
$\delta_{QCD}(s)$ is QCD correction. 
The perturbative QCD (pQCD) calculations has limited applicability at the resonances 
and quark-antiquark production threshold energy regions,
therefore experimentally measured values of $R(s)$ at $\sqrt{s}<2$~GeV are used
via dispersion integrals in many applications for prediction of
various physical quantities such as the running fine structure constant
$\alpha_{QED}(M_Z)$\cite{Jegerlehner:2006ju},
the hyperfine splitting in muonium~\cite{Czarnecki:2001yx},
the anomalous magnetic moment of the muon $a_{\mu}
=(g_{\mu}-2)/2$\cite{Keshavarzi:2019abf,Davier:2019can}.
The last one demonstrates long standing $3\hyph4\sigma$ deviation between the
experimental measurement and the standard model (SM) prediction.
There are efforts by several groups to
compute the leading order hadronic vacuum polarization contribution
$a_\mu^{HVP}$ to the anomalous magnetic moment of the muon
using the lattice QCD approach
\cite{Aoyama:2020ynm}. The recent calculation
by the BMW collaboration reaches the subpercent precision and reduces
the deviation between the experimental $(g_{\mu}-2)/2$ value and the SM prediction \cite{Borsanyi:2020mff}.
Most recent evaluations in different sub-regions by the other groups 
additionally increase tension between 
the dispersive, $R(s)$-based, and the lattice QCD predictions
\cite{Colangelo:2022vok,Ce:2022kxy,ExtendedTwistedMass:2022jpw,Blum:2023qou,ExtendedTwistedMassCollaborationETMC:2022sta}.
Hadronic $\tau$ decays with the spectral function
measurements can be also used as input for $R(s)$, but this requires a
challenging accounting of isospin-breaking corrections~\cite{Miranda:2020wdg}.
The other novel approach has been proposed to determine $a_\mu^{HVP}$
measuring the effective electromagnetic coupling in the spacelike
region via $\mu e$ scattering data by the the MUonE experiment
\cite{CarloniCalame:2015obs,Abbiendi:2016xup,Abbiendi:2677471}. It would provide another independent determination
of the hadronic contribution to muon g-2.

The $\pi^+\pi^-$ channel as part of $R(s)$ is used in the calculation 
of the muon anomaly and gives the major part of the hadronic
contribution, which at leading order (LO) is $506.0 \pm 3.4 \times 10^{-10}$ out of 
the full $a_{\mu}^{HVP,LO}=693.1\pm4.0 \times 10^{-10} $ value for all
hadronic states~\cite{Aoyama:2020ynm}.
The hadronic vacuum polarization (HVP) contributions at next-to-leading (NLO) and next-next
to-leading orders are $a_{\mu}^{HVP,NLO}=-9.83\pm0.07 \times 10^{-10} $
and $a_{\mu}^{HVP,NNLO}=1.24\pm0.01 \times 10^{-10} $, while the
light-by-light contribution is $a_{\mu}^{HLbL}=9.2\pm1.8 \times 10^{-10} $~\cite{Aoyama:2020ynm}.
%
The precision of $\pi^+\pi^-$ channel measurement also determines the overall
uncertainty $\Delta a_{\mu}= \pm 4.3\times 10^{-10}$ of the standard
model prediction of muon $g-2$ in the data-driven calculation.
To conform to the ultimate target precision of the ongoing Fermilab experiment\cite{Muong-2:2015xgu,Muong-2:2021ojo} 
$\Delta a ^{exp}_{\mu}[E989] \approx \pm 1.6 \times 10^{-10}$ and the future J-PARC muon g-2/EDM experiment\cite{Saito:2012zz},
the $\pi^+\pi^-$  production cross section needs to be known with the relative 
overall systematic uncertainty about 0.2\%.

Several sub-percent precision measurements of the $e^+e^-\to\pi^+\pi^-$ cross section exist.  
The energy scan measurements were performed at VEPP-2M collider by the CMD-2
experiment (with the systematic precision of 0.6--0.8\%) \cite{CMD-2:2003gqi,CMD-2:2005mvb,Aulchenko:2006dxz,CMD-2:2006gxt}
and by the SND experiment (1.3\%) \cite{Achasov:2006vp}. 
These results have somewhat limited statistical precision. 
There are also measurements based on the initial-state radiation (ISR) technique by 
KLOE (0.8\%)~\cite{KLOE:2008fmq,KLOE:2010qei,KLOE:2012anl,KLOE-2:2017fda},
$BABAR$(0.5\%)~\cite{BaBar:2012bdw} and BESIII(0.9\%)~\cite{BESIII:2015equ}. 
Due to the high luminosities of these $e^+e^-$ factories, the accuracy of the 
results from the experiments are less limited by statistics, meanwhile they are 
not fully consistent with each other within the quoted systematic uncertainties. 

One of the main goals of the CMD-3 and SND experiments at the new VEPP-2000 $e^+e^-$ collider at BINP, Novosibirsk, 
is to perform the new high precision high statistics measurement of the $e^+e^-\to\pi^+\pi^-$ cross section. 
Recently, the first SND result based on about 10\% of the collected statistics 
was presented with a systematic uncertainty of about 0.8\%~\cite{SND:2020nwa}. 
Here we present the first CMD-3 result.

In comparison with the previous CMD-2, the next generation CMD-3 experiment 
has collected between one and two orders of magnitude more statistics,
depending on the beam energy range.
This allows to study possible sources of systematic errors of the cross section measurement
in much more detail. The CMD-3 detector has much better performance compared with its predecessor: 
the drift chamber has twice better momentum resolution and 2--5 times smaller tracking inefficiency;
the unique multilayer LXe calorimeter with the tracking capabilities gives information on a shower profile. 
The improved momentum resolution allowed to use this
information for a particle separation in the most important
$\rho$-resonance energy range,
which provides an additional independent method to the previously used separation based on the
energy deposition information.

The paper is structured as follows. First, we briefly describe the detector,
the collected data and the analysis strategy. 
Then we discuss in detail particle separation of the selected events,
possible sources of background,
evaluation of the detection and trigger efficiencies,
calculation of the radiative corrections. A special section is devoted to the measurement of
the forward-backward charge asymmetry. The final sections include the discussion of the sources of
the systematic error, the presentation of the final result, the VDM fit of the measured cross section, 
the comparison with the other experiments and the evaluation of the corresponding
contribution to the $a_\mu^{HVP}$.

\section{VEPP-2000 and CMD-3}
The electron-positron collider VEPP-2000~\cite{Shatunov:2016bdv,Shwartz:2016fhe}
has been operating at Budker Institute of
Nuclear Physics since 2010. The collider was designed to provide instantaneous luminosity 
up to $10^{32} \rm cm^{-2}\rm s^{-1}$ at the maximum center-of-mass (c.m.) energy 
$\sqrt{s} = 2 \;\rm GeV$. 
Two detectors \mbox{CMD-3}~\cite{Aulchenko:2001je,Khazin:2008zz} 
and SND~\cite{Achasov:2009zza} are installed at the interaction regions of
the collider. In 2010 both experiments started  
data taking. 
The physics program~\cite{Eidelman:2006cn} includes high precision 
measurements of the $e^+e^- \rightarrow  hadrons$ cross  
sections in the wide c.m.~energy range up to 2~GeV, where the rich
intermediate dynamics is of particular interest, studies of known and
searches for new vector mesons and other intermediate states, 
studies of the $n\bar{n}$ and $p\bar{p}$ 
cross sections near their production thresholds and searches for exotic
hadrons. It requires a detector with a high efficiency for multiparticle events 
and good energy and angular resolutions for charged particles as well as for 
photons.

\begin{figure}[t]
\centering
\begin{minipage}{.45\textwidth}
\centering
\includegraphics[width=0.75\linewidth,keepaspectratio]{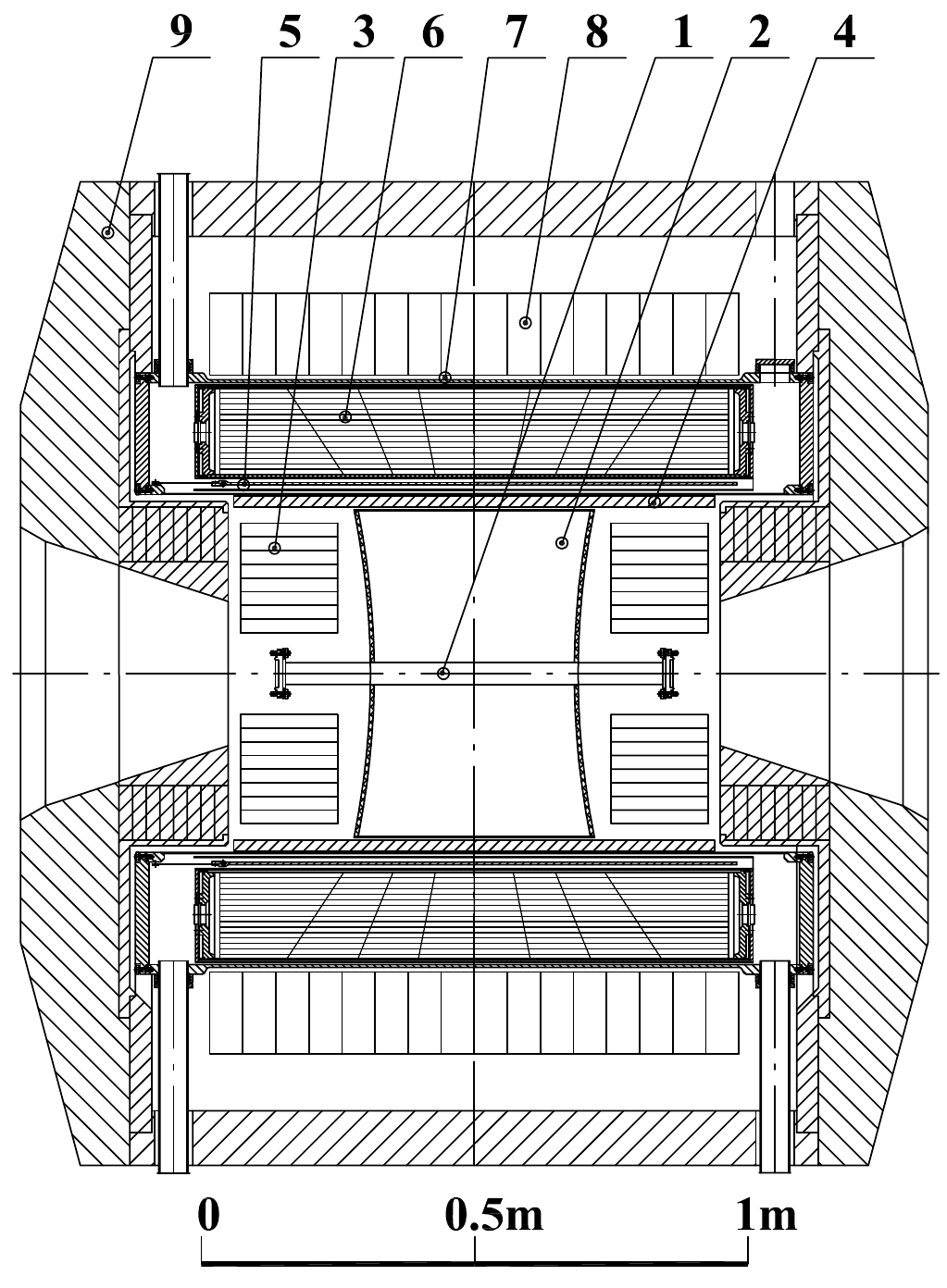}
\caption{CMD-3 detector: 1 -- beam pipe, 2 -- drift chamber, 3 -- BGO
crystal calorimeter, 4 -- $Z$-chamber, 5 -- superconducting solenoid ($0.13X_0$, $13 \rm\, kGs$),  6 -- LXe calorimeter,
7 -- TOF system,  8 -- CsI crystal calorimeter, 9 -- yoke,
not shown is the muon range system.}
\label{cmd3}
\end{minipage}%
\hspace*{.05\textwidth}
\begin{minipage}{.5\textwidth}
%
\centering
\includegraphics[width=1.1\linewidth]{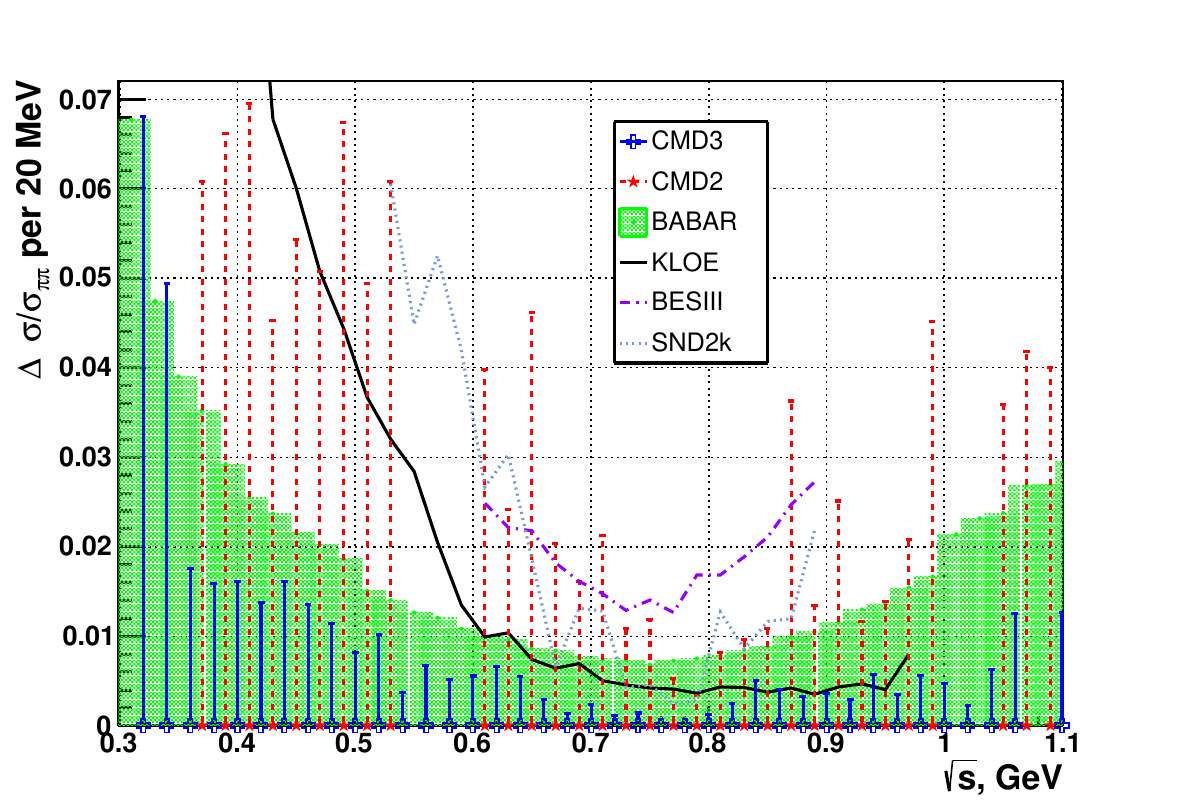}
\caption{Relative statistical precision of $|F_{\pi}|^2$ from 
the CMD-3 data in comparison with the CMD-2, $BABAR$, KLOE, BESIII and
SND@VEPP-2000 results. Integrated statistic over 20 MeV bin is shown.}
\label{fpi_stat} 
\end{minipage}
\end{figure}

\mbox{CMD-3} (cryogenic magnetic detector) is a general-purpose
detector, as shown in Fig.~\ref{cmd3}.  Coordinates, angles and momenta of
charged particles are measured by 
the cylindrical drift chamber (DCH) with a hexagonal cell for a high efficient
reconstruction of tracks in the uniform 1.3~T magnetic field,
where $\sigma_{P}/P \sim \sqrt{0.6^2+(4.4\cdot p[\text{GeV}])^2}\%$ is typical momentum resolution.
$z$-coordinates (along the positron beam direction) of the DCH hits used
for the track polar angle evaluation are determined by the charge division method.
For the purpose of the precise fiducial volume calibration, the $Z$-chamber, the MWPC with a strip
readout placed outside of the DCH, was in operation until 2017. 
The calorimetry is performed with the end cap BGO crystal
calorimeter and the barrel calorimeter, placed outside of the superconducting
solenoid. The barrel calorimeter consists of the
two systems: the liquid Xenon (LXe) ionization calorimeter 
surrounded by the CsI scintillation calorimeter.
The total thickness of the barrel calorimeter is about $13.5 X_0$,
where the inner LXe part constitutes $5.4 X_0$.  The LXe
calorimeter has seven layers with strip readout which give
information about a longitudinal shower profile and allow to measure coordinates
of photons with about 2 millimeter precision\cite{Anisenkov:2014xpa}.
The energy resolution in the barrel calorimeter is
$\sigma_{E}/E \sim 0.02\oplus 0.034/\sqrt{E[\text{GeV}]}$.

The first energy scan below 1~GeV for the $\pi^+\pi^-$ measurement 
was performed at the VEPP-2000 collider in 2013 (labeled in the text below as RHO2013), 
when the integrated luminosity of 17.8 pb$^{-1}$ was collected in 66 energy points. 
In 2014-2016 there was a long shutdown for the collider and detector upgrades. 
In particular, a new electron and positron injector facility was commissioned, 
which allowed to increase luminosity significantly. 
The next energy scan in the $\rho$-meson c.m.~energy region was carried out during
2017-2018 season (labeled as RHO2018), where about 45.4~pb$^{-1}$ were collected in 93 energy points. 
During these two data taking seasons also the integrated luminosity of 25.7~pb$^{-1}$ was recorded in 37 points at the c.m.~energies 
near the $\phi$-meson resonance.
At the end of 2019, the additional 1~pb$^{-1}$ data sample was collected in 13 points at the c.m.~energies $\sqrt{s}<0.6$~GeV 
(LOW2020 scan), 
which increased $\pi^+\pi^-$ statistics by a factor of 2--5 at the $\pi^+\pi^-$ production threshold energy region.
In total, about $3.4\times10^{7}$ $\pi^+\pi^-$, $3.7\times10^{6}$
 $\mu^+\mu^-$ and $4.4\times10^{7}$ $e^+e^-$ events at
 $\sqrt{s}<1$~GeV were used in this analysis.
The $\pi^+\pi^-$ data sample collected by the CMD-3 detector is higher than that in the
previous CMD-2 experiment and in the ISR measurements of
the $BABAR$\cite{BaBar:2012bdw} and KLOE\cite{KLOE:2008fmq,KLOE:2010qei} experiments, as shown in Fig.~\ref{fpi_stat}. 


\section{Data analysis}

\subsection{Overview}

\begin{figure*}[t]
\centering
\includegraphics[width=0.38\linewidth]{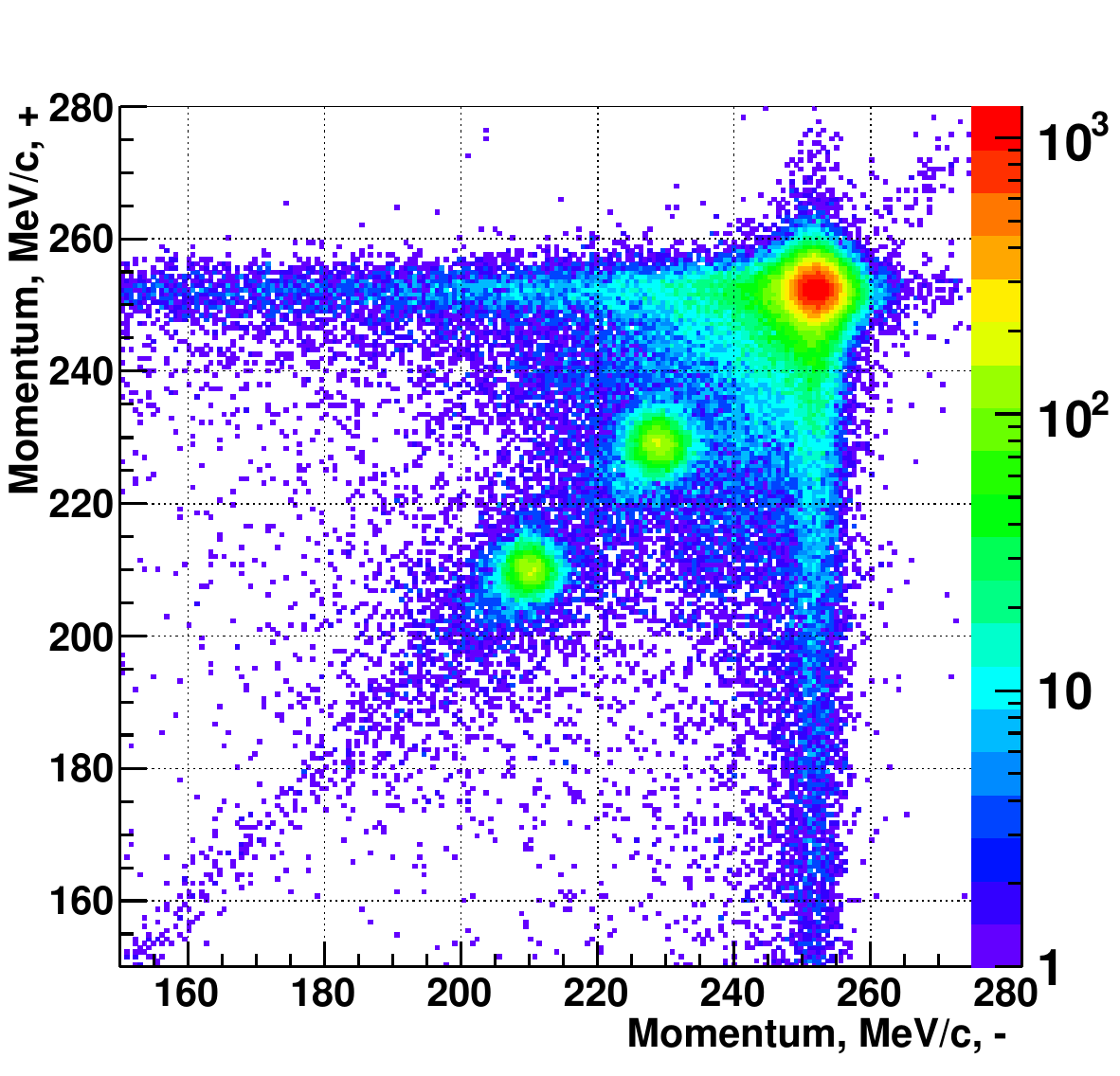}
\hspace{1cm}
\includegraphics[width=0.395\linewidth]{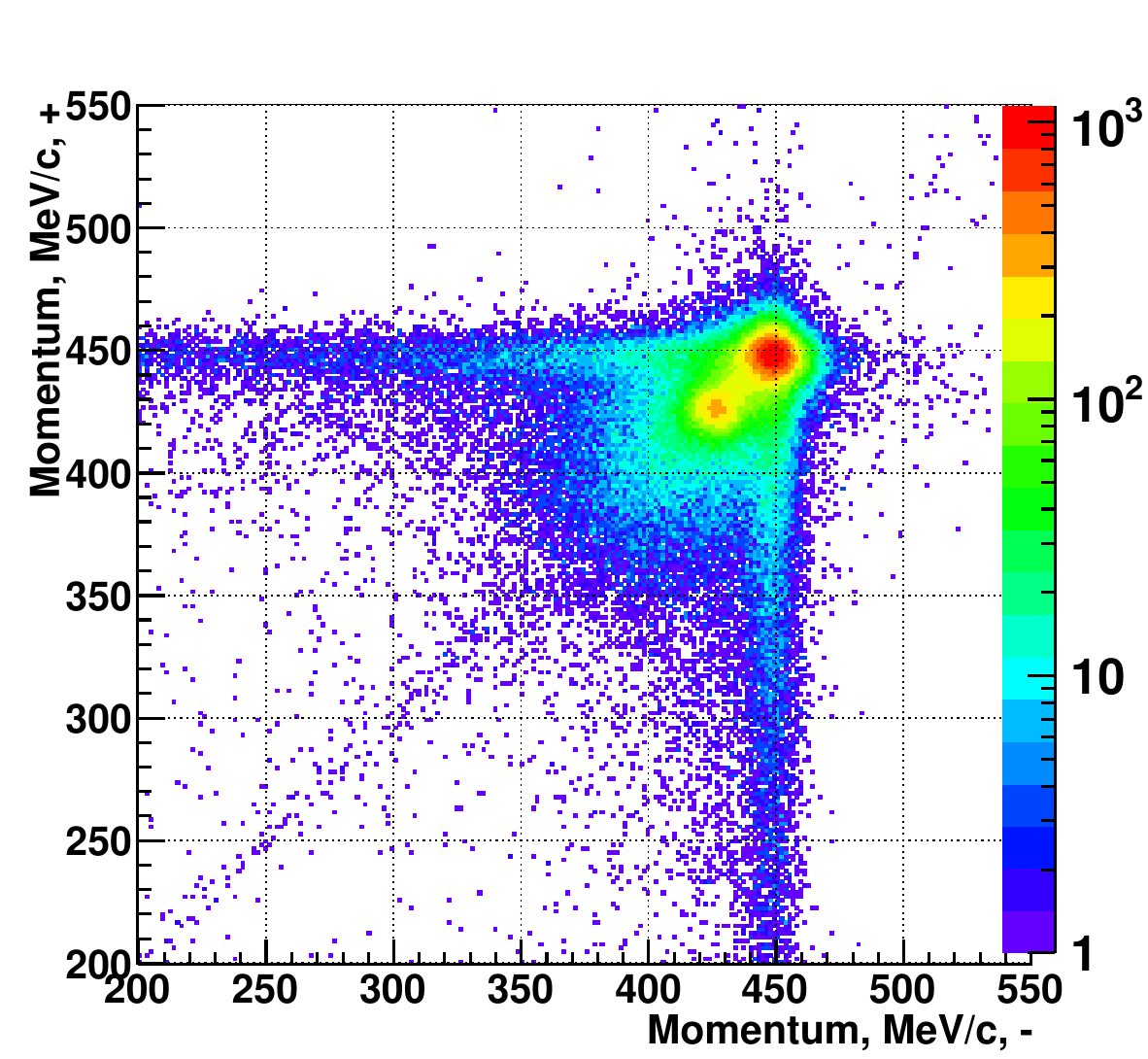}
\vspace{-0.2cm}
\caption{Distribution of the measured momenta in
  the DCH for positive vs negative tracks in the collinear events at the beam energy $E_{beam}$=250~MeV (left) and
  at the highest beam energy $E_{beam}$=450~MeV (right) still used in
  the momentum-based separation. The peaks from left to right
  correspond to $\pi^+\pi^-,\mu^+\mu^-$ and $e^+e^-$ events
  respectively, the tails from main $e^+e^-$ peak  are from the radiative and
  bremsstrahlung loss, the events
  distributed along the diagonal are a cosmic background. 
}
\label{figmomen} 
\centering
\includegraphics[width=0.38\linewidth]{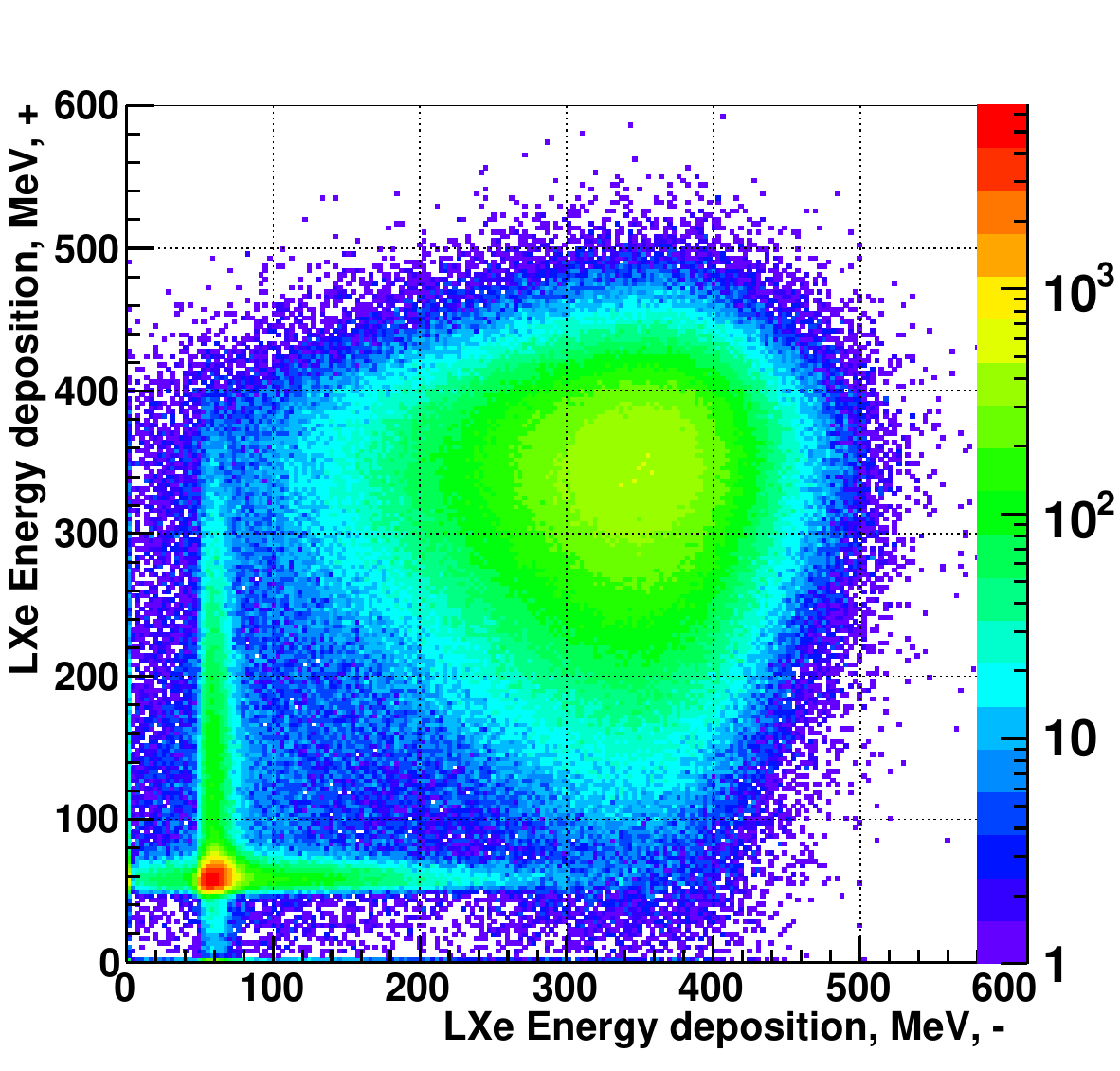}
\hspace{1cm}
\includegraphics[width=0.38\linewidth]{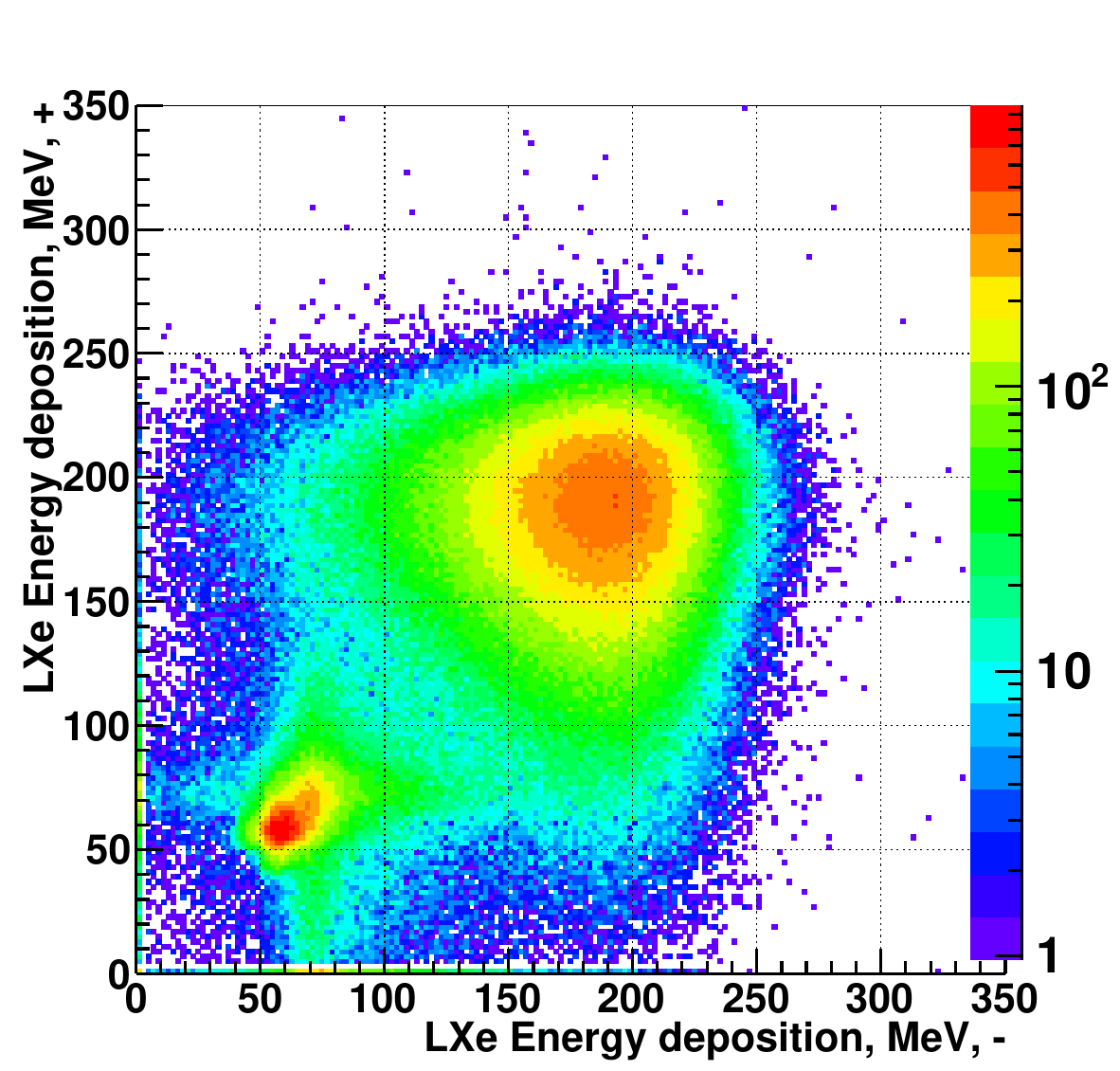}
\vspace{-0.2cm}
\caption{Distribution of the measured energy
  depositions in the LXe calorimeter for positive vs negative tracks in the collinear events at the beam energy $E_{beam}$=478~MeV (left)
  and at the lowest beam energy $E_{beam}$=274~MeV (right) still used in the
  energy deposition-based separation. The peak around 50~MeV
  correspond to minimum ionizing loss (MIP) of  $\pi^+\pi^-$ and $\mu^+\mu^-$, tails
  from the MIP peak come from energy deposition of pions after nuclear
  interaction, the wide distribution on the top right is the
  electromagnetic shower of $e^+e^-$ particles.
}
\label{figmomen2} 
\end{figure*}

The $e^+e^-\to\pi^+\pi^-$ process has a simple event signature with two back-to-back 
charged particles. Such collinear events can be selected by using the following criteria:
two back-to-back well reconstructed oppositely charged tracks are presented
in the detector, originated close to the interaction point, 
registered in a high efficient region of the DCH. The selection cuts
are based only on the information obtained from the DCH and require
presence of two tracks with following conditions:
\begin{subequations}
  \label{eq:cuts}
   \begin{enumerate}[label=\alph*)]
  \item quality of tracks:
\begin{gather}
\chi^2/ndf < 10,\quad N_{hits} \ge 10,
\end{gather}
\item opposite by the charge, close by the track time:
\begin{gather}
    q_1+q_2  = 0,\quad |t^+-t^-| < 20~\text{nsec}, 
\end{gather}
\item close to the beam vertex:
\begin{gather}
    |\rho_{average} = (\rho^++\rho^-)/2| < 0.3~\text{cm},\nonumber\\
    |\Delta\rho = (\rho^+-\rho^-)| < 0.3~\text{cm},  \nonumber \\
    |Z_{average}| < 5~\text{cm},\quad |\Delta Z| < 5~\text{cm},  \label{eq:cutsvtx}
\end{gather}
\item filtration of low momentum tracks and high momentum cosmic background:
\begin{gather}
    0.45E_{\text{beam}}<p^{\pm}<
   \text{min}(E_{\text{beam}}+100~\text{MeV},4./3E_{\text{beam}}),\nonumber\\
    p^{\pm} > 1.15p_{K^{\pm}},  \label{eq:cutsmom}
\end{gather}
\item collinearity:
\begin{gather}
    \left||\phi^+-\phi^-|-\pi\right|<0.15~\text{rad},\nonumber\\
    |\theta^+  - (\pi -\theta^- )|<0.25~\text{rad},   \label{eq:cutscoll}
\end{gather}
\item in the good fiducial volume of the detector: 
\begin{gather}
   1<\theta^{event} = (\theta^+ + \pi -\theta^-)/2<\pi - 1~\text{rad}, \label{eq:cutsth}
\end{gather}
  \end{enumerate}
\end{subequations}
where $\chi^2$ of the track fit and the number of hits on
a track ($N_{hits}$) describe the track quality;
$q_{i}, t^{\pm}, \rho^{\pm}, Z^{\pm}, p^{\pm}, \phi^{\pm},
\theta^{\pm}$ -- the charge, the detection time, the signed impact
distance to the beam axis, $z$-coordinate of the impact point, the momentum,
the azimuthal and polar angles are reconstructed track parameters in the
DCH; $p_{K^{\pm}}$ -- kaon momentum in the $K^+K^-$ process at beam
energy $E_{\text{beam}}$, which is used for an additional kaons
filtration above $\phi$-resonance c.m.~energies.
Momentum and polar angle track parameters were corrected for 
an additional beam and common vertex constraint. This constraint helps us
to improve a momentum resolution, important for the particle separation,
as well as to reduce a possible contribution to the systematic error from the average polar angle cut. 

The selected data sample includes signal $\pi^+\pi^-$ pairs, accompanied by $e^+e^-, \mu^+\mu^-$ pairs 
and single cosmic muons, passed near the interaction point and reconstructed as $\mu^+\mu^-$ pair. 
There is practically no any other significant physical background at energies
$\sqrt{s}<$ 1 GeV.

These final states can be separated using either the information about energy deposition of particles 
in the calorimeter or information about particles momenta, measured in
the drift chamber, as demonstrated in Fig.~\ref{figmomen} and Fig.~\ref{figmomen2}. 
At low energies the momentum resolution of the drift chamber is sufficient to separate different
types of particles. The pion momentum distribution is well separated from the
electron one up to energies of about $\sqrt{s}\lesssim0.9\,\rm GeV$, while the
$\mu^+\mu^-$ events are separated from the others up to $\sqrt{s}\lesssim0.7\,\rm GeV$.
Above this energy the number of muons was fixed relative to the 
number of electrons according to the QED prediction as discussed later
for Eq.~(\ref{eqnmm}) and in Sec.~\ref{sectionradcor}.
At higher c.m.~energies the energy deposition distribution of electrons (positrons) in the calorimeter, 
determined by the electromagnetic shower, is well separated from the similar distributions 
for the minimum ionizing particles (muons and pions). 
In order to have the performance of the separation procedure to be stable in the whole energy range, 
the energy deposition only in the inner LXe calorimeter was used for the particle
separation. It helps to keep the peak positions of energy distributions 
for electrons and pions or muons at larger distance.
Usage of a thinner calorimeter also helps to reduce the probabilities of nuclear interactions of pions 
and muons stops in the calorimeter and makes the distribution of energy depositions narrower 
and more predictable without more complicated structures. 
In this approach it is impossible to separate the minimum ionization part of the
energy deposition signals from $\mu^+\mu^-$ and $\pi^+\pi^-$, 
therefore the ratio of the number of muons to the number of electrons was fixed at all energies. 

The momentum-based particle separation works better at lower energies, 
while the energy deposition-based separation performs the best at higher energies and becomes
less robust at lower energies.
The performance of both methods is nearly matched in the dominant $\rho$-resonance production c.m.~energy range 
(0.6--0.9~GeV), which gives the possibility for the cross
checks between two methods.
As it will be shown further, an additional consistency check comes from the analysis 
of the polar angle distribution, 
which can be used as the third separation method for an independent evaluation 
of the ratio of the numbers of the different final states.

\subsection{Pion form factor determination}

The cross section of the process $e^+e^-\rightarrow\pi^+\pi^-$ can be
written as:
\begin{eqnarray}
\label{fpitosigma}
\sigma_{e^+e^-\to\gamma\to\pi^+\pi^-}=&&
\sigma^{0}_{e^+e^-\rightarrow\pi^+\pi^-}\cdot|F_\pi|^2\nonumber\\
=&&\frac{\pi\alpha^2}{3s}\beta_\pi^3\cdot|F_\pi|^2,
\end{eqnarray}
where $\sigma^{0}_{e^+e^-\rightarrow\pi^+\pi^-}$ is the lowest order cross section of the pointlike pion pair production.
The pion form factor is evaluated from the experimental data using the following
expression:
\begin{equation}
\label{formdef}
\left| F_{\pi }\right| ^{2}=
(\frac{N_{\pi^+ \pi^- }}{N_{e^+e^-}} - \Delta^{bg})
\cdot \frac{\sigma^{0}_{e^+e^-}\cdot (1+\delta_{e^+e^-})\cdot\varepsilon_{e^+e^-}}
{\sigma^{0}_{\pi^+\pi^-}\cdot (1+\delta_{\pi^+\pi^-})\cdot\varepsilon_{\pi^+\pi^-}}
\end{equation}
The ratio $N_{\pi\pi}/N_{ee}$ is obtained from the event separation
procedure, $\Delta^{bg}=N_{bg}^{sim}/N_{ee}^{sim}$ -- the correction for background processes 
(applied only in the energy deposition-based event separation, where only cosmic 
events background are taken into account in the likelihood minimization),
$\sigma^{0}_{i}$ ($i=e^+e^-,\,\pi^+\pi^-$) -- the lowest order cross section of the corresponding pair production in the selected
polar angles, $\delta_i$ -- radiative correction for this cross section within chosen
selection criteria Eq.~(\ref{eq:cuts}),
$\varepsilon_i$ -- detection efficiency which accounts for
a track reconstruction, a trigger efficiency, bremsstrahlung energy loss of electrons,
nuclear interactions of pions, pion loses due to decays and etc.
The detection efficiency is obtained as much as possible from the
experimental data and it is applied in the multiplicative way
$(1-\delta^{trg}_{eff})\cdot(1-\delta^{base}_{eff})\cdot(1-\delta^{Z cut}_{eff})\cdot...\cdot()$ as
corrections to the ``ideal'' detector case with $\varepsilon=1$.
The expression similar to the above one was used to predict the ratio of a number of $\mu^+\mu^-$
pairs to a number of $e^+e^-$ pairs:
\begin{equation}
\label{eqnmm}
\frac{N_{\mu^+ \mu^- }}{N_{e^+e^-}} = 
\frac{\sigma^{0}_{\mu^+\mu^-}\cdot(1+\delta_{\mu^+\mu^-})\cdot\varepsilon_{\mu^+\mu^-}}
{\sigma^{0}_{e^+e^-}\cdot (1+\delta_{e^+e^-})\cdot\varepsilon_{e^+e^-}},
\end{equation}
further, this ratio was fixed in the event separation procedure. When the procedure
allows to measure the number of muon pairs at the c.m.~energies $\sqrt{s} \lesssim 0.7$~GeV, 
the measured ratio is compared to the calculated one as a cross-check.
It should be also noted that the radiative corrections 
$\delta_{ee}$ and $\delta_{\mu\mu}$ include the contribution from the vacuum
polarization in a photon propagator, while in case of the $\pi^+\pi^-$ process
it is included in the definition of the pion form factor.

The typical values of the lowest order cross section ratios are $\sigma^{0}_{\pi^+\pi^-}/\sigma^{0}_{e^+e^-}  =
0.027$ and $\sigma^{0}_{\mu^+\mu^-}/\sigma^{0}_{e^+e^-}=0.081$ at the
$M_\rho$ c.m.~energy within used $1<\theta<\pi-1$~rad polar angles, and it gives
$N_{\pi^+ \pi^- }/N_{e^+e^-}\sim 1.205$ and $N_{\mu^+ \mu^-}/N_{e^+e^-}\sim 0.083$
together with the pion form factor ($|F_\pi|^2\sim 45$ at the $\rho$ peak), radiative corrections and efficiencies.

\subsection{Event type separation}
\label{separationsection}

The numbers of $e^+e^-$, $\mu^+\mu^-$, $\pi^+\pi^-$ pairs in the selected sample of collinear events are determined by
the minimization of the likelihood function constructed for the two dimensional distribution 
of selected observables for two particles in the pair:
\begin{equation}
 -ln L = - \sum_{events} ln \left[ \sum_i N_i f_i(X^+, X^-) \right] + \sum_i N_i,
\end{equation}
where $i$ denotes the event type, $N_i$ is the number of events of
type $i$ and $X^\pm$ denotes the measured observable for the positive/negative track.
Two independent approaches are used, where $X$ is the particle's momentum or $X$ is the particle's energy deposition 
in the LXe calorimeter. Probability density functions (PDFs) $f_i(X^+, X^-)$ describe the corresponding 2D distribution 
for each type of the final state. 

In case of the event separation by particle momentum, 
the construction of PDF functions $f_i(p^+, p^-)$ starts from the ideal momentum spectra for $e^+e^-,
\mu^+\mu^-, \pi^+\pi^-$ events obtained from the MC generators, 
in which the collinear signal selection criteria are applied to the generated events. 
Two independent generators, MCGPJ~\cite{Arbuzov:2005pt} 
and BabaYaga@NLO~\cite{Balossini:2006wc}, were used, as it will be described later 
in Sec.~\ref{sectionradcor}. The ideal generated distributions are then convolved with the
detector response function which accounts for the momentum resolution, the bremsstrahlung 
of electrons at the beam pipe and the inner DCH wall, and the pion decay in flight.
These resolution apparatus functions are defined in a general way,
where the most of their parameters are kept free in the minimization.
This allows to take into account long term changes in the detector
performance. The functional form describing the bremsstrahlung loss is
added by the power function with 2 free parameters reflecting $X/X_{0}$, and the spectra of
reconstructed tracks after the pion decay in flight is fixed from the
full simulation while keeping 2 free parameters to describe ratios
of different types of such events. The consistency checks of particle specific
components inside of the detector response function are described in more
details in Sec.~\ref{ptspecificloss}.

\begin{figure}[t]
\centering
\includegraphics[width=0.5\linewidth,keepaspectratio]{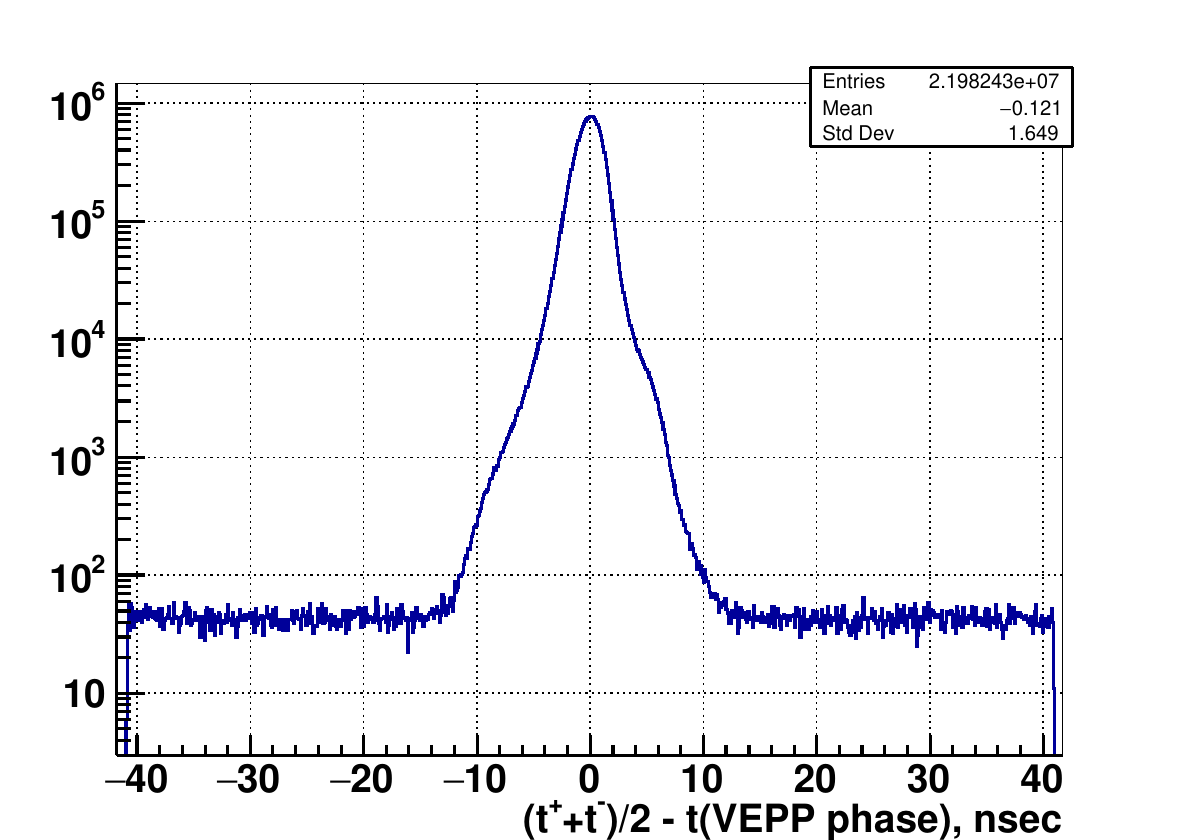}
\caption{Reconstructed time in the DCH of the event as relative to the beam collision time. 
  The uniform baseline corresponds to
  the cosmic events, Gaussian-like peak -- to the events from the beam collisions. 
  Full collinear event sample for the RHO2013 season is shown.}
\label{figtevent}
\end{figure}

The 2D momentum spectra for the cosmic muon events is constructed from
the experimental data by selecting events
with the reconstructed time in the DCH away from the beam collision time, as shown in
Fig.~\ref{figtevent}. The time
resolution of the drift chamber for the two track events is about 1.--1.5~nsec,
which should be compared to the 81.7~nsec time interval between beam
collisions.
The asymmetrical behavior originates from asymmetrical
distribution of variations of the drift times of DCH hits along a traversed track.
The very clean sample of cosmic events can be obtained by applying 
$|t^\text{event}-t^\text{beam}|>20.4$~nsec selection cut.
The data sample over each data taking season provides enough
statistics for accurate evaluation of the cosmic background spectra. 
At threshold energy region $\sqrt{s}\leq0.381$~GeV the detector was operated with reduced magnetic
field for a very limited time, having $B_{field} = 0.65$~T for RHO2013 and 1~T for LOW2020 data taking seasons
instead of 1.3~T normally used.
Different magnetic fields lead to the different momentum resolutions and
even momentum distributions because of a geometrical selection, so it is
necessary to construct the specific PDF function for the reduced magnetic field case.
Increased PDF uncertainty due to the limited statistics of the cosmic data here, 
enhanced by the higher ratio of the numbers of cosmic events to the $\pi^+\pi^-$ events, 
leads to the additional 0.5\% systematic uncertainty to
the pion form factor measurement at these c.m.~energies.
 
Momentum spectra of the $e^+e^-\to\pi^+\pi^-\pi^0$ and $e^+e^-\to e^+e^-e^+e^-,\mu^+\mu^-e^+e^-$ ($4\ell$) 
background events are obtained from the full MC simulation using
corresponding generators as discussed later in Sec.~\ref{bgsection}. Numbers of events for these background
processes passed selection cuts relative to the number of $2\pi$ events are $\sim 0.8\%$ for
$3\pi$ at the $\omega$ peak and $\sim 0.8\%$ for the two-photon four-lepton production events at the lowest energy points.
The actual effect on the measured pion form factor is even smaller as their momentum
spectra are quite different from that of the signal process.
At the other energies the relative number of background events is negligible. 

\begin{figure}[t]
\begin{minipage}[t]{.475\textwidth}
\centering
\includegraphics[width=1.1\linewidth]{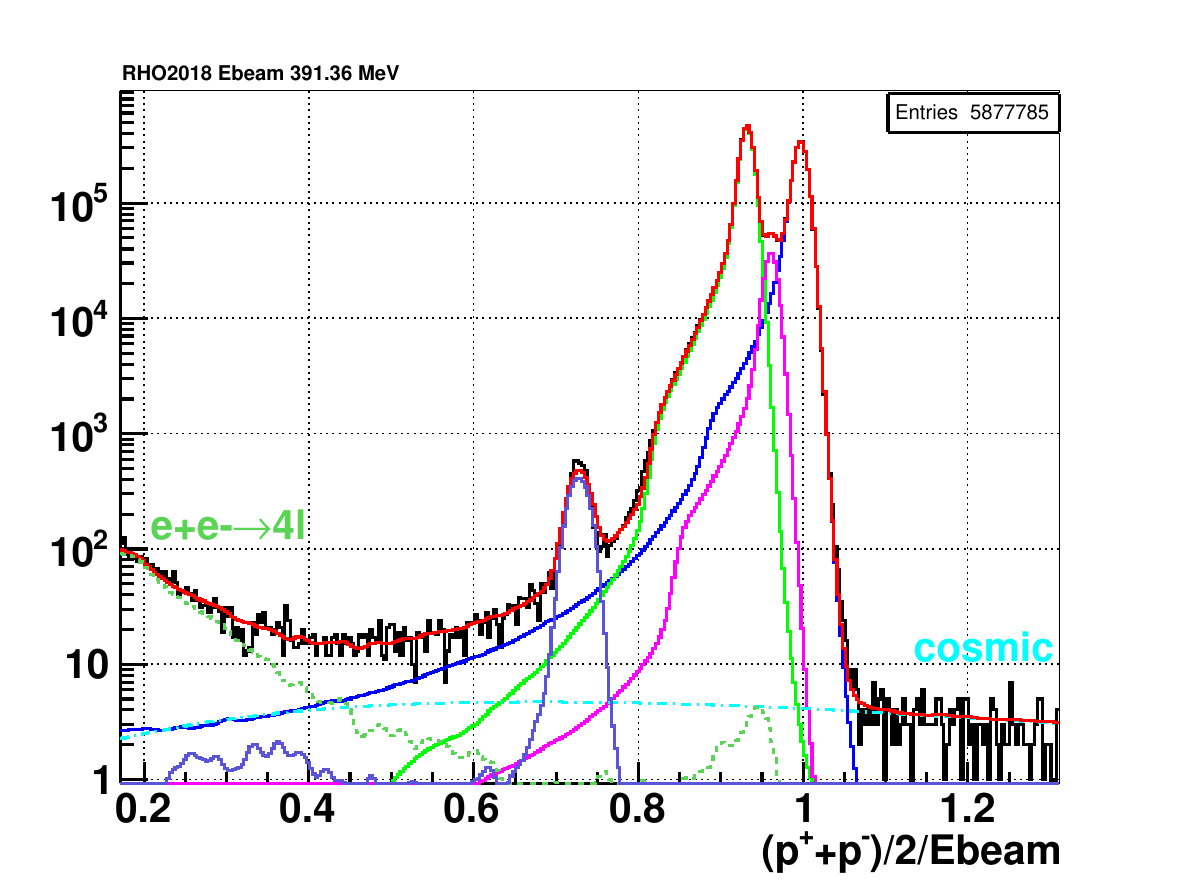}
\caption{Distribution of the average momentum 
  $(P^++P^-)/2$ of the opposite charge particles
  for the events with $|P^+-P^-|/E_{beam}<0.038$, $E_{beam}=391.36\,\rm MeV$, 
  where the black histogram -- data,  
  red line -- fit result after minimization, the other colored
  lines -- projections of different components: the peaks from left to
  right correspond to $3\pi,2\pi,2\mu,e^+e^-$
  events (gray blue, green, magenta and blue respectively), dot-dashed cyan line -- cosmic events, dashed green line -- $e^+e^-\to 4\ell$.}
\label{figpiavmom} 
\end{minipage}
\hspace*{.05\textwidth}
\begin{minipage}[t]{.475\textwidth}
\centering
\includegraphics[width=1.1\linewidth]{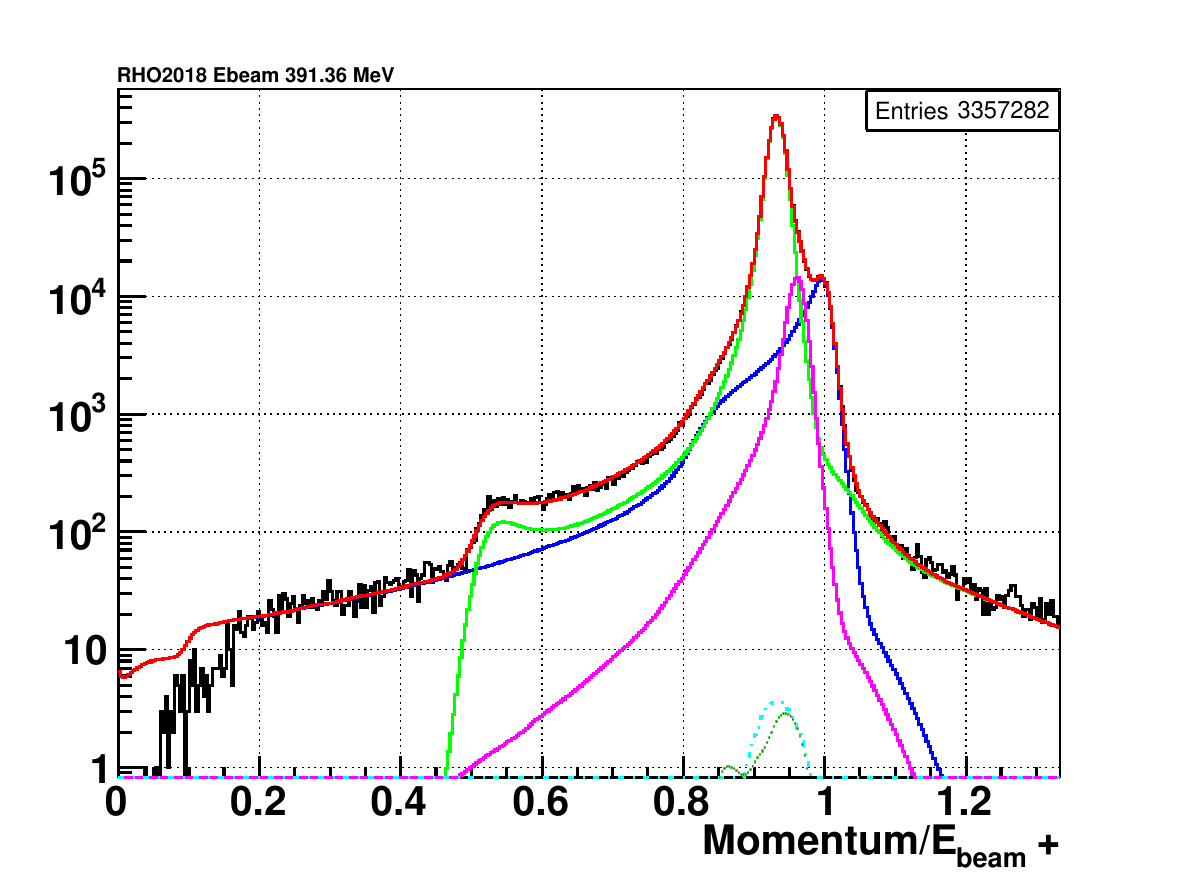}
\caption{Momentum distribution of positively charged
  particles with the momentum of negative within $|P^- - P_\pi| < 10$~MeV, $E_{beam}=391.36\,\rm MeV$,
  where the black histogram -- data,
  red line -- fit result after minimization, the other colored
  lines -- projection of different components: the peaks from left to right --
  $\pi^+ \text{(green)}, \mu^+ \text{(purple)}, e^+ \text{(blue)}$ particles.
}
\label{figpimom} 
\end{minipage}
\end{figure}

The result of the likelihood minimization based on the distributions of particle momenta 
for the $E_{beam}=391.36$~MeV beam energy point is demonstrated in Figs.~\ref{figpiavmom},~\ref{figpimom} for different momentum projections.

In case of the event separation by energy depositions of particles,
the PDF distributions $f_i(E^+,E^-)$ are taken mostly from the experimental data.
The PDFs are assumed to be factorized over two charges: $f_i(E^+,E^-)=f^+_i(E^+)f^-_i(E^-)$.
The possible correlations between $E^+$ and $E^-$ are introduced as
corrections to the main PDF functions.
The primary source of the correlation is the dependence of the calorimeter thickness and, 
correspondingly, the energy deposition on the polar angle of the track. 
For collinear events the inclinations of polar angles for two particles are nearly the same. 
The correlation between $E^+$ and $E^-$ is reduced by applying correction on the energy deposition, 
which takes into account the corresponding dependencies with angle.
Not corrected, this correlation leads to the systematic bias of the ratio
$N_{\pi\pi}/N_{ee}$ of about $\sim 0.25\%$, and after the correction applied the bias is negligible.

The electron PDF is described by a generic functional form, sum of asymmetric Gaussian-like with long tails, 
where all 11 parameters per projection are free during minimization. 
The small correlation coming from the double initial photon radiation, 
which gives correlated tail toward lower momenta of both particles, 
was evaluated from the simulation.
The corresponding correlated terms, describing this tail behavior
along $P^+\sim P^-$ momenta of the opposite charge particles, 
was added to the $f_{ee}(E^+,E^-)$ with fixed parameters. 
Accounting of this correlation leads to about 0.2\% effect on the
measured pion form factor in the central $\rho$-meson peak c.m.~energy region and
decreasing with higher energies.

The muon PDF, as for minimum ionizing particles (MIP),
is taken from the clean cosmic sample selected by the time of the event as
was described before. In case of the $\mu^+\mu^-$ process, an additional
momentum selection was applied to the full cosmic sample to have momenta
in the range of $(0.9\hyph1.1)p_{\mu}(E_{beam})$, where $p_{\mu}(E_{beam})=\sqrt{E^2_{beam}-m^2_\mu}$.
Number of cosmic muons was estimated 
from the time distribution for each energy point according to
Fig.~\ref{figtevent} by rescaling from the sidebands ($20.4\,\text{nsec}<|t^\text{event}-t^\text{beam}|<40.8\,\text{nsec)}$ 
to the signal region ($|t^\text{event}-t^\text{beam}|<20.4\,\text{nsec}$)
assuming the flat distribution, and was fixed for the likelihood minimization.

The pion PDF is described as a sum of several contributions.  
The long tails of the energy deposition of pions, which undergo nuclear interactions in the detector, 
are described by a generic monotonous function as sum of 
Gaussians with constant width and fixed positions distributed
uniformly up to the right
edge of energy deposition tail and with varied heights, 5 free parameters per projection.
The functional form for the MIP pions, 
which passed calorimeter without nuclear interactions, was taken the 
same as for the muons with a released set of parameters (such as the average energy, width etc.).

The MIP parts of the energy deposition of $\pi^+\pi^-$ and $\mu^+\mu^-$
events overlap strongly, which may introduce a significant systematic
effect on the retrieved number of events after the minimization due to imprecise descriptions of the overlapping PDFs.
In order to reduce this effect, the flat 2D-PDF functions for all particle types were used
in the range of $10<E^+ \& E^-<100$~MeV. The flat PDF does not rely
on a distribution of events underneath and gives the exact number of events
as fitted normalization after the likelihood minimization,
which effectively accounts the number of particular events in this region in the unbiased way.


Background events, which will be described in the next Sec.~\ref{bgsection}, 
mostly contribute to the signal $\pi^+\pi^-$ events 
(some of them give the same response and  
some ones directly contribute to the energy deposition signal region), 
except for the $e^+e^-\to e^+e^-e^+e^-$ process which contributes mostly to the $e^+e^-$ events.
Thus, in contrast to the fit of the momenta distribution, 
the noncollinear background terms (except for the cosmic events) were not added 
to the likelihood function, 
but rather the ratio $N_{\pi\pi}/N_{ee}$ was corrected for the background after
the minimization as $\Delta^{bg}=N_{bg}^{sim}/N_{ee}^{sim}$
correction according to Eq.~(\ref{formdef}).


The full likelihood function has 36 or 56 free parameters in total to be adjusted during the minimization
for the momentum-based or energy deposition-based approaches respectively. 
Most of these parameters describe the details of the detector response functions 
in a functionally general way. 
The separation based on the momentum information is used for the energies up to 
$\sqrt{s}\leq 0.9$~GeV, while the separation based on the energy deposition information is
used starting from $\sqrt{s}\geq 0.54$~GeV.

\begin{figure}[t]
\centering
\includegraphics[width=0.6\linewidth,keepaspectratio]{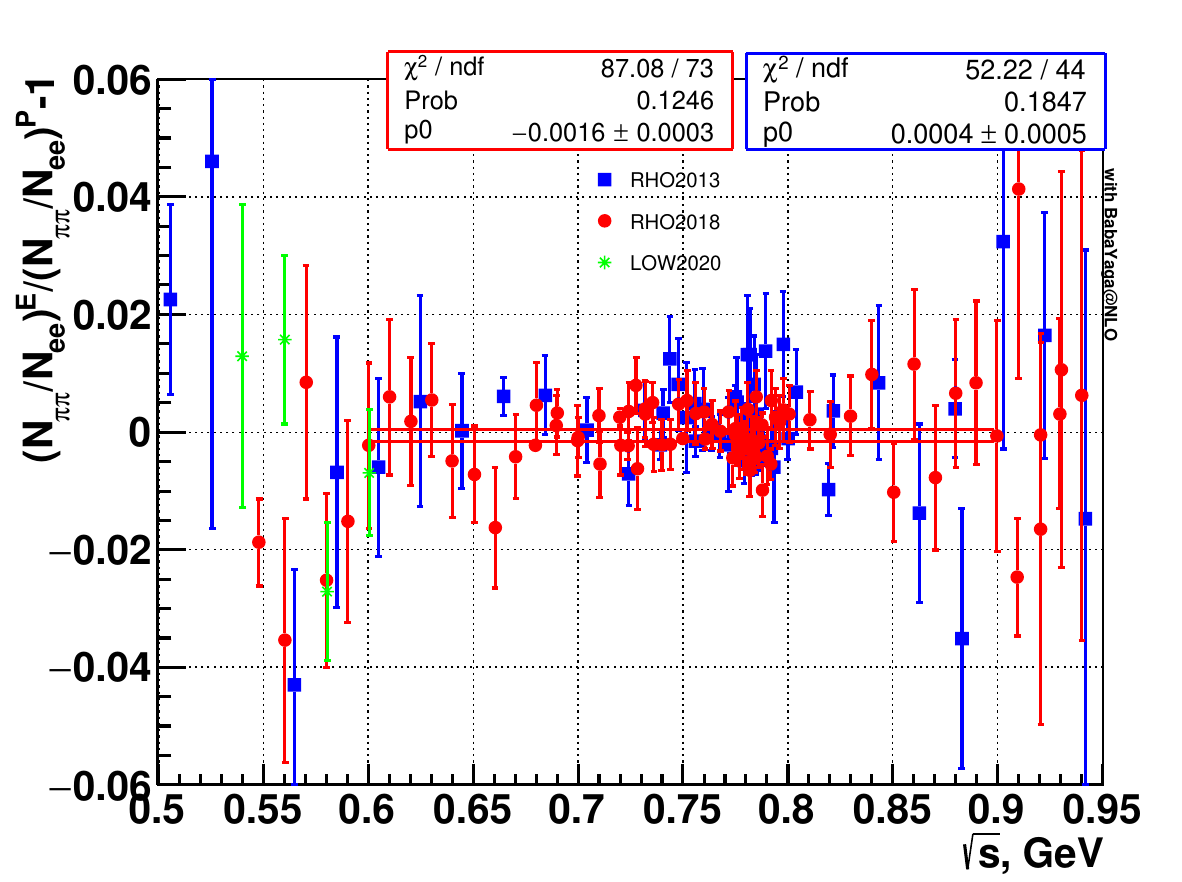}
\caption{Comparison of the results of the momentum-based and energy deposition-based event separations. 
         The different data taking seasons are shown separately. The
         average differences between both separation methods are shown
         as the fit parameters for each season.}
\label{figPvsE}
\end{figure}

It is known that the maximum likelihood method could provide a biased
estimation of the parameters, introducing some systematics.
For both separation methods possible biases were studied with the
help of the full simulation using mixed statistics of different
processes of the same size as for the data. 
In the $\rho$-meson resonance c.m.~energy region the systematic uncertainty of the obtained
$N_{\pi\pi}/N_{ee}$ ratio is estimated to be below 0.2\% for both methods. At the higher energies, 
the overlap of the momentum distributions of electrons
and pions increases, which degrades the momentum-based
separation capability and leads to the growth of the systematic error 
from 0.2\% at $\sqrt{s}=0.8$~GeV to 1.5\% at $0.9$~GeV. 
On the contrary, the distributions of energy deposition of electron and MIPs start 
to overlap significantly at lower energies, as can be seen in Fig.~\ref{figmomen2},
when the electron peak position is shifted below 200~MeV at lower energies.
At the same time muons start to stop in the LXe
calorimeter at the energies below $E_\mu < 200$~MeV and pions stop at $E_\pi < 250$~MeV. 
The distribution of energy deposition of the stopped particles becomes very distorted 
and it extends well beyond the MIP 2D-PDF box.
All these effects degrade the energy deposition-based separation
capability leading to the growth of the estimated systematic uncertainty from 0.2\% at $\sqrt{s}=0.6$~GeV to 2\% at $0.54$~GeV. 
The observed biases of the $N_{\pi\pi}/N_{ee}$ ratios after the
likelihood minimization on the simulated data are not
applied to the pion form factor, they are considered as a part of the systematic uncertainty.
The comparison of $N_{\pi\pi}/N_{ee}$ ratios obtained by two
separation methods is shown in Fig.~\ref{figPvsE}.
Only specifically for this plot, for a demonstrative purpose,
the result of the momentum-based separation was corrected above 0.82~GeV
according to the simulation, with growing bias up 1.5\% at 0.9~GeV as mentioned above.
Figure \ref{figPvsE} shows the comparison for the case when 
the electron's and muon's PDFs for the momentum-based separation 
were constructed using the BabaYaga@NLO MC generator as input. 
It can be seen that two fully independent separation methods are
well compatible within $\leq0.2\%$, and the average
difference between both methods is $(-0.12\pm0.03)\%$ in the central 0.7--0.82~GeV c.m.~energy range.
Looking at the right side of the plot, good compatibility of the two separation methods 
on the level below $1\%$ is seen (the fit of the data in the c.m.~energy region $\sqrt{s}=0.85\hyph 0.95$~GeV 
gives the average difference of $(-0.22\pm 0.34)\%$~).
The mentioned above bias raise in $(N_{\pi\pi}/N_{ee})^{P}$
up to $\sim-1.5\%$ at $\sqrt{s}\sim0.9$~GeV for the momentum-based separation minimization 
was already corrected for this plot from the MC simulation.
This also indicates good agreement between 
experimental data and MC simulation for the bias c.m.~energy dependence.

For the final result, the experimental ratios $N_{\pi\pi}/N_{ee}$
together with their separation-related statistical uncertainties
(excluding the common Poisson part of errors)
were averaged between two independent separation methods in the $0.54-0.9$~GeV c.m.~energy range with weights, equal to the 
corresponding separation-related systematic uncertainties inverse squared.
The merged values were used for the pion form factor determination.
Out of the overlap region, it was used either 
the momentum-based separation at lower energies or the energy deposition-based separation
at higher energies.

A further method, which is out of the scope of this analysis, is under
development to exploit the full power of the layered
barrel calorimeter. The available independent measurements in 7~strip layers of the LXe calorimeter, 
energy deposition in the CsI calorimeter and transversal cluster sizes give a better discrimination 
power between different particles due to the different interaction processes involved (electromagnetic
shower, ionization process, nuclear interaction). The analysis based on the full
information from the combined calorimeter is planned to be
the major method for the pion form factor measurement above 1~GeV,
where the ratio of the $\pi^+\pi^-$ events to the others is strongly decreasing with the c.m.~energy.

%

\subsection{Background subtraction}
\label{bgsection}

\begin{figure}[t]
\begin{minipage}[t]{.495\textwidth}
\centering
\includegraphics[width=1.04\linewidth]{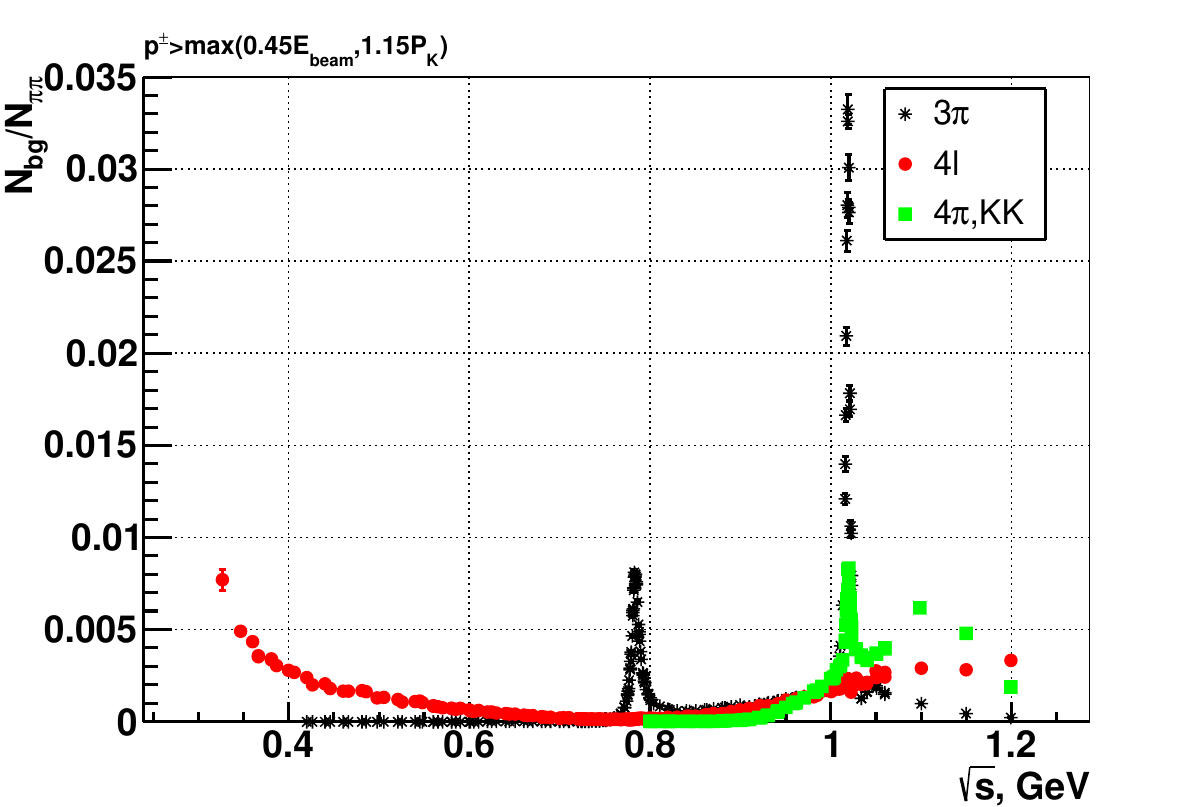}
\caption{Background contributions in selected collinear events as
  ratio to number of $\pi^+\pi^-$ events.}
\label{figbg}
\end{minipage}\hfill
\begin{minipage}[t]{.495\textwidth}
\centering
\includegraphics[width=1.\linewidth]{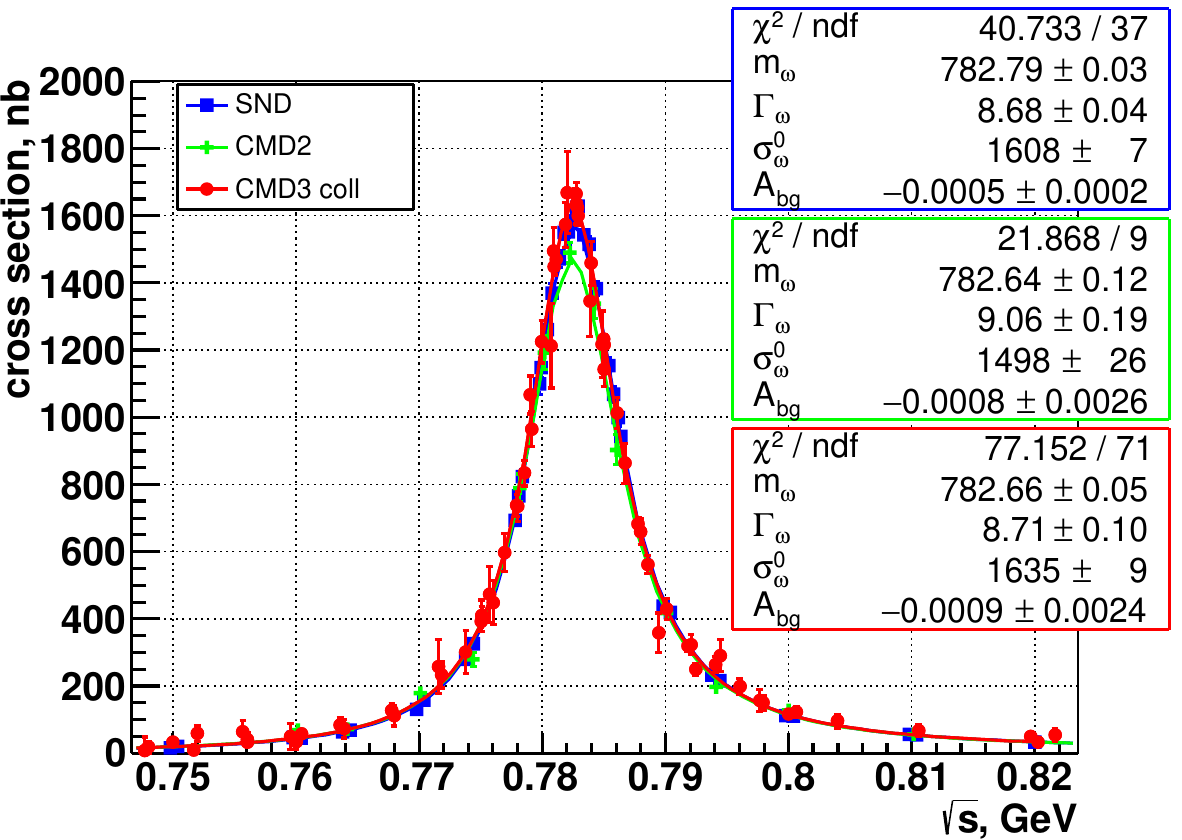}
\caption{$e^+e^-\to \pi^+\pi^-\pi^0$ cross section extracted from
  collinear events in this analysis (red markers, fitted line and fit
  result box) in comparison with CMD-2 (green) and SND
  (blue) results.}
\label{fig3pi} 
\end{minipage}
\end{figure}

Possible background sources were tested 
by the multihadronic Monte Carlo generator \cite{Czyz:2013sga}, which
covers most of the $e^+e^-\to X$ processes properly weighted by their 
experimental cross sections.
The reactions of $e^+ e^-$ annihilation with the production of the final states:
$\pi^+\pi^-\pi^0$, $K^+K^-$, $K_{S}K_{L}$, 
$2\pi^+2\pi^-,\pi^+\pi^-2\pi^0$, as well as QED two-photon $e^+e^-e^+e^-$ and $\mu^+\mu^-e^+e^-$ 
production processes  contribute to the $\pi^+\pi^-$ background. Their contributions were evaluated 
by the dedicated MC generators: $3\pi$ according to the $\rho\pi$ model,
kaon production -- using the MCGPJ generator~\cite{Arbuzov:2005pt},
four pion events -- with the most advanced existing model including various
intermediate components~\cite{Kozyrev:2019ial}, four lepton events --
by the DIAG36 generator~\cite{Berends:1986ig}. 

The full MC simulation of the detector was performed.
The number of the simulated events of the particular background 
process normalized to the number of simulated Bhabha events in the used the collinear selection criteria
was estimated based on the measured cross section of the process 
and the calculated Bhabha cross section. 
In such an approach the common inefficiencies are partially canceled out. 
The ratios of the numbers of events for various background processes to the number of $\pi^+\pi^-$ events
are shown in Fig.\ref{figbg}. The largest background is observed at the c.m.~energies around
$\phi$-meson peak with the following contribution from the different final states:
$3\pi$ -- 3.32\%, $K^+K^-$ -- 0.29\%, $K_{s}K_{L}$ -- 0.07\%, $2\pi^+2\pi^-$ -- 0.14\%, 
$\pi^+\pi^-2\pi^0$ -- 0.21\%, $e^+e^-e^+e^-$ -- 0.12\%, $\mu^+\mu^-e^+e^-$ -- 0.11\%.
The dominant $3\pi$ cross section is known with the precision better than 5\%.


The momentum distribution of $e^+e^- \to 4\,\text{lepton}$ events only 
partially overlaps with the distribution of the signal $\pi^+\pi^-$ events.
This background contributes mostly to the cosmic
or $\mu^+\mu^-$ events as shown in Fig.~\ref{figpiavmom}. 
Implementation of the corresponding PDF functions to the momentum-based likelihood function 
results in a $0.2\hyph0.05\%$ change of the extracted $N_{\pi\pi}/N_{ee}$
ratio for energy points in the range of 0.32--0.38~GeV and negligible in the others.

Since the momentum distribution for the $3\pi$ events is well separated from the distributions 
of the other final states (as seen in Fig.~\ref{figpiavmom}), 
the number of $3\pi$ events is also extracted from the momentum-based likelihood fit.  
Such an approach allows to obtain the $\sigma(e^+e^-\to \pi^+\pi^-\pi^0)$ 
cross section as a byproduct of this analysis based on the sample of
the selected collinear events. 
It should be noted that this measurement is based on a small subset of full $3\pi$ 
sample since the cuts for the collinear events are applied. 
The $3\pi$ cross section is calculated using the simplified
efficiency correction, based on the efficiency analysis for the $2\pi$ events, as follows:
\begin{multline}
\label{3picross}
  \sigma(e^+e^-\to \pi^+\pi^-\pi^0)= \\
  \frac{N_{\pi^+\pi^-\pi^0}}{N_{e^+e^-}}\times
  \frac{\sigma^{0}_{e^+e^-}\cdot (1+\delta_{e^+e^-})}{
    (1+\delta_{3\pi})}\times(1+\delta^{en.spr.}) \times 
  \left(\frac{\varepsilon_{e^+e^-}}{\varepsilon_{3\pi}^{ISR}}\right)^{sim}
  \times
  \frac{(\varepsilon_{e^+e^-}/\varepsilon_{\pi^+\pi^-})^{exp}}
  {(\varepsilon_{e^+e^-}/\varepsilon_{\pi^+\pi^-})^{sim}}.
\end{multline}
In this approach we take into account: a difference in the pion specific losses
between $3\pi$ and $2\pi$ events, radiative corrections $\delta_{3\pi}^{ISR}$ and $\delta_{e^+e^-}$ 
and the beam energy spread correction $\delta^{en.spr.}$.
The beam energy spread of about 200~keV results in at most 0.42\% correction
at the peak of $\omega$-meson. The radiative correction for the
$3\pi$ process includes only the initial state radiation (ISR)
according to works~\cite{Kuraev:1985hb,WorkingGrouponRadiativeCorrections:2010bjp}, and
the final state radiation (FSR) is included in the measured cross section.
The detection efficiency of the $3\pi$ process ($\varepsilon_{3\pi}$)
includes the solid angle acceptance factor, in
contrast to the other efficiencies in Eqs.~(\ref{3picross}),~(\ref{formdef}),
which are defined in the polar angle range of $1<\theta<\pi-1$~rad.

The measured $3\pi$ cross section in the region of the $\omega$-resonance is shown 
in Fig.~\ref{fig3pi} together
with the CMD-2~\cite{CMD-2:2000ixx,CMD-2:2003gqi} and SND~\cite{Achasov:2003ir} results.
The fit of the cross section by the same function with the contributions 
from the $\omega$ and $\phi$ meson resonances as well as the nonresonant background
is performed for all experiments.
Free parameters of the fit are $m_\omega,
\Gamma_\omega,\sigma^{0}_\omega$ and $A_{bg}$ as shown in Fig.~\ref{fig3pi}, while the other parameters are 
fixed at their world average values from the PDG2022 compilation\cite{Workman:2022ynf}. The peak
cross section corresponds to the branching fraction of the $\omega$ decay via
$\sigma^{0}_\omega=12\pi(\mathcal{B}_{\omega\to e^+e^-}\cdot \mathcal{B}_{\omega\to 3\pi})/m_\omega^2$, where both $m_\omega$ and
$\sigma^{0}_\omega$ parameters are obtained from the fit.
The fit of data yields the cross section peak value
$\sigma(e^{+}e^{-}\to\omega\to 3\pi) = 1635 \pm 9 \pm 54$~nb
(or $\mathcal{B}_{\omega\to e^+e^-}\cdot \mathcal{B}_{\omega\to 3\pi}=(6.82\pm 0.04\pm 0.23) \times 10^{-5}$)
\label{omega3pi}, 
which is in good agreement with the SND measurement $1608 \pm 7 \pm 55$~nb
and about $2.2\sigma$ away from the CMD-2 result $1498 \pm 26 \pm 19$~nb. 
The obtained result also agrees well within the systematic uncertainty
with the results by BESIII~\cite{BESIII:2019gjz} and $BABAR$~\cite{BABAR:2021cde} experiments.
The main sources of the systematic uncertainty specific for the $3\pi$ process are: 
a possible difference of the track reconstruction inefficiencies
for the $2\pi$ and $3\pi$ samples (0.5\%),
a limited MC statistics used for the $3\pi$ PDF description (2.0\%),
an uncertainty of the $\rho\pi$ model used in the $3\pi$ MC generator to extract selection efficiency (2.4\%).
Some sources with the smaller systematic uncertainty contributions are the same as for the $\pi^+\pi^-$ process, 
they will be summarized later in Table~\ref{systfpi}.
The total systematic uncertainty of the $3\pi$ cross section in this analysis is
estimated as $3.3\%$ at the $\omega$ c.m.~energies.

The model uncertainty was estimated as a possible contribution from
the $\rho'\pi$ intermediate state in the vector meson dominance (VMD) description.
It comes from the $\omega(1650)$ decay
according to the paper~\cite{SND:2020ajg}, and it should be suppressed at
c.m.~energies near the $\omega(782)$ mass. A non-$\rho\pi$ state at the $\phi$-meson resonance energies was measured 
in the papers~\cite{Achasov:2001hb,Achasov:2003ir,KLOE:2003kas,Akhmetshin:2006sc} and the following estimation 
of the normalized value of the contact amplitude was obtained:
$a = 0.10 \pm 0.002$ (Table 3 from ~\cite{Akhmetshin:2006sc}). This corresponds to the
$\sigma_{\rho'\pi}/\sigma_{\rho\pi} < 0.025$ ratio at the $\omega$ resonance, if the same
contact term is taken and no additional suppression between $\phi$ and $\omega$ energies is applied.
The additional $\rho'\pi$ amplitude could interfere with the main $\rho\pi$
with different relative phases, the worst case leads up to about 2.4\% possible change 
in the total efficiency of collinear events
selection [Eq.~(\ref{eq:cuts})] for $3\pi$ events.
More general approach to describe three-body decay dynamics using the so-called
Khuri-Treiman (KT) dispersion relations was discussed in the papers~\cite{Niecknig:2012sj,JPAC:2020umo,Stamen:2022eda}.
The Dalitz plot of $\omega\to3\pi$ decay have been studied
by the WASA-at-COSY~\cite{WASA-at-COSY:2016hfo} and
BESIII~\cite{BESIII:2018yvu} experiments.
Using fit results of the BESIII measurement and of theoretical
dispersive analysis of the $3\pi$ decay amplitude by the JPAC collaboration
(2 and 3 parameters fit results in Table 1 from~\cite{JPAC:2020umo})
lead to a decrease of the collinear events selection efficiency by
$0.7\hyph1.9\%$ relative to using of the $\rho\pi$ module. This
numbers are within the estimate given above using the simplified VMD consideration.

The effect from the inaccuracy of the $3\pi$ PDF description was estimated using different parametrizations.
The variation of the constructed PDF gives the
consistent changes in the fit results for both data and MC simulation.
Thus, the related correction was taken according to
the simulation, and the value of the correction is conservatively considered as
the corresponding systematic uncertainty.

While the most of the contributions to the $3\pi$ cross section systematic uncertainty can be improved with
more advanced treatment within the same analysis procedure, but the dominant model uncertainty
comes from a limited knowledge of the possible intermediate states of the $3\pi$
production and will be improved in a dedicated analysis of the full $3\pi$ event sample. 


To test the accuracy of the background subtraction in the $\pi^+\pi^-$ process,
the more stringent cut on minimal momentum was applied: 
$p^{\pm}>0.6E_{\text{beam}}$ (instead of $0.45E_{\text{beam}}$) 
and $p^{\pm}>1.2p_{K^{\pm}}$ (instead of $1.15p_{K^{\pm}}$).
It reduces the numbers of background $3\pi$ events by
$\sim 30\%/50\%$ at the $\omega/\phi$-meson peaks, respectively, 
and more than by a factor of 5 for the other channels around the $\phi$-meson resonance c.m.~energies.
The pion form factor obtained with the new cuts, changes by $(-0.020\pm 0.004)\%$ and
$(+0.05\pm 0.01)\%$ in average
at the c.m.~energy points near $\omega$ and $\phi$ meson resonances, respectively. 
This ensures that the background estimation itself is consistent at the level of $\lesssim 5\%$.
Thus, the corresponding background knowledge contribution to the systematic error of the pion form
factor measurement is estimated as 0.05\% at
the $\omega$-meson peak, 0.2\% at the $\phi$-meson peak.
The systematic error increases linearly from
0 to 0.15\% for the $\sqrt{s} = 0.9 \hyph 1.2$~GeV outside the $\phi$-meson resonance.

\section{Detection efficiency}
\subsection{Selection efficiency}
\label{sectioneff}
The sample of collinear events used in this analysis is selected 
by only the tracking information from the drift chamber in accordance with Eq.~(\ref{eq:cuts}).
Assuming that the reconstruction of the drift chamber 
data and calorimeter data are independent, it becomes possible to measure the efficiency 
of the tracking-based selection criteria by using a sample of the test events independently selected 
in the barrel calorimeter.
For this purpose, the test sample of collinear events was preselected with only two
back-to-back clusters in the barrel calorimeter in the same
fiducial volume as for the primary sample:
\begin{gather}
  |||\phi^{cl}_{1}-\phi^{cl}_{2}|-\pi|-\Delta\phi_{B}^{e/\mu/\pi}(\theta_{av}) |<0.1~\text{rad},  \nonumber \\
  |Z_{average}^{clusters}~\text{or}~Z_{good}^{track}| < 5~\text{cm}, \nonumber \\
   1<(\theta_1^{cl} + \pi -\theta_2^{cl})/2<\pi - 1~\text{rad}, \label{eq:testcuts}
\end{gather}
where $\phi^{cl}, \theta^{cl}$ -- the azimuthal and polar angles of the
detected clusters, $Z_{average}^{clusters}=(Z_{1}^{cl}+Z_{2}^{cl})/2$ --
the average $z$-coordinate of the clusters.   
The $\Delta\phi_{B}^{e/\mu/\pi}$ correction describes the particle deflection 
in the magnetic field of the detector. 
In order to suppress the $3\pi$ background, it is
required that there are no additional energetic clusters in the end cap part.
To reduce the contamination from other backgrounds, 
the presence of at least one good track (by the
$N_{hits}$ and $\chi^2$) originated from the beam interaction point 
and connected to one of the
clusters by the angle and the impact point on the inner surface of the calorimeter is required. 
The correlated losses of both tracks were studied separately, as will be described below.

\begin{figure}[t]
\begin{minipage}[t]{.495\textwidth}
\centering
\includegraphics[width=1.\linewidth]{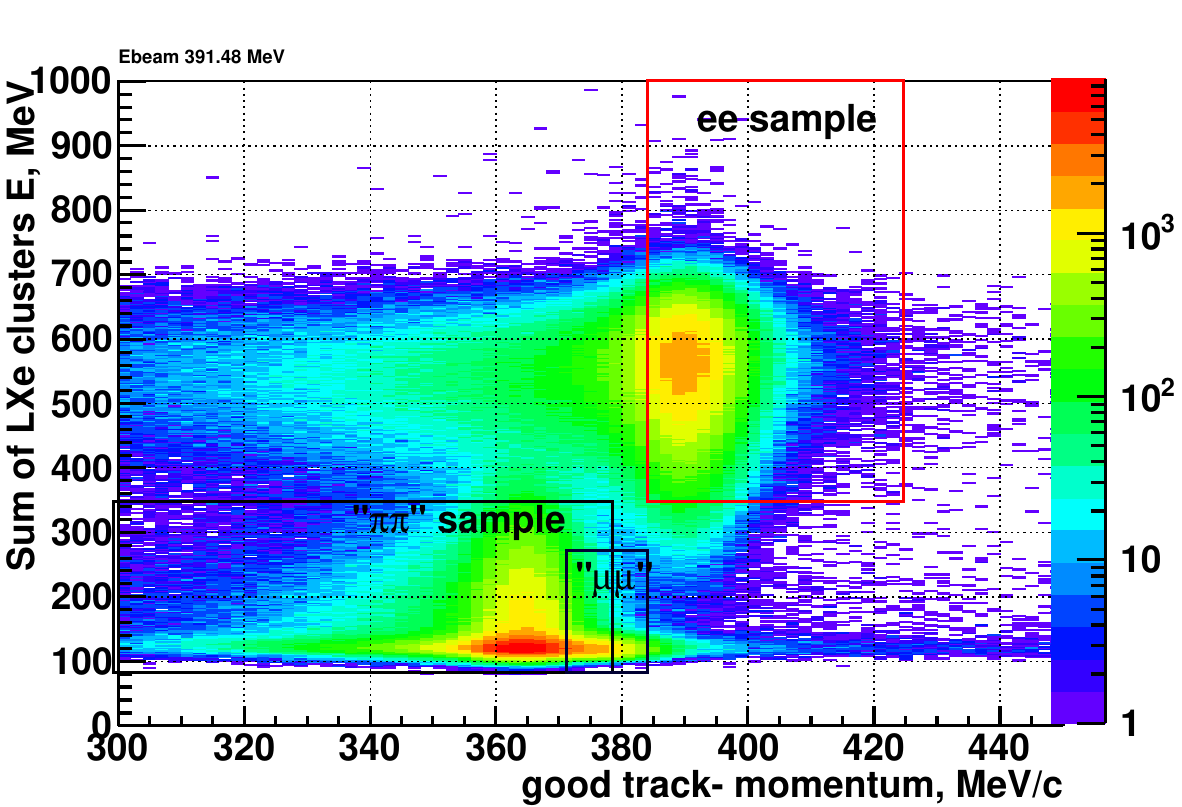}
\caption{Test event types selection for the efficiency determination,
  the $E_{beam}=391.48$~MeV beam energy point.}
\label{figtestsel}
\end{minipage}\hfill
\begin{minipage}[t]{.495\textwidth}
\centering
\includegraphics[width=1.\linewidth]{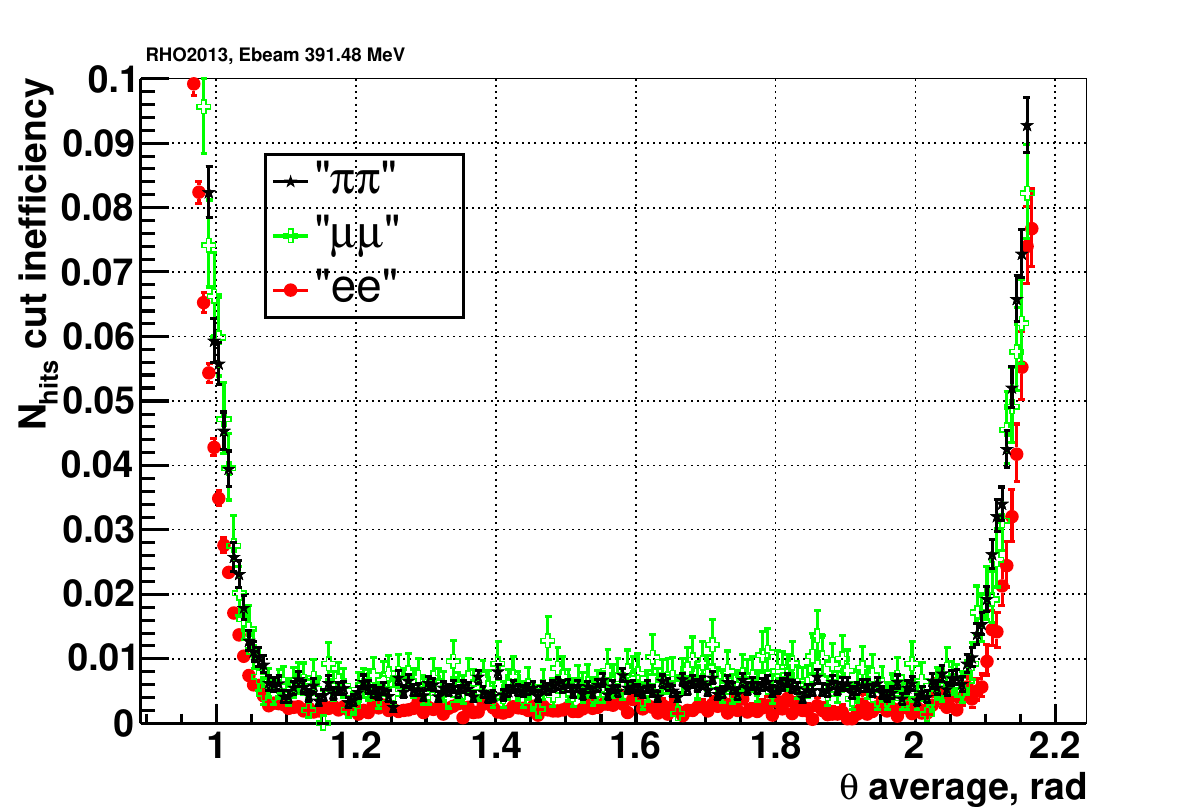}
\caption{Event inefficiency due to $N_{nhits}$ selection with polar
  angle of event, ($E_{beam}=391.48$~MeV (RHO2013)).}
\label{fignhiteff} 
\end{minipage}
\end{figure}

Then, this test event sample is subdivided into three type classes
according to the energy deposition and momentum of the good track as
shown for the $E_{beam}=391.48$~MeV beam energy point in Fig.~\ref{figtestsel}.
This gives a clean sample of Bhabha events and mixed sample of ``MIP'' 
events ($\pi^+\pi^-$ and $\mu^+\mu^-$).
At lower energies $\sqrt{s}\lesssim 0.7$~GeV the energy-momentum selection 
also allows to separate the samples of $\pi^+\pi^-$ and $\mu^+\mu^-$ events.

For the purpose of the efficiency analysis the selection cuts Eq.~(\ref{eq:cuts}) are 
split into the basic set, to find compatible tracks with $\Delta \rho$, $\Delta Z$ and charge cuts, and all the rest. 
The efficiency of the basic set (base efficiency in the following discussion) 
was directly measured using the test sample as the probability
to find two good tracks in the test event, where the both tracks should be compatible by the basic set of cuts. 
The efficiencies of the remaining cuts were studied separately. 

The $Z_{average}$, $\rho_{average}$ and $\Delta t$ cuts inefficiencies
were studied using the data itself.
The typical resolution of the impact distance of a track to a beam axis $\sigma_\rho$
is about 0.25--0.3~mm for a single track, 
therefore the cut on the average impact parameter $\rho_{average} = (\rho^++\rho^-)/2 <3$~mm is very weak, 
designed mainly to reduce cosmic events background.
The cut on the $\Delta\rho$ suppresses the long resolution tails of the $\rho_{average}$ distribution.
The remaining inefficiency of the applied cuts on the $\rho_{average}$,
beyond included in the base efficiency and
which could arise when both tracks have same value on long tails in the $\rho^\pm$ distributions,
is negligibly small.

The same thing is for the $|\Delta t|<20$~nsec cut, the main purpose of which is to have
possibility to select the clean sample of cosmic events by the time of an event $t_{average}$ 
as shown in Fig.~\ref{figtevent} (or, reversely, to reduce in addition the cosmic background).
The core resolution of the time of the track $t_{trk}^{\pm}$ is about 2~nsec and
the measured inefficiency of the $\Delta t$ selection is at the level of $(0.15-0.5)\%$ that comes only from the tails of
the time distribution. Moreover, the difference 
between the $\pi^+\pi^-$ and $e^+e^-$ inefficiencies, which is only relevant for the $|F_{\pi}|^2$ determination 
according to Eq.~(\ref{formdef}), is at the level of $(0.05-0.15)\%$.

The $\theta^{event}$, $\Delta\theta$, $\Delta\phi$ and momentum cuts 
determine the lowest order cross section $\sigma^0$ and the radiative
corrections $(1+\delta)$ calculation in Eq.~(\ref{formdef}). 
Any bias in these cuts lead to corrections to the $\sigma^0\cdot(1+\delta)$,
which can also be considered as additional contributions to the
inefficiency and it will be discussed later in Sec.~\ref{systematicsection}. 
Several possible effects, such as an additional resolution smoothing effect or energy losses in the detector, 
were estimated using the full MC simulation, in which the detector conditions such as the resolutions, the
correlated noises, etc. were reproduced as in the experimental data at each energy point.
The typical one track angle resolutions are
$\sigma_{\theta}=15\hyph30$~mrad and $\sigma_{\phi}=9\hyph 6$~mrad, which
should be compared to the $\Delta\theta$ and $\Delta\phi$ cuts of 0.25 and 0.15~rad in Eq.~(\ref{eq:cutscoll}), respectively.
The overall effect from angle resolution, obtained using the
full detector simulation, gives about (0.10--0.25)~\% inefficiency correction for the $e^+e^-$
events, with the largest correction observed for the 2013 data taking conditions.
Nevertheless, the difference of the $e^+e^-$ and $\pi^+\pi^-$ inefficiencies is less than
(0.05--0.10)~\%, except for the narrow energy range $\sqrt{s}=0.82-0.88$~GeV, where
the additional correction for the $\pi^+\pi^-$ events grows up to 0.25\%.
This is because of the ISR radiative return from these energies to the $\rho$-meson cross section peak, 
which produces the noncollinearity of $\Delta\theta\simeq 0.25$~rad, that affects the 
$\Delta\theta$ distribution at the edge of the applied selection cut.

The CMD-3 drift chamber has high efficiency for particles with the polar
angle in the range $1<\theta <\pi - 1$~rad, 
when a track passes all wire layers. At lower angles 
 the number of hits on tracks starts to drop rapidly, worsening also the resolution of the track parameters.
To properly account for the edge effects it is important to evaluate
the inefficiency of the cut on $N_{hits}$ in exactly the same fiducial
volume restriction within the same $\theta^{event}$ definition as used for the pion form factor measurement. 
This inefficiency is studied using the full collinear event sample selected 
with the standard set of cuts by Eq.~(\ref{eq:cuts}) with $N_{hits}$ cut released. 
The dependence of this inefficiency on the $\theta^{event}$ for
different types of events is shown for one of the energy points in
Fig.~\ref{fignhiteff} ($E_{beam}=391.48$~MeV, RHO2013). 
The edge effects are different for the $\pi^+\pi^-$ and $e^+e^-$ events, 
mainly because of the different ionization energy losses ($dE/dx$) in the DCH.
The base efficiency, described at the beginning of the section, was
adjusted to avoid double counting; 
if the event was rejected only by the cut on the $N_{hits}$, this cut was ignored and the event 
was considered as passed the selection criteria.

To increase luminosity, the collider was operated with the long beams
during the last part of the season in 2013 and afterward 
with the typical size of the beams interaction region of $\sigma_Z \sim 2.3-3$~cm
along the beam axis (with $\sigma_Z \sim 1.3-1.7$~cm in the first part of the 2013 season).
The long beams have comparable length to the length of the drift chamber with $L/2\sim20$~cm.
To ensure that the selected tracks in the used range of polar angles were still inside the good fiducial volume of the DCH,
the strong enough cut on the longitudinal coordinate of the event vertex $|Z_{average}|<5$~cm was
applied as listed in Eq.~(\ref{eq:cutsvtx}), which filters up to ~10\% of events.
This inefficiency was extracted from the analysis of the $Z_{vtx}$ 
distribution of selected tracks on the same collinear sample as used
in the analysis by Eq.~(\ref{eq:cuts}), except that the $Z_{vtx}$ selection itself and $N_{hits}$ cut were released.  
Energy deposition from the combined calorimeter, including the BGO end cap part, was used
for the event type tagging. These modifications on the signal sample and tagging were necessary 
to avoid possible distortion of the tails of the $Z_{vtx}$ distribution due to inefficiencies.
The cosmic background, originated not from beam interactions, is well predicted
and was subtracted using the event time as shown in Fig.~\ref{figtevent}.
While the $Z_{vtx}$ selection inefficiency has sizeable values,
originally from the collider it is exactly the same for all processes.
Possible differences may come from the different angular dependence of the efficiencies 
and edge effects for the $\pi^+\pi^-$ and $e^+e^-$ events. But as long as we stay 
in the highly efficient volume of the detector, this difference is negligible. 
The additional detector related effect from the track $Z$-vertex resolution
$\sigma_{Z_{vtx}}\lesssim 3$~mm contributes to the efficiency much smaller, of the order of
$\lesssim0.5\%$. And the difference
between the $\pi^+\pi^-$ and Bhabha events is even smaller with the typical values of the
$\varepsilon^z_{ee}/\varepsilon^{z}_{\pi\pi} -1 \lesssim (0.05 \hyph 0.10)\%$, with at most
up to 0.3\% at the $\sqrt{s}<1$~GeV when the enormous correlated noise
situation was observed in the detector during the part of the 2018 season 
and about 0.4\% at the $\sqrt{s}>1$~GeV for the 2013 season.
This was studied using the full detector MC simulation and cross checked by the data 
using samples of events of different types as mentioned above (see Fig.\ref{figtestsel}).
The additional cross-check was performed by realizing the $Z_{vtx}$
cut from 5~cm to 8~cm in Eq.~(\ref{eq:cutsvtx}) and redoing the full analysis to extract the
pion form factor (redoing all efficiencies, the particle separation
and etc). The $Z_{vtx}$ selection inefficiency changes from ~10\% to
$\lesssim 1\%$ for the RHO2018, while the measured pion form factor is consistent with baseline selections at the
level $(-0.05\pm0.01)\%$, $(0.04\pm0.01)\%$ and $(-0.12\pm0.05)\%$ for the
RHO2013, RHO2018 and LOW2020 seasons respectively.

The described above procedure to reconstruct efficiency assumes that 
sources of the inefficiencies treated separately are not statistically correlated within used acceptance.
High precision goal of this study requires to prove this assumption.
This assumption may not be valid if a detector performance is strongly degraded,
especially when the inefficiencies are large and the second-order effects become significant. 
Either the events with large $Z_{vtx}$ are used, which adds short tracks 
with the deteriorated parameters at the edge of the tracking system.
Possible level of the correlations was studied with the help of the full MC simulation, 
which shows an unaccounted correlation between inefficiencies of
$(0.05\hyph 0.25)\%$ at the $\sqrt{s}=0.6\hyph 1$~GeV for $\pi^+\pi^-$ events and
$(0.02\hyph 0.10)\%$ for $e^+e^-$ in the RHO2013 data taking season.
The effect is 2-3 times smaller for the RHO2018 season, when the DCH was in much better conditions.
The main sources of these inefficiencies originate from the correlations between
$\Delta \theta$, $p^\pm$ cuts and $N_{hits}$, $\Delta\rho$, $Z_{vtx}$ selection criteria.
The difference of the correlated inefficiencies between $\pi^+\pi^-$ and $e^+e^-$ samples is about 0.1\% at
the $\rho$-meson c.m.~energies and 0.15\% at the $\sqrt{s}>1$~GeV for the RHO2013 season. They were considered as
the corresponding contributions to the systematic uncertainty of the pion form factor measurement.

As it was discussed earlier, at least one good track was required for the test event selection.
However, the loss of both tracks could not be fully independent.
The probability of the loss of both tracks in an event in a correlated way is also
studied with the specially selected test events without requirement of the presence of a good track.
To suppress a higher background level from cosmic events, the
additional compatibility requirement was introduced for the collinear
topology of the track segments reconstructed in the calorimeter, 
which were solely determined by the ionization coordinates in the LXe strip layers.
The number of events originated from the beam interaction point 
with both tracks lost was extracted by the analysis of 
the $Z_{average}^{clusters}=(Z_{1}^{cl}+Z_{2}^{cl})/2$ distribution.
The typical value of the correlated inefficiency to lose both good tracks are $0.1\hyph0.5\%$ and increases with the beam energy. 
The difference of the inefficiencies between $\pi^+\pi^-$ and $e^+e^-$ events is about 0.05\% at the $\sqrt{s}<1$~GeV and
$\lesssim0.3\%$ at the c.m.~energies above 1 GeV. These values were taken as the contribution to
the systematic error of the trigger efficiency, as the main reason of the correlated
loss of both tracks is related to the out-of-sync trigger issue. 

\subsection{Particle specific losses}
\label{ptspecificloss}

\begin{figure}[t]
\begin{minipage}[t]{.475\textwidth}
\centering
\includegraphics[width=1.1\linewidth]{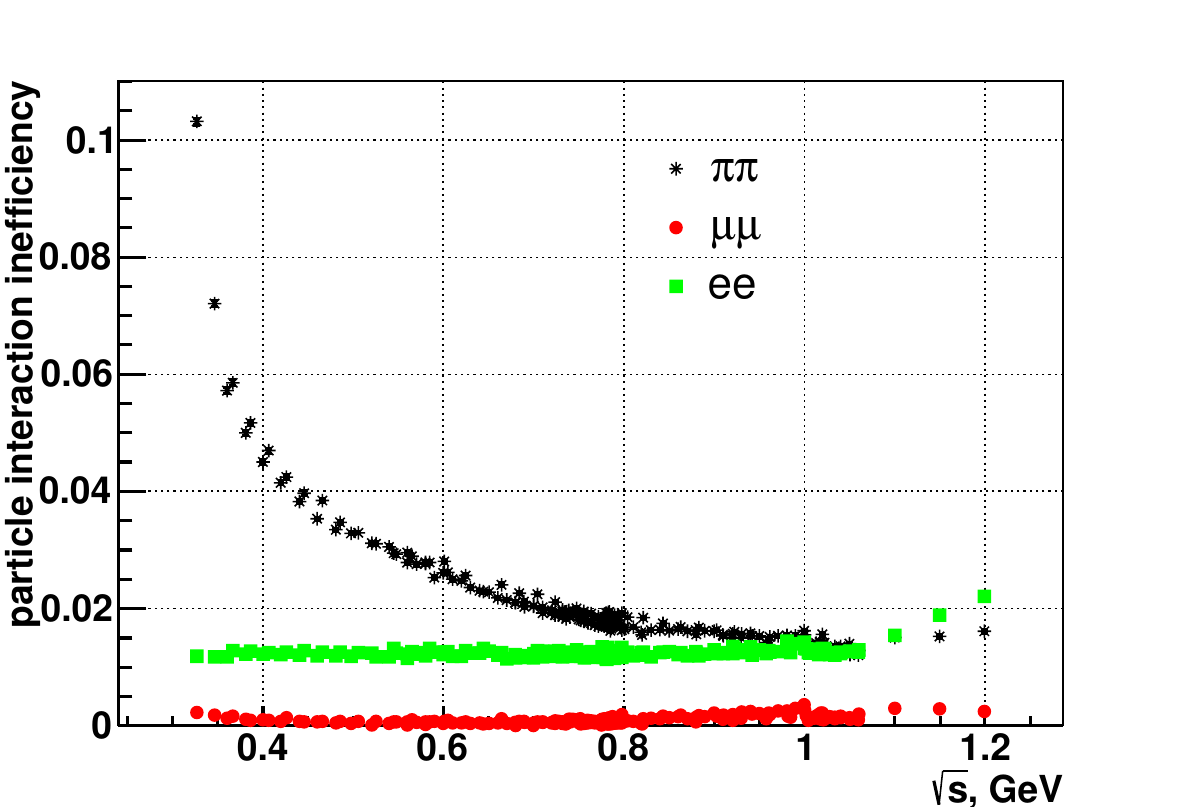}
\caption{Inefficiencies specific for different types of
  events coming from bremsstrahlung, nuclear interactions, decay in
  flight, etc.}
\label{figbrem} 
\end{minipage}
\hspace{.03\textwidth}
\begin{minipage}[t]{.475\textwidth}
\centering
\includegraphics[width=1.1\linewidth]{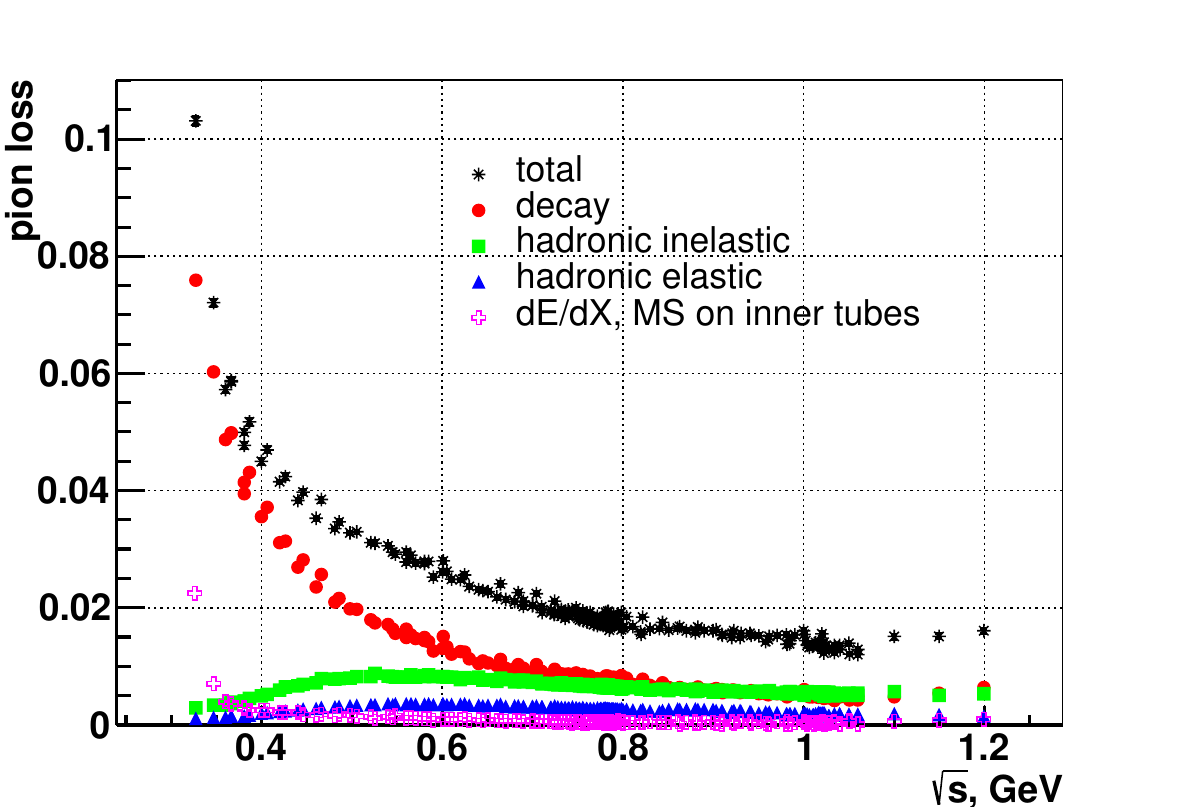}
\caption{Contributions from different effects for the $\pi^+\pi^-$ event specific losses.}
\label{figbrempion} 
\end{minipage}
\end{figure}

The events under study may be lost not only from the track reconstruction
inefficiencies in the DCH, but also because of the particle specific losses like the decay in
flight, bremsstrahlung, the nuclear interaction with 
the detector materials, and multiple scattering (MS) on the wall of the beam vacuum pipe. 
Some of these lost events without collinear cluster in the 
calorimeter are not included in the test samples. Hence, they are not taken into 
account in the efficiency determination described in the previous section. 
The already accounted part of these losses for the $\pi^+\pi^-$ events is about $30\%$ 
at the $\rho$-meson resonance c.m.~energies, for the $e^+e^-$ events $\sim5\%$ and for $\mu^+\mu^-$ less than 10\%.
The particle specific losses mentioned above are extracted by using the full MC
simulation of the detector.
The corresponding correction is taken as the ratio between the full efficiency obtained in
the MC simulation and the efficiency obtained with the subset of events when no any mentioned 
above process is happened in the tracking volume.
The MC efficiency was defined as the ratio of the number of events
passed selections after the full reconstruction to the number of events within acceptance at the generator level.
The efficiency also takes into account the effect of the finite 
detector resolution, when the filtrated events after the generators are returned 
back to the selected sample after the reconstruction.
The overall particle specific inefficiencies are shown in
Fig.~\ref{figbrem} and the different contributions to the $\pi^+\pi^-$
event loss are shown in Fig.~\ref{figbrempion}.
Changes in the inefficiency trends at the $\sqrt{s}>1.1$~GeV
come from the additional momentum selection to filter kaons, 
which start to contribute here as listed in Eq.~(\ref{eq:cutsmom}).
To exclude double counting, the corresponding particle specific losses parts were subtracted
from all efficiencies described in the previous Sec.~\ref{sectioneff}. 
These corrections were determined for the each specific test data samples 
which are used in the estimation of the corresponding inefficiencies mentioned above. 
It was also taken into account that the selected test samples have an admixture from the 
different types, for example, the ``MIP'' sample includes about 5\% of
the $\mu^+\mu^-$ events at the $\rho$-meson c.m.~energies and up to 50\% near the $\phi$-meson resonance.

The efficiency correction due to the bremsstrahlung for electrons is about $1.2\%$,
where the main part comes from photon radiation on the wall of the beam vacuum
pipe (with the thickness of 0.00575$X_{0}$) and the inner wall of the DCH (0.00241$X_{0}$), 
while the contribution from the material in the DCH gas volume is about 0.15\%.
The related material budget is known with the precision better than 5\% of the total $X/X_{0}$.
Moreover, its effect can be extracted from the analysis of the momentum spectra of electrons obtained in the
particle separation. The functional addition from the bremsstrahlung to the base momentum tail
from the generator was incorporated in the PDF definition for the $e^+e^-$ events,
where one of the free parameters is related to
an effective $X/X_{0}$. The part of the bremsstrahlung inefficiency, when events don't
pass only the momentum cuts in Eq.~(\ref{eq:cutsmom}), is about 0.9\%.
These numbers obtained from  the fitted PDF on the experimental data 
and on base of the full MC sample are consistent at the level of $(0.013\pm 0.005)\%$.
The additional uncertainty $<0.015\%$ to this comparison comes 
from the different initial momentum spectra provided by BabaYaga@NLO or MCGPJ generators.
The momentum tail for electrons comes almost equally from the radiative corrections and the bremsstrahlung
process in the detector. The consistency check above comparing
the data with the MC prediction includes both effects.
The systematic uncertainty related to the bremsstrahlung loss for the
$e^+e^-$ events is estimated as 0.05\%.

The nuclear interaction of a pion leads to its loss with a probability of less than 1\%, 
mostly coming from the interaction with the wall of the beam vacuum pipe and the inner wall of the DCH.
The precision of this correction was conservatively
estimated as 20\%, as a knowledge of the hadronic cross sections implemented in the GEANT4 toolkit,
in accordance with the previous studies~\cite{CMD-2:2003gqi,SND:2020nwa}.


The strongest pion loss comes from the decay in flight,
where the number of events decaying in the DCH volume
changes from 3\% at $\rho$-meson resonance c.m.~energies to 13\% at the lowest energy point.
While this process is precisely known, complications come from the
produced broken parts of the track inside the DCH volume, making
the reconstruction to be highly dependent on the detector performance condition.
The momentum spectrum of pions is shown in Fig.~\ref{figpimom}, where the
left tail comes from the muon spectrum after the pion decays at the beginning of the tracking volume, 
and the right tail is produced when the decay vertex appears in the middle of the DCH. 
The latter case gives confusion for the reconstruction of a broken trajectory, which can be
found as a single track. It can be viewed in
Fig.~\ref{figpimom}: the green line shows $2\pi$ PDF and magenta line
shows $2\mu$ PDF without effect of the subsequent muon decay.
The forms of these left and right parts of the momentum spectrum 
were fixed separately from the full simulation, while the numbers of events in these
tails were free parameters during the likelihood minimization. 
It provides the possibility to control the
reconstruction efficiency of such decay in flight events in comparison with full MC simulation. 
The consistency between data and MC simulation on
the number of events in the pion decay tails is about (1--3)\% (as averaged over
the statistics of the different seasons). 
The worst inconsistency was observed in the 2013 season when a few layers in the middle of the
DCH were not operating.
It should be noted that to improve the consistency,
the most recent CMD-3 detector MC simulation includes different
detector specific effects in the DCH, like: amplitudes variations, wires
inefficiencies, correlated noises on both ends of the wire readout, 
$r-\phi$ resolution dependencies with layers and etc, which were conditioned with time per
each collected energy point. And the bare DCH description,
without account of all these effects, was given inconsistency of about 15\% in
the reconstruction efficiency of pion decayed tracks,
which can give a feeling of the upper possible range of the level of complications
on the reconstruction of pion decayed tracks in the case of a naive description of the detector.
The other check of this pion decay inefficiency correction was performed with repeating the full analysis 
with the relaxed cuts:
$N_{hits} \ge 10 \to 8$, $\chi^2<10 \to 20$, $|\Delta \rho| <0.3 \to 0.6 ~\text{cm}$. 
The looser cuts make the pion decay
inefficiency smaller by a factor of 2--2.5 at the $\sqrt{s}>0.54$~GeV, 
but the pion form factor variation is only at the level of 
$(0.04\pm 0.01)\% / (0.01 \pm 0.01)\%$ for the RHO2013/RHO2018 seasons.

Based on the data-MC inconsistency, the corresponding systematic uncertainty 
in the pion form factor measurement from the pion decay
inefficiency was estimated as 0.1\%
at the $\rho$-meson resonance c.m.~energies and 0.2\% at the lowest energy points. 
The dependence of this estimation on the c.m.~energy was taken from a comparison of the comprehensive drift chamber
simulation and the bare DCH MC simulation case. Or in the other words,
the ratio of the events with the pion decays in the DCH volume is about 3\%
at the $\rho$-meson resonance c.m.~energies, and 3\% inconsistency in the reconstruction
of the decayed pions results in about $0.1\%$ of
the full efficiency uncertainty. While at the lowest energies 
the reconstructible pion decayed events fraction of the full collinear sample,
with tracks that could potentially be reconstructed and
selected (excluding the extreme cases when, for example, the angle of
the decayed track disobeys the collinearity condition), grows up to
7\%, which after multiplying by 3\% data-MC inconsistency in such events
corresponds to 0.2\% systematic uncertainty.

\begin{figure}[t]
\begin{minipage}[t]{.475\textwidth}
\centering
\includegraphics[width=1.1\linewidth]{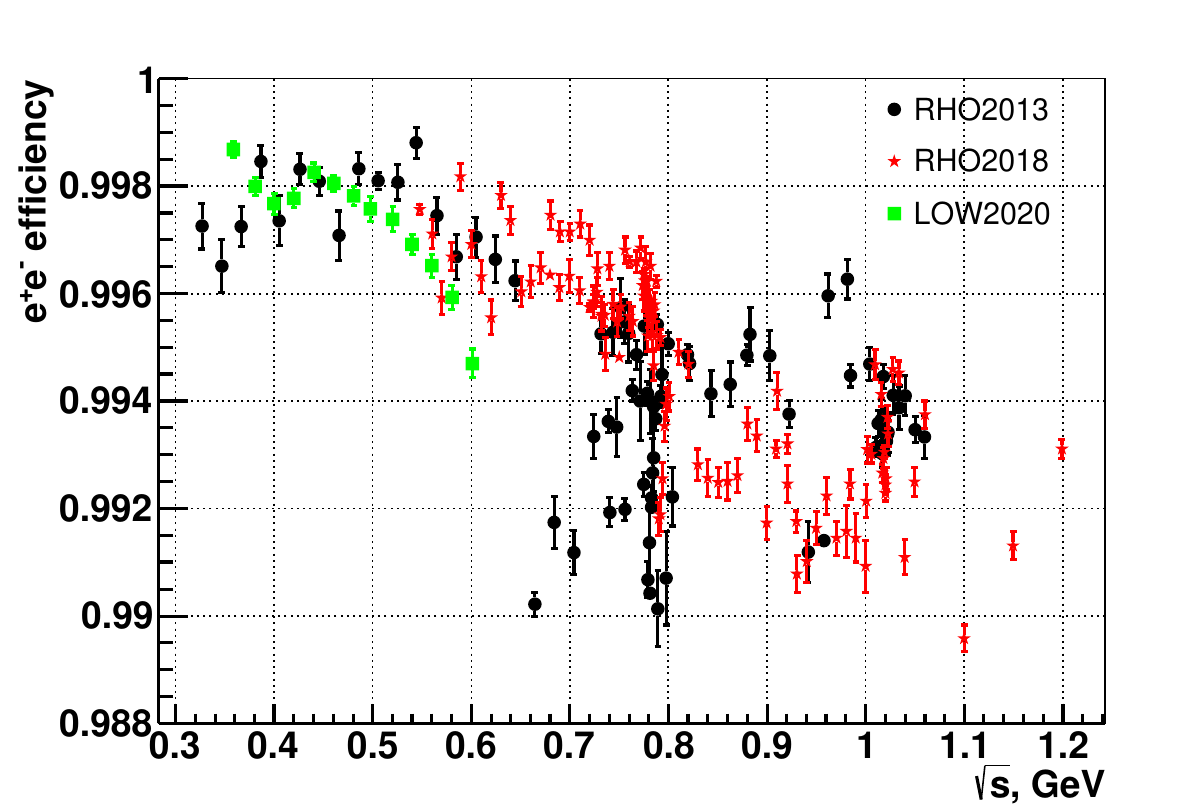}
\caption{The $e^+e^-$ efficiency, where the $Z_{vtx}$ selection inefficiency, $\theta$ angle resolution, and
  bremsstrahlung corrections are not applied. The data of the RHO2013, RHO2018, and LOW2020 seasons are 
  shown by the filled circles, stars and squares, respectively.}
\label{figeffee} 
\end{minipage}
\hspace{.03\textwidth}
\begin{minipage}[t]{.475\textwidth}
\centering
\includegraphics[width=1.1\linewidth]{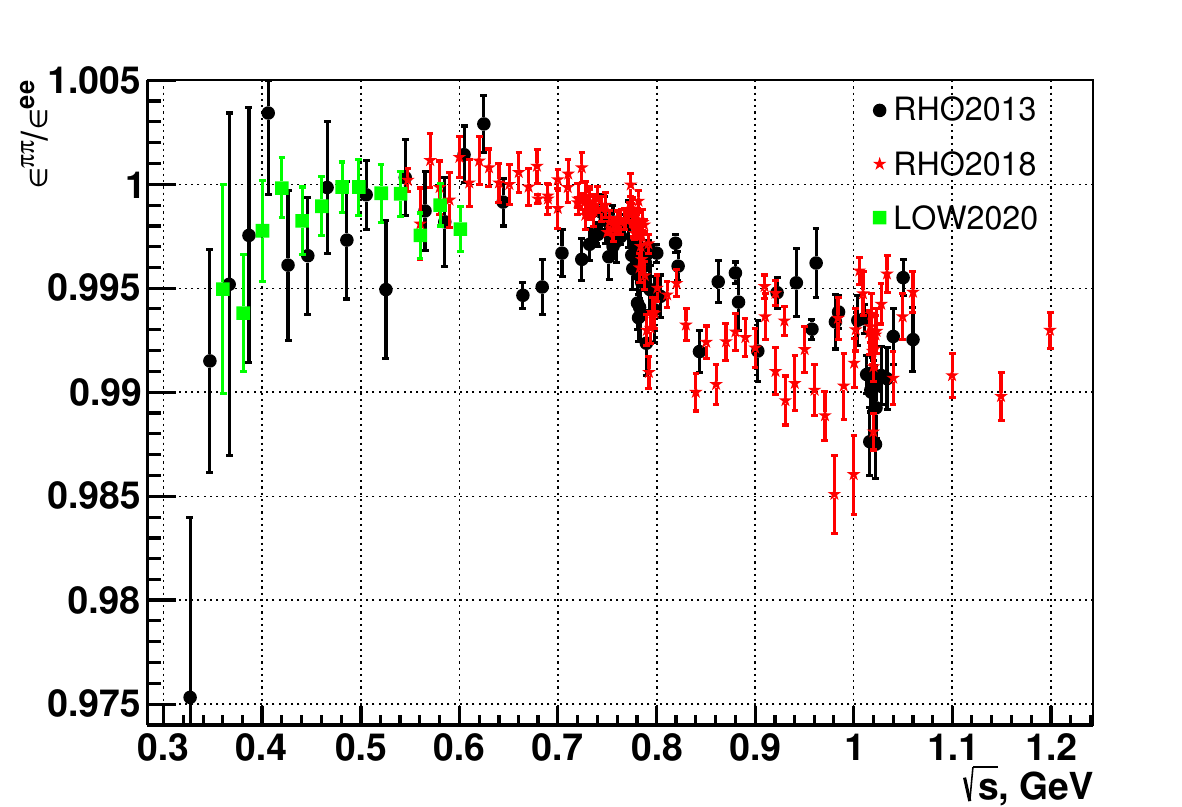}
\caption{Ratio of the $\pi^+\pi^-$ and $e^+e^-$ efficiencies without
  particle specific corrections, shown in Fig.~\ref{figbrem}. The data of the RHO2013, 
  RHO2018, and LOW2020 seasons are shown by the filled circles, stars, and squares, respectively.}
\label{figeffpieeratio} 
\end{minipage}
\end{figure}
The overall $e^+e^-$ efficiency (without $Z_{vtx}$ selection inefficiency, polar angle resolution and
bremsstrahlung corrections) is shown in Fig.~\ref{figeffee}. While
the ratio $\varepsilon^{\pi\pi}/\varepsilon^{ee}$ of the $\pi^+\pi^-$ and $e^+e^-$ efficiencies is shown in
Fig.~\ref{figeffpieeratio} and the particle specific corrections shown in
Fig.~\ref{figbrem} are not included in this plot. 
The severe change in the $\Delta\varepsilon = \varepsilon^{\pi\pi} - \varepsilon^{ee}$ at the $\sqrt{s} =0.79$~GeV for
the RHO2018 season comes from the deplorable noise
conditions in the detector during scanning from higher to lower c.m.~energies.
And during the RHO2013 season, the DCH was operated without few layers in the middle of the
drift chamber, that strongly affects the tracking performance.

\subsection{Trigger efficiency}
Two triggers are used during the data taking: ``charged'' and
``neutral'', which are assumed to be independent at the first order.
The signal of the charged trigger is produced by the track finder (TF) module. 
TF collects the information from the tracking system and 
searches for a signature of at least one track in the drift chamber
(as 3 nearest groups of wires, which are fired in 6 DCH outer layers).
The signal of the neutral trigger is produced by the cluster finder (CF) module.
CF processes the information from the calorimeter about the energies and positions of 
the clusters and searches for neutral patterns above some energy deposition thresholds. 
Having two independent trigger signals at the same time allows to study the efficiency
of the certain module with the data sample having signal from another one.

\begin{figure}[t]
\begin{minipage}[t]{.475\textwidth}
\centering
\includegraphics[width=1.1\linewidth]{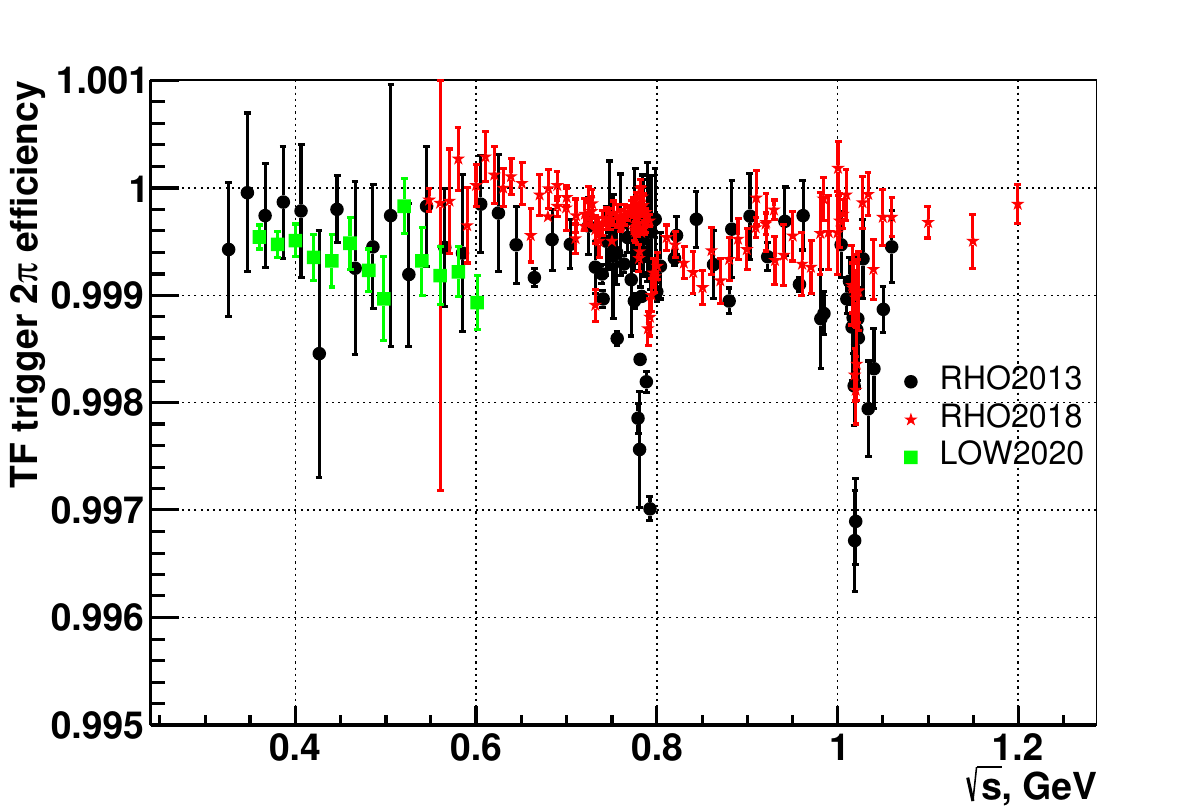}
\caption{The charged trigger (TF) efficiency for the $\pi^+\pi^-$ events.}
\label{figTFeffpipi} 
\end{minipage}
\hspace{.03\textwidth}
\begin{minipage}[t]{.475\textwidth}
\centering
\includegraphics[width=1.1\linewidth]{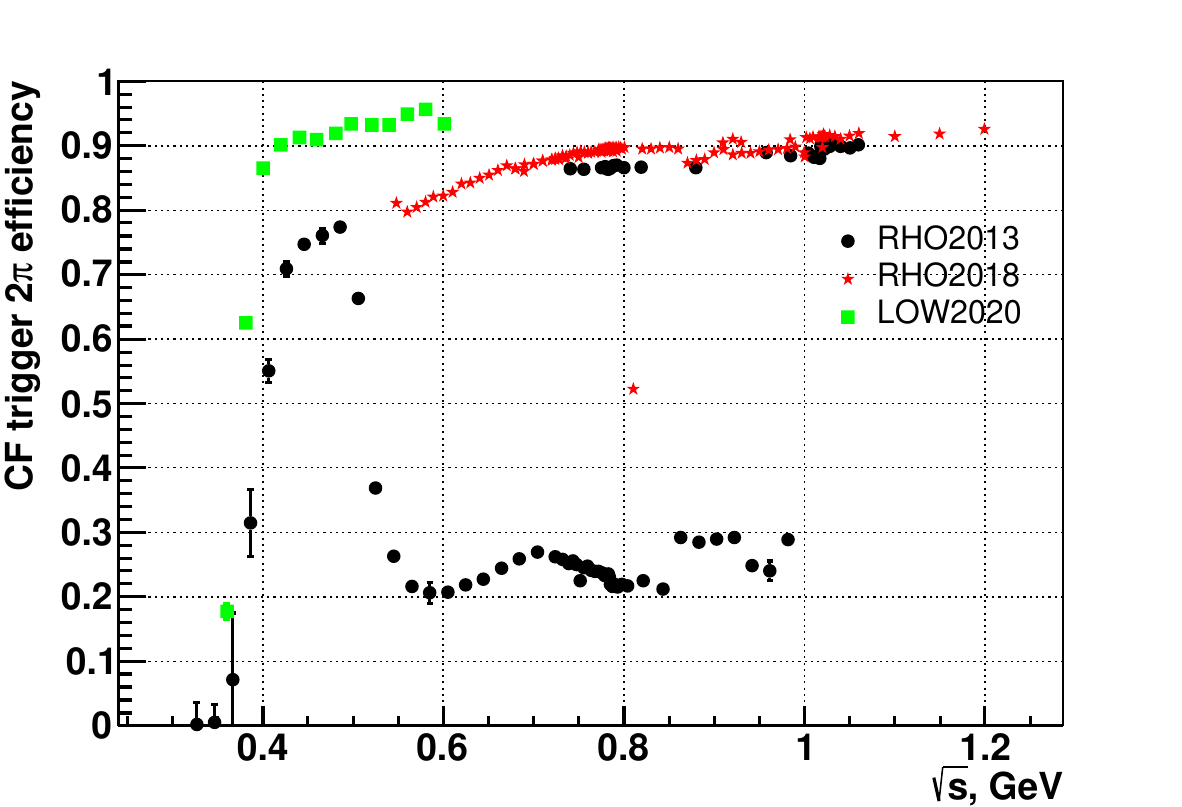}
\caption{The neutral trigger (CF) efficiency for the $\pi^+\pi^-$ events.}
\label{figCFeffpipi} 
\end{minipage}
\end{figure}

In case of the TF, the trigger efficiency as a function of the polar angle is obtained 
for the sum of all collinear tracks, and then it is convolved with
the angle distribution of the particular process. 
The inefficiency of the TF for a single track is about 1\% when the track
passes all DCH layers. It gives 
better than 0.1\% inefficiency, $1-\varepsilon^{trig}_{TF}$, for the events with two
tracks within the polar angles $1.1<\theta^{event}<\pi-1.1$~radians and 
increases to around $0.1\hyph 1.\%$ at the $\theta^{event}\sim 1$~radian. 
The overall charged trigger
inefficiency for the selected collinear events is about (0.05--0.15)\% 
with at most 0.2--0.4\% for a few energy points 
as shown specifically for the $\pi^+\pi^-$ events in Fig.~\ref{figTFeffpipi} for the different datasets.
As a cross check, the TF efficiency is also obtained in an integral way
using the event classes with different energy depositions, in the
similar way as it was described in the previous section~\ref{sectioneff}.
A consistent result has been obtained. 

The CF trigger efficiency is evaluated as a function of 
the energy deposition of two clusters, and then it is convolved with 
the energy deposition profile of the specific process.
The CF efficiency is usually as high as (97--99)\% for Bhabha events at
the c.m.~energies around the $\rho$-meson cross section peak, and it is decreasing 
down to 92\% for the c.m.~energy points below the $\sqrt{s}<0.6$~GeV.
The CF efficiency for the $\pi^+\pi^-$ events is around 90\% at the $\rho$-meson resonance peak, 
but part of the 2013 data were collected with a higher energy deposition threshold, 
decreasing the CF efficiency to about 20\%, as it is shown in Fig.~\ref{figCFeffpipi}.

Nevertheless, assuming the independence of the TF and CF triggers, 
the total trigger efficiency (TF or CF) for the $\pi^+\pi^-$ events is well above 99.94\%
everywhere in this analysis, and it is even higher for Bhabha events.

An additional analysis is performed to study possible correlations
between the two trigger signals due to a difference in time or in the event topology. 
Generally, two triggers are not fully independent as they have  
the very different time response, as the different physical processes and detector electronics
are involved. It can give different probabilities 
to lose an event due to a out-of-sync time for the different detector channels.
The TF trigger has faster response time as compared to
the CF, and it gives a predominance of the TF over the CF to fire the
event when both trigger signals are present.
This could introduce a correlation between the triggers via different
event efficiencies in the presence of either only one trigger or both of them.
It is worth noting that the track reconstruction efficiency depends 
on the type of the trigger. For the charged trigger which utilizes the 
information from the DCH, the track reconstruction efficiency is higher than 
that for the neutral trigger based on the information from the calorimeters. 
The existence of the charged trigger implies that the signatures of the 
tracks have been already reconstructed in the DCH, hence the track reconstruction 
efficiency is naturally high. In contrary, in case of the neutral trigger, signatures 
of the tracks in the DCH are not guaranteed and also the trigger starting time is not optimal 
for the DCH electronics. As a result, the track reconstruction 
efficiency is lower for the events passed only by neutral trigger. 
This effect is taken into account by the detection reconstruction efficiencies calculated 
for the specific triggers combination. 
They are used in the evaluation of the trigger efficiency as:
$\varepsilon^{trig}_{TF}=N_{TF\&CF}/N_{CF}\times\varepsilon^{rec}_{CF}/\varepsilon^{rec}_{TF\&CF}$,
where $N_i$ and $\varepsilon^{rec}_i$ -- number of the selected events and
the reconstruction efficiency for the data sample with the presence of the
$CF$ trigger signal (regardless of the TF presence) or with the presence of both $TF\&CF$ signals in the event.
The account of the ratio of the trigger conditional efficiencies
gives the correction factor of $1.5-2.0$ for the TF inefficiency in comparison with the simplified 
$(1-N_{TF\&CF}/N_{CF})$ estimation.
Impact on the CF inefficiency is more moderate on a percentage level for the $\pi^+\pi^-$ events.
This effect was already included in the values of the TF and CF efficiencies listed at the beginning 
of this subsection and shown in Figs.~\ref{figTFeffpipi},~\ref{figCFeffpipi}.

The other issue, related to the trigger system operation, is to produce
a starting time too late, which leads to a correlated loss of both tracks in
the drift chamber due to a limited DCH digitizer time window,
while an energy deposition in the calorimeter is still
present in an event. This was discussed at the end of
Sec.~\ref{sectioneff} and the corresponding trigger related systematic uncertainty,
0.05\% at the $\sqrt{s}<1$~GeV and 0.3\% at the c.m.~energies above 1~GeV, 
is assigned.

Few other hypothesis on possible hidden correlations were checked.
For example, if the neutral trigger signal could somehow spoil the final
trigger response, it should gives a different effect on the total efficiency in case of
a different rate of the CF trigger presence.
A comparison of the above mentioned two datasets from the RHO2013 season,
with the very different 90\% or 20\% CF efficiencies 
for the two pion events, shown in Fig.~\ref{figCFeffpipi}, 
can be such a cross-check of hidden correlations.
The measured form factors for both CF trigger conditions are
statistically compatible with the average relative difference $(0.12\pm0.17)\%$.


\subsection{Total efficiency vs polar angle}
\label{sectioneffvstheta}
\begin{figure*}[t]
  \centering
\includegraphics[width=.495\linewidth]{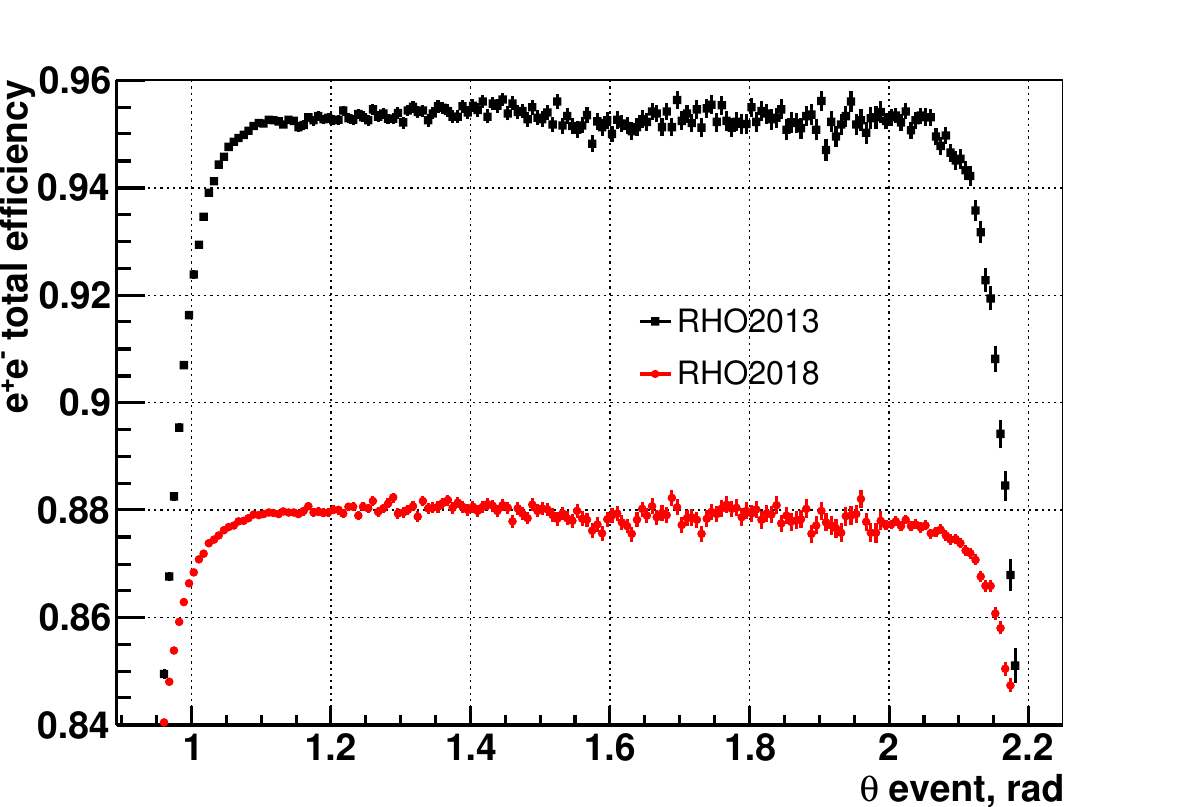}
\includegraphics[width=.495\linewidth]{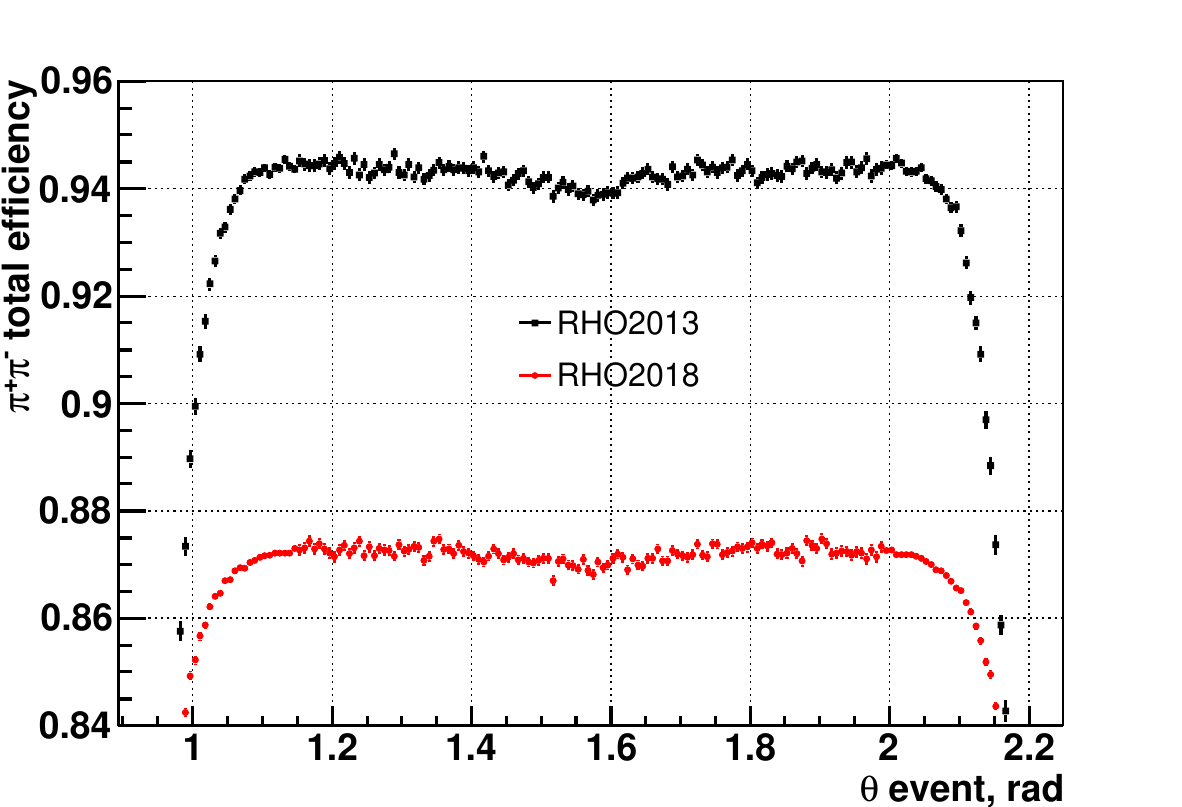}
\caption{The total selection efficiency versus polar angle for the $e^+e^-$ (left) and $\pi^+\pi^-$ (right) 
  events summed over $\sqrt{s}=0.7\hyph0.82$~GeV energy points for the two
  data-taking seasons. 
}
\label{figeffpi} 
\end{figure*}
The total detection efficiency for the $\pi^+\pi^-$ and $e^+e^-$ events versus
the polar angle averaged over the $\sqrt{s}=0.7\hyph0.82$~GeV energy points is shown in
Fig.~\ref{figeffpi}. It includes all the known effects described above and
corresponds to the differential behavior of the $\varepsilon$ defined in Eq.~(\ref{formdef}).
The main efficiency loss comes from the $Z$ vertex selection with
the average efficiencies of 97.0\% and 89.2\% for the RHO2013 and RHO2018 data samples, 
respectively, and it is nearly the same for the $\pi^+\pi^-$ and $e^+e^-$ events. 
The efficiency dependence is well symmetric over
$\theta =\pi/2$ radian, with a small dependence at the level $(0.2\hyph0.3)\%$ for
the $e^+e^-$ events because of not symmetric differential cross section. 
The drop of the efficiency at the level $(0.4-0.5)\%$ around
$\theta \sim \pi/2$~radians comes from the $Z$ vertex selection and the polar angle
resolution effects. The angular resolution changes by a factor of 2 from the $\theta = $~1 to
$\pi/2$~radians due to the charge screening effect reducing the amplitudes 
for the tracks perpendicular to the wires in the DCH.
The drop of the efficiency by 2\% and 4.5\% at
the edges of the used angle range comes from the requirement on the number of hits in the DCH,
as shown in Fig.~\ref{fignhiteff}. 

\section{Radiative corrections}
\label{sectionradcor}

The radiative corrections for the $\pi^+\pi^-/\mu^+\mu^-$ final states are calculated
by the MCGPJ~\cite{Arbuzov:2005pt} generator,
while for the $e^+e^-\to e^+e^-\gamma$ process it is preferable to use
the BabaYaga@NLO~\cite{Balossini:2006wc} generator.
All these generators are based on the next-to-leading order (NLO) calculations, and
the higher order terms in some approximations (the parton shower approach, collinear structure
functions, etc). The declared precisions are 0.2\% for the MCGPJ and 0.1\% for the BabaYaga@NLO generators.
Both generators are consistent in the integrated
cross section at the level better than 0.1\% for Bhabha process~\cite{Abbiendi:2022liz}, but the BabaYaga@NLO somewhat better
describes the differential distributions as it will be shown later.
In case of the $\mu^+\mu^-$ process, the initial momentum spectrum to
construct the PDF, used in the momentum-based separation, is taken from the BabaYaga@NLO,
while the integrated radiative correction is calculated by the MCGPJ.
In case of the BabaYaga@NLO generator, the muon mass term
is missed in the FSR virtual correction~\cite{Abbiendi:2022liz}, 
which results in the 0.4\% underestimation of the $\mu^+\mu^-$ cross section at the
lowest energy point used in the analysis.

\begin{figure}[t]
  \centering
\includegraphics[width=.495\linewidth]{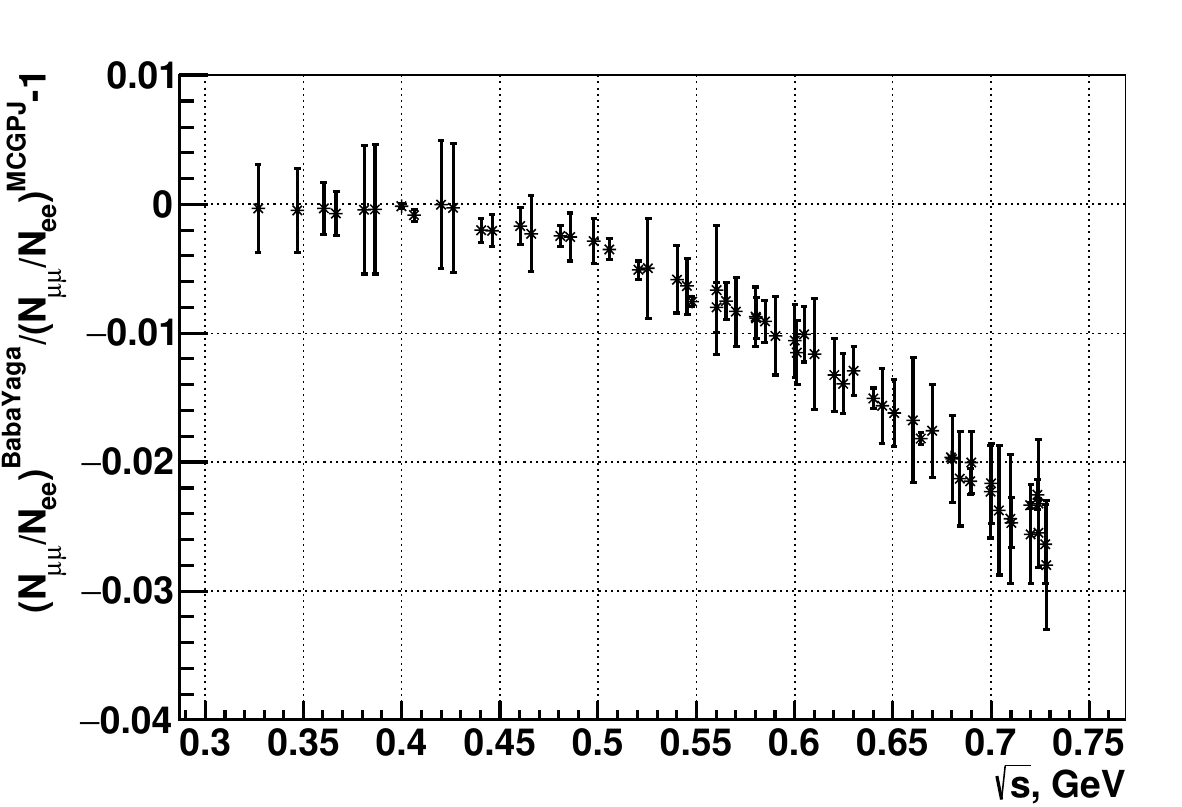}
\includegraphics[width=.495\linewidth]{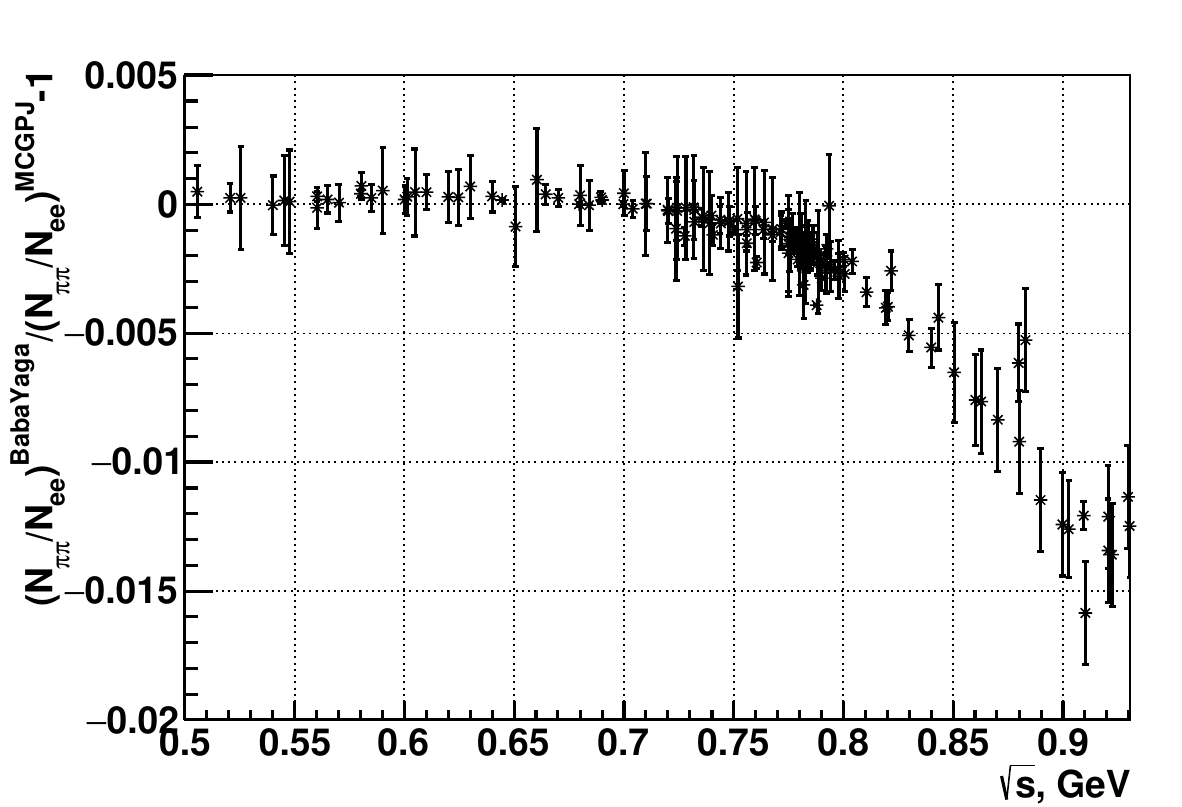}
\caption{The relative effect on the $N_{\mu\mu}/N_{ee}$ (top) and on
  the $N_{\pi\pi}/N_{ee}$ (bottom) ratios from using the $\mu^+\mu^+$, $e^+e^-$
  momentum spectra from either
  the BaBaYaga@NLO or the MCGPJ generators as input for the momentum-based separation approach. 
}
\label{fignmmnppnee_radcor} 
\end{figure}

The high statistics collected with the CMD-3 detector allowed to observe a
discrepancy in the momentum distribution between the experimental data and the theoretical
spectra from the original MCGPJ generator based on the paper~\cite{Arbuzov:2005pt}. 
For example, notable excess of the fitted function over the data was observed 
in the momentum range $P_{av}=(0.4\hyph 0.7) E_{beam}$ in the Fig.~\ref{figpiavmom} before special 
MCGPJ modification. 
The source of the discrepancy was understood: the collinear approximation for the photon jets 
was not good enough to describe the differential
cross section in the $P^+\times P ^-$ momentum distributions when energies of both final leptons 
are much smaller than the initial values (when two energetic photon jets are involved in the kinematic).
Therefore, the modified MCGPJ generator is used in this analysis with several
improvements, where the main one was an inclusion of the photon jet angular
distribution as for the one photon approximation.
Even after these modifications, the difference still remains in the momentum
far tails at the level of about $10\%$ between the modified MCGPJ and BabaYaga@NLO generators.
The effect of the usage of the different $e^+e^-/\mu^+\mu^-$ momentum
distributions from different generators indicates a
possible systematic uncertainty from the differential cross section prediction.
The momentum spectra from the generators are essential only for the particle
separation based on the momentum information for the PDF construction, and the corresponding impact
on the extracted $N_{\mu\mu}/N_{ee}$ and $N_{\pi\pi}/N_{ee}$ ratios are
shown in Fig.~\ref{fignmmnppnee_radcor}. This gives the sizeable effect on
the measured $\sigma_{e^+e^-\to\mu^+\mu^-}$ cross section, and as it
will be shown later, the data are consistent with the prediction when the BabaYaga@NLO
generator is used. This is one of the reasons, along with others that will be mentioned later,
why the BabaYaga@NLO is preferable to use for $e^+e^-\to e^+e^-\gamma$ process.

\begin{figure}[t]
\begin{minipage}[t]{.475\textwidth}
\centering
\includegraphics[width=1.1\linewidth]{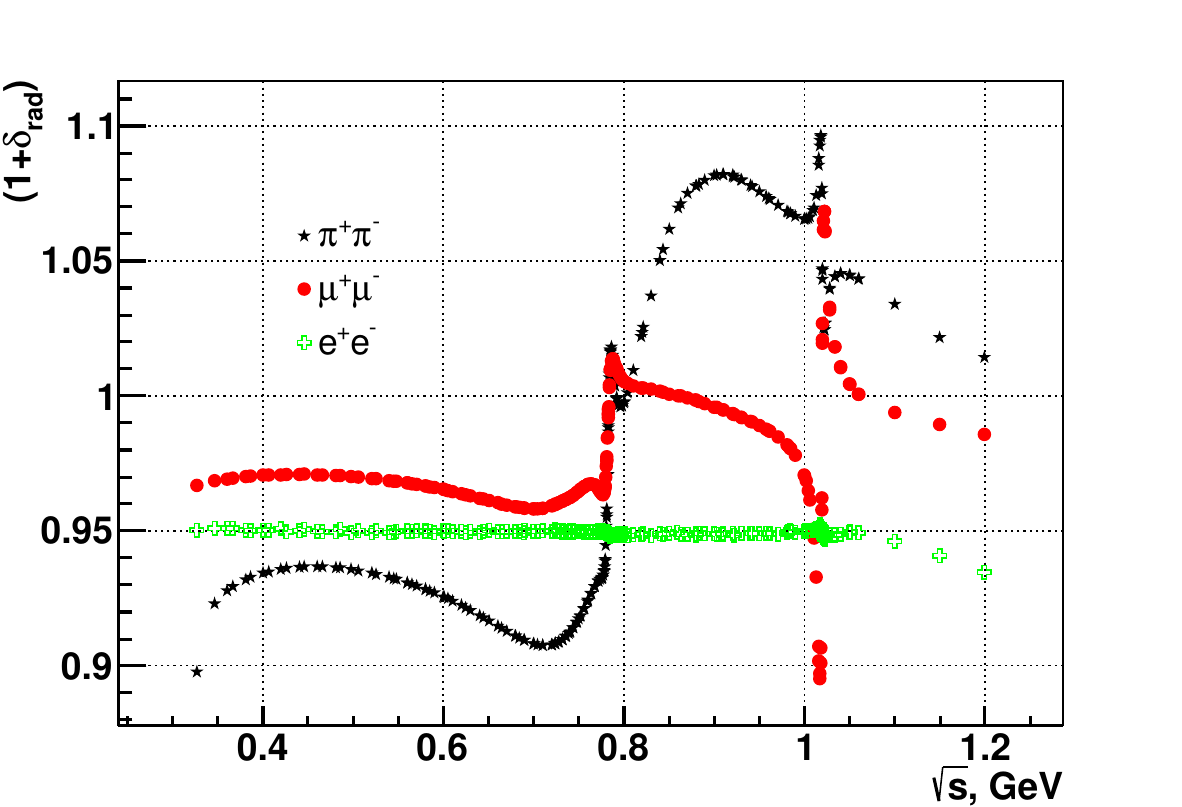}
\caption{The radiative corrections $1+\delta_{rad}$ for collinear events
  within selection criteria Eq.~(\ref{eq:cuts}). }
\label{figradcor}
\end{minipage}
\hspace{.03\textwidth}
\begin{minipage}[t]{.475\textwidth}
\centering
\includegraphics[width=1.1\linewidth]{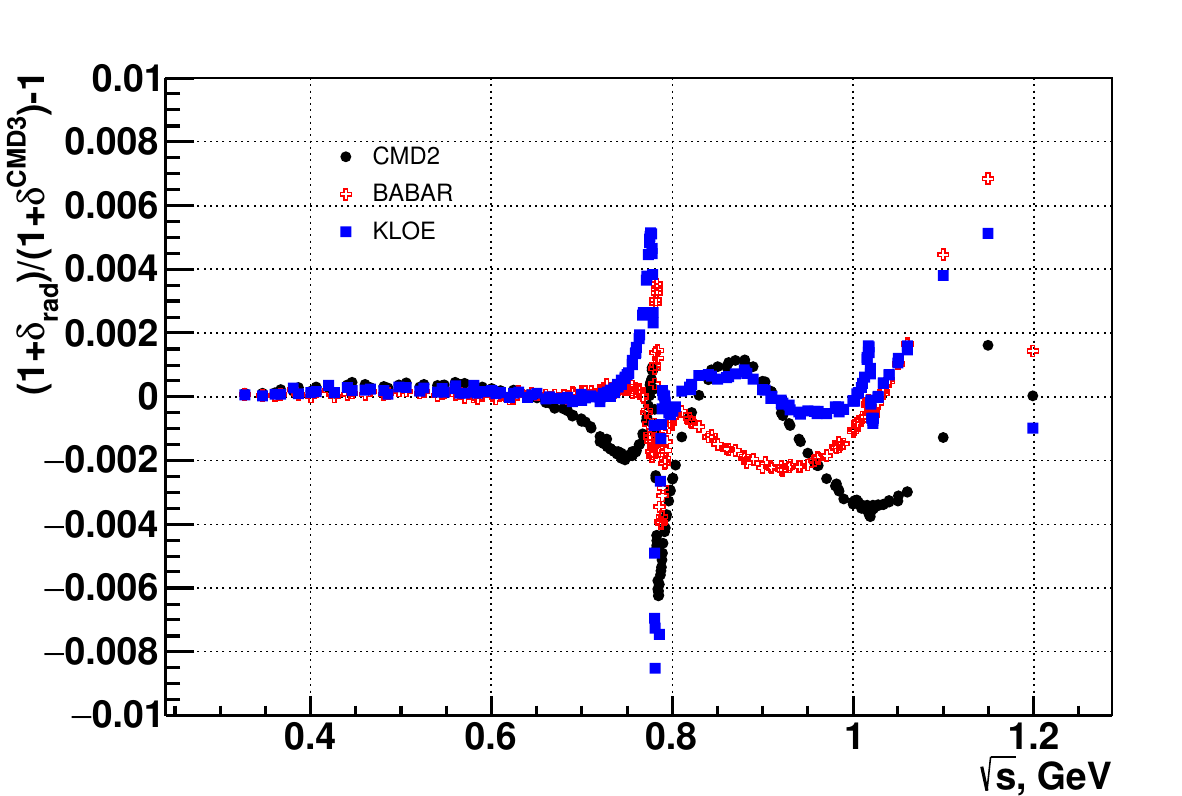}
\caption{The relative difference of the radiative corrections for the $\pi^+\pi^-$  events 
         with the pion form factor parametrizations based on the different experimental datasets
         and with the parametrization obtained in this analysis for the CMD-3 data.}
\label{figradcorparam} 
\end{minipage}
\end{figure}

The radiative correction for the $e^+e^-\to\pi^+\pi^-\gamma$ process 
does not include the vacuum polarization term in the photon propagator,
as it is considered to be an intrinsic part of the hadronic cross section and
is already included in the definition of the pion form factor.
The radiative correction for the $\pi^+\pi^-$ process depends on the used pion
form factor parametrization as input, so the iterative calculation is necessary.
Three iterations are enough to reach the precision $\lesssim 0.02\%$ except
for the c.m.~energy region around $\phi/\omega$-meson resonances, 
and another two to get the same value
convergence at the fast changing cross section near these narrow resonances.
The obtained $(1+\delta_{rad})$ radiative corrections for the collinear events
passed the selection criteria in Eqs.~(\ref{eq:cutsmom}--\ref{eq:cutsth}) are shown in
Fig.~\ref{figradcor}, where there are only relevant
cuts on the $\Delta \phi$, $\Delta \theta$, $\theta^{event}$ and $p^{\pm}$ are applied 
at the generator level.

To estimate a possible systematic effect on the radiative correction from the different pion form-factor
parametrizations, the various
experimental datasets from the $BABAR$, KLOE and CMD-2 measurements are used and
they are fitted separately according to the form factor parametrization as
will be described in Sec.~\ref{resultsection}. Parameters of
the $\phi\to\pi^+\pi^-$ decay are fixed from the CMD-3 measurement.
The relative differences of the calculated $1+\delta_{rad}$ in case of the $BABAR$, KLOE and CMD-2 
parametrizations and the CMD-3 case are shown in Fig.~\ref{figradcorparam}.
The fast change near the $\omega$-meson resonance is coming from a possible energy
shifts in different experiments, and it should be accounted for in
the corresponding systematic uncertainty, related to the beam energy scale.
The CMD-2 measurement is limited by the statistics
at higher c.m.~energies, what results in a larger radiative correction uncertainty.
The influence of the pion form factor behavior on the radiative
correction is added as an additional systematic uncertainty with 0.2\% at the c.m.~energies $\sqrt{s}>0.74$~GeV and
$0.5\%$ at the highest few energy points $\sqrt{s}\geq 1.1$~GeV.

\section{The forward-backward charge asymmetry}
\subsection{Asymmetry measurement}
\label{secasym}

\begin{figure*}[t]
\begin{minipage}[t]{.475\textwidth}
\centering
\includegraphics[width=1.\linewidth]{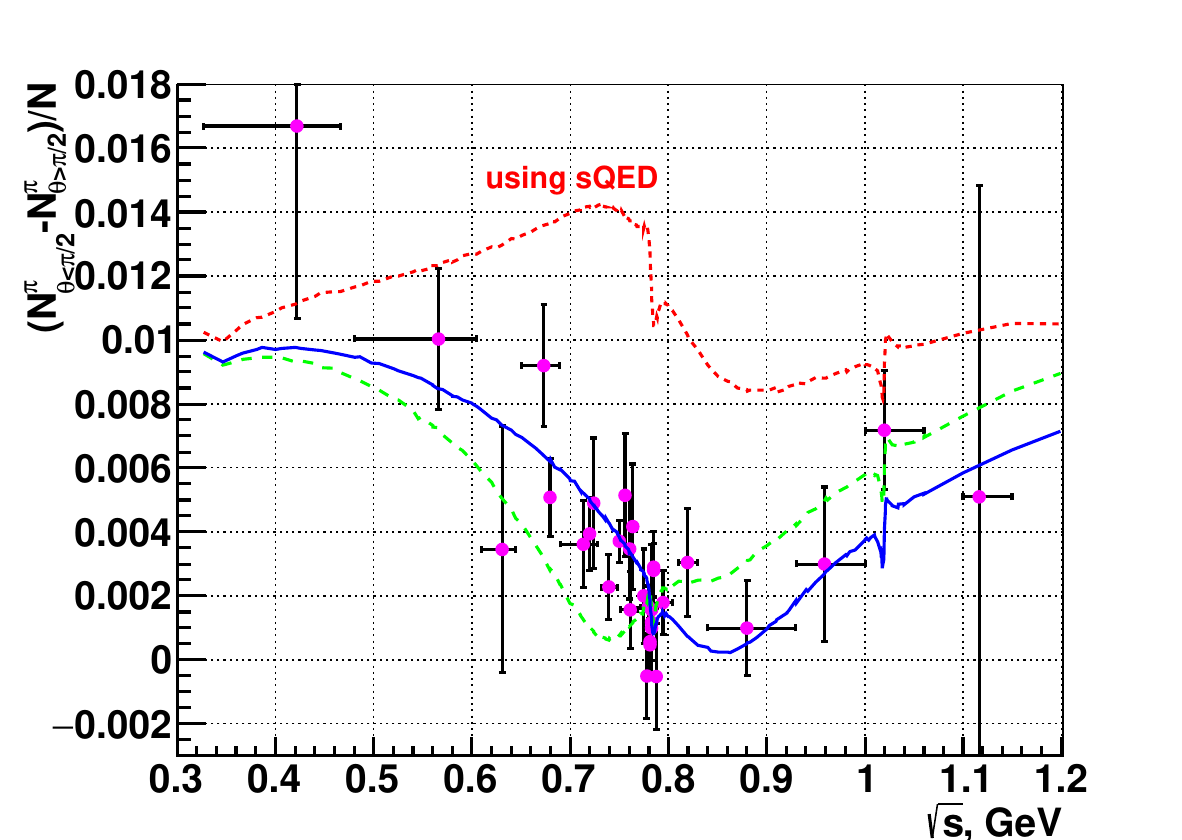}
\caption{The measured asymmetry in the $\pi^+\pi^-$ process at the CMD-3 in comparison with
        the prediction based on the commonly used scalar QED (sQED) approach (red dotted line),
        the GVMD model~\cite{Ignatov:2022iou} (blue line) and the
        dispersive calculation~\cite{Colangelo:2022lzg} (green dashed line). }
\label{asym} 
\end{minipage}
\hspace{.03\textwidth}
\begin{minipage}[t]{.475\textwidth}
\centering
    \includegraphics[width=1.\linewidth]{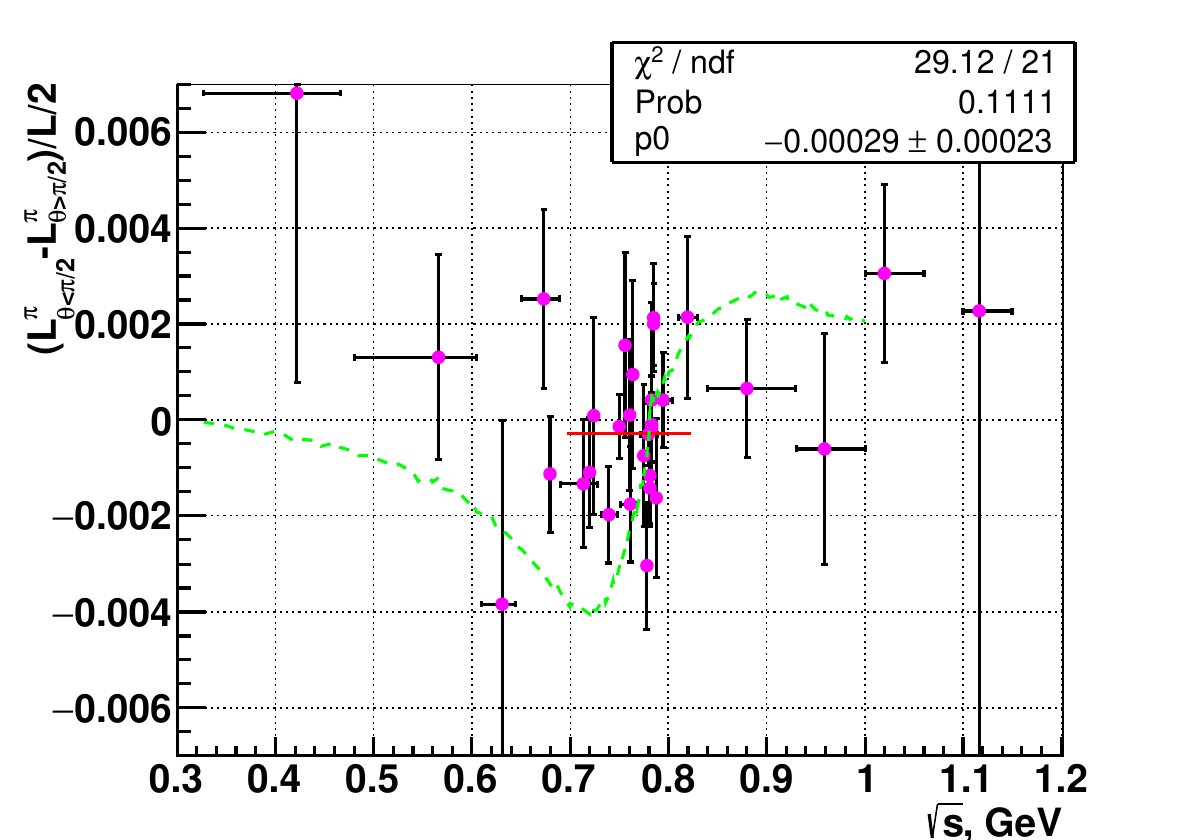}
    \caption{The difference between the measured asymmetry in the $\pi^+\pi^-$ data
      (points) or the dispersive calculation~\cite{Colangelo:2022lzg} (green dashed line) 
      and the prediction based on the GVMD
      model~\cite{Ignatov:2022iou} (GVMD corresponds to Y=0).}
    \label{asympp} 
\end{minipage}

\begin{minipage}[t]{.475\textwidth}
\centering
    \includegraphics[width=1.\linewidth]{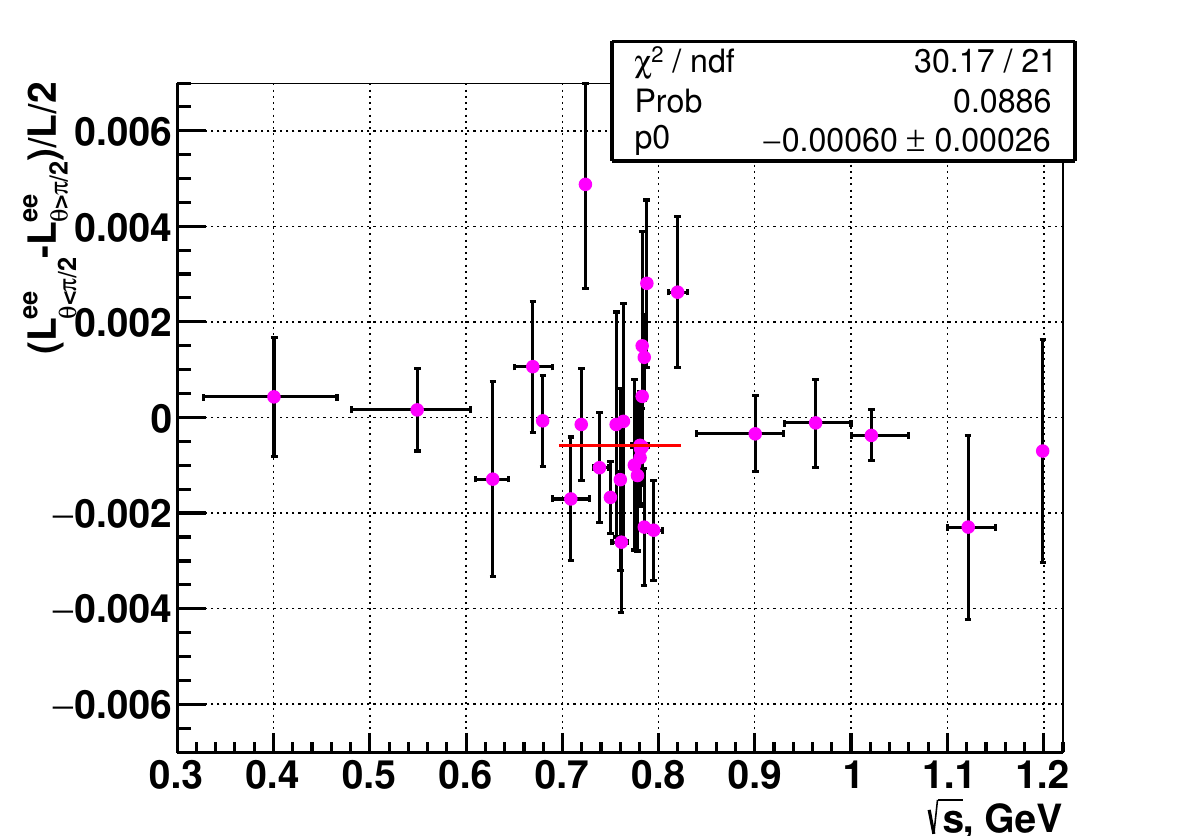}
\end{minipage}
\hspace{.03\textwidth}
\begin{minipage}[t]{.475\textwidth}
    \includegraphics[width=1.\linewidth]{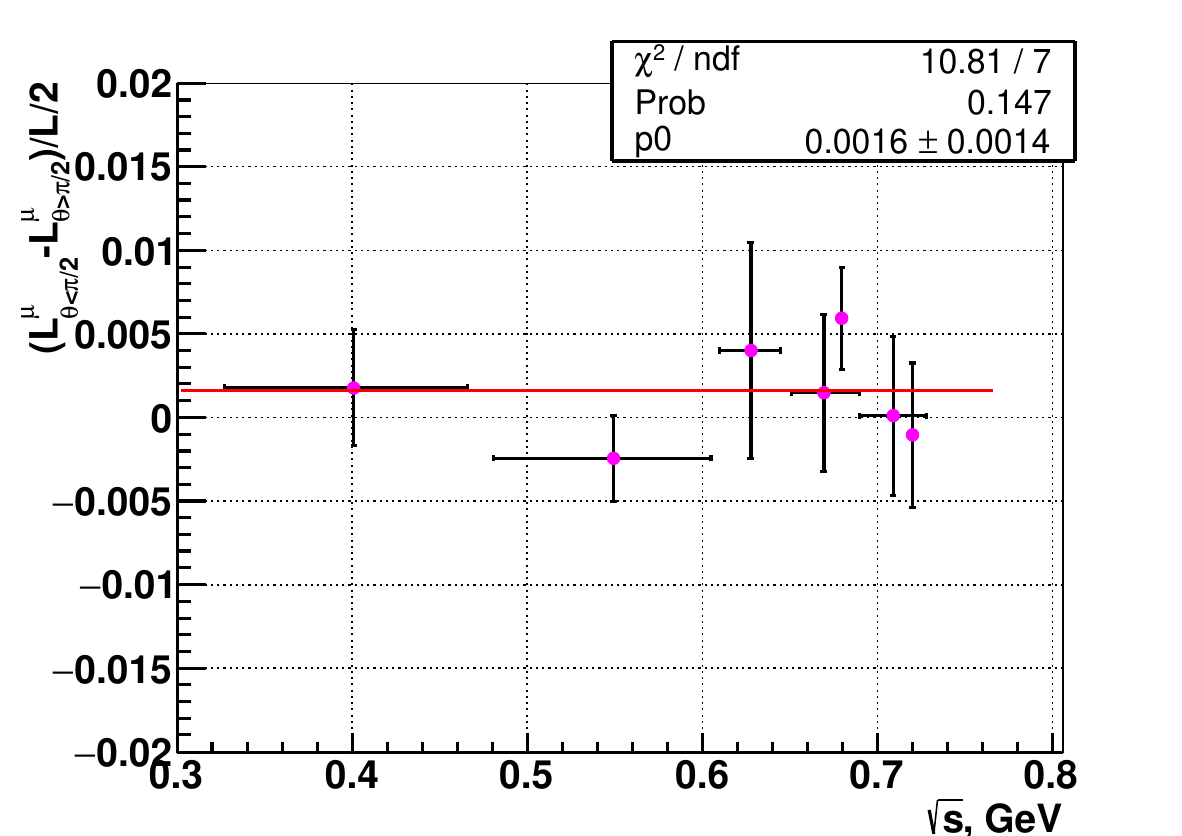}
\end{minipage}
    \caption{Difference between the measured and predicted asymmetries
      for the $e^+e^-$ events (left plot) and $\mu^+\mu^-$ events (right plot).}
    \label{asymeemm} 
\end{figure*}

One of the most significant sources of the systematic uncertainty comes from the fiducial
volume determination, and good understanding of the observed angular distributions is important
for the investigation and control of this contribution. 
Quantitative characteristic of the angular differential cross section is the
forward-backward charge asymmetry. 
The asymmetry is strongly affected by used models for the description of the pion with photon interaction
in the calculation of the radiative corrections for the
$\pi^+\pi^-$ channel. The comparison of the experimental asymmetry with the theoretical predictions
can be rather sensitive test for the models~\cite{BaBar:2015onb,KLOE:2005jxf,KLOE:2009fgr}.

The measured asymmetry is defined as a difference between the
detected numbers of events to the forward and backward regions of the detector:
\begin{equation}
    A=\frac{N_{\theta<\pi/2}-N_{\theta>\pi/2}}{N_{\theta<\pi/2}+N_{\theta>\pi/2}}\,,
\end{equation}
where $\theta = (\theta^+ + \pi -\theta^-)/2$ is defined as the average polar angle 
of an event, and event selection criteria are the same as for the main
$\pi^+\pi^-$ cross section analysis in Eq.~(\ref{eq:cuts}).
The numbers of events in this expression were corrected for detector effects like inefficiencies, resolution smearing,
etc, so that the measured asymmetry can be directly compared with
the predictions obtained with a Monte-Carlo generator. The applied total
selection efficiency versus the polar angle was discussed in Sec.~\ref{sectioneffvstheta}.

The obtained experimental $\pi^+\pi^-$ asymmetry together with
the theoretical predictions are shown in Fig.~\ref{asym}.
The points correspond to the overall statistics collected at this moment
with the CMD-3 detector at the $\sqrt{s}<1$ GeV. On this plot, some of
energy points are merged together, where the mean energy is the weighted by the numbers of events
averaged and the horizontal error bars show the energy range of the merged points. 
Different theoretical predictions are also shown on this plot, where the dotted red line corresponds to the
conventional approach based on the scalar QED (sQED) assumption~\cite{Arbuzov:2020foj},
which is usually used for the radiative correction calculations. The
discrepancy of this prediction with the experimental data is much larger
than the statistical uncertainty of the measurement. The dependence of the
asymmetry on the invariant mass of two pions shows that this
deviation comes from the virtual corrections, when $M_{\pi\pi}\simeq\sqrt{s}$, and gave a clue that 
the source of the discrepancy is related to the
calculation of the box like diagrams within the sQED assumption.
The improved approach using the generalized
vector-meson-dominance (GVMD) model in a loop integral was proposed
in the paper~\cite{Ignatov:2022iou}. This theoretical result is
shown by the blue line in Fig.~\ref{asym}. In addition
to the original paper, the $\rho-\omega$ interference is taken into account in the
calculation, which slightly improves the description near the $\omega$
mass. As it is seen, the GVMD model prediction agrees well with the experimental asymmetry. 
The difference between the measured asymmetry and the GVMD
prediction agrees with zero within statistical precision, as shown in Fig.\ref{asympp}. 
The average value of this difference in the c.m.~energy range $\sqrt{s}=0.7\hyph0.82$ GeV is
$\delta A = (-2.9\pm 2.3)\cdot 10^{-4}$ which should be compared with the $(-105\pm 2.3)\cdot 10^{-4}$
asymmetry difference obtained with the conventional sQED approach. The other calculation for the C-odd
radiative correction in a dispersive formalism was later presented in
the paper~\cite{Colangelo:2022lzg}, which is shown by
the green dashed line as relative to the GVMD result in Fig.\ref{asympp}.
Good consistency of the asymmetry is observed for both models at the c.m.~energies near the $\omega$ mass, but
some moderate discrepancy between these models is seen at the energies $\sqrt{s}\sim0.7$~GeV. 
The average difference between the
measured asymmetry and the prediction using dispersive calculation
is $\delta A = (+25.2\pm 3.9)\cdot10^{-4}$ at the $\sqrt{s}= 0.7\hyph0.77$~GeV, and 
$\delta A = (-3.1\pm2.8)\cdot10^{-4}$ at the $\sqrt{s} = 0.77\hyph0.82$~GeV. 
The average asymmetry difference over the whole c.m.~energy diapason $\sqrt{s}=0.7\hyph0.82$~GeV is
$\delta A = (+6.3\pm 2.3)\cdot10^{-4}$.


The same asymmetries between the detected numbers of events in both halves
of the detector normalized to the predicted cross sections are also considered for
the $e^+e^-\to e^+e^-$ and $\mu^+\mu^-$  processes. 
They are calculated as $\delta A=(L_{\theta<\pi/2}-L_{\theta>\pi/2})/(2L)$,
where the $L_i = N_i/(\sigma^0_i\cdot(1+\delta_i)\cdot\varepsilon_i)$
corresponds to the effective luminosity based on specific event type defined in the corresponding $i$th polar
angle sector, and $L=(L_{\theta<\pi/2}+L_{\theta>\pi/2})/2$.
These relative to the prediction asymmetries are shown in
Fig.~\ref{asymeemm} (the number of the $\mu^+\mu^-$ events is possible to extract
separately only at the lowest c.m.~energies).
The average values of the asymmetry differences are:
$\delta A = (-6.0\pm 2.6)\cdot 10^{-4}$
for the $e^+e^-$ events in the same c.m.~energy range $\sqrt{s}=0.7\hyph0.82$ GeV 
and $\delta A = (16\pm 14)\cdot 10^{-4}$ for the $\mu^+\mu^-$ process for the c.m.~energies 
$\sqrt{s}<0.74$~GeV. Good agreement between measured and predicted 
asymmetries for the $e^+e^-$ and $\mu^+\mu^-$ events is observed. 





The $e^+e^-$ calculation is done by the BabaYaga@NLO generator, while
the asymmetry difference in the case of the MCGPJ generator prediction becomes
$\delta A = (-14.0\pm 2.6)\cdot 10^{-4}$.
The angular dependence of the differential cross section obtained by BabaYaga@NLO is
confirmed by the exact fixed order NNLO calculation with the McMule
framework~\cite{Banerjee:2021mty}. The comparison between three
packages at the $\sqrt{s}=0.76$~GeV (with the vacuum polarization effect switched
off) gives the asymmetry differences $\delta A = (-6.0\pm 0.2)\cdot 10^{-4}$ between the BabaYaga@NLO 
and MCGPJ predictions, and $\delta A =(-6.6 \pm 0.3)\cdot 10^{-4}$ between the McMule and MCGPJ
(errors correspond to the statistical precision of calculations).

\subsection{Particle separation based on polar angle distribution}

\begin{figure}[t]
  \centering
\includegraphics[width=.495\linewidth]{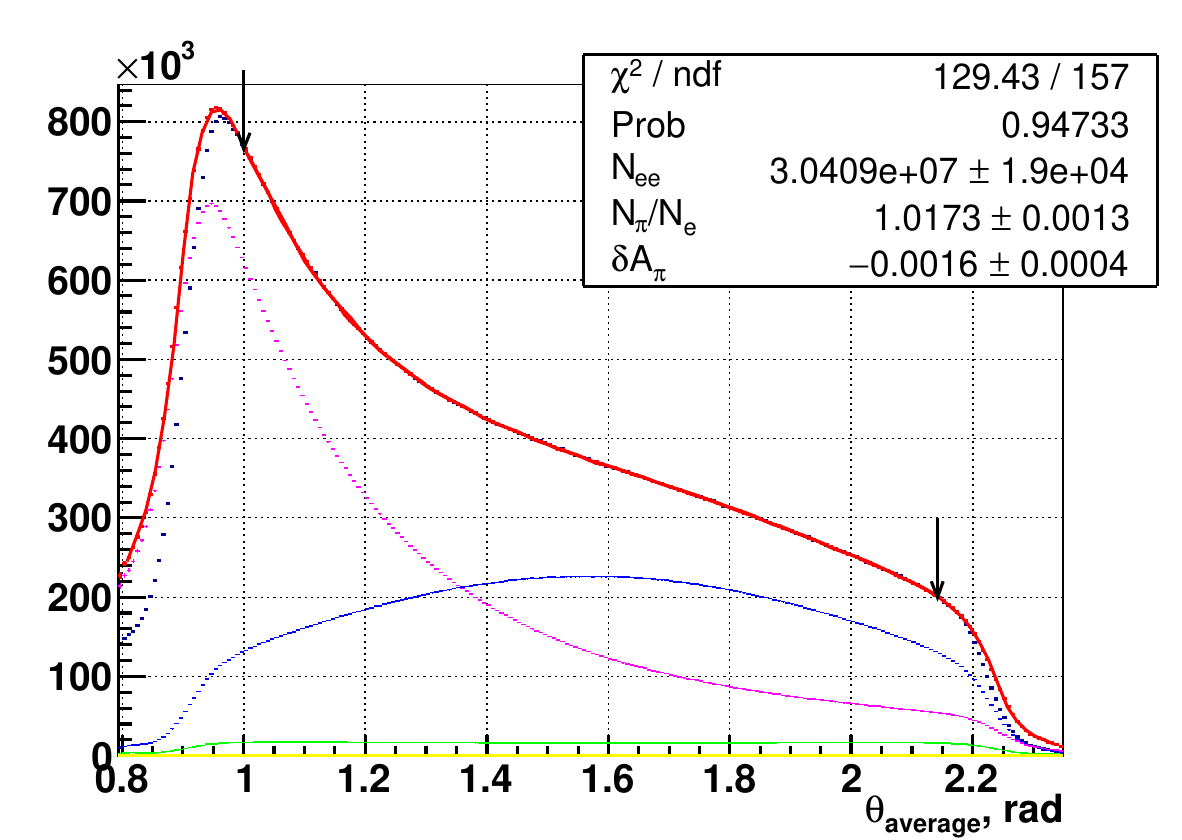}
\includegraphics[width=.495\linewidth]{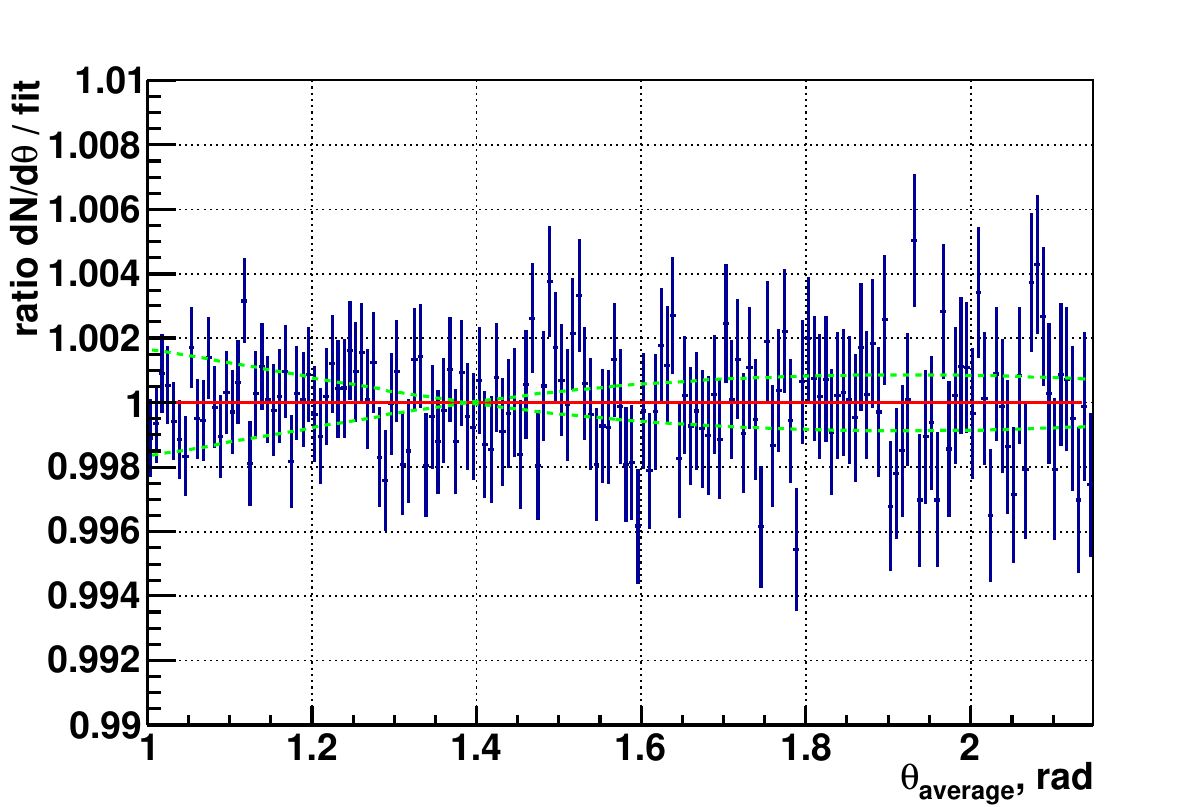}
\caption{ On the top: the polar angle distribution (black points) of the selected collinear events
  from  $\sqrt{s}=0.7\hyph0.82$~GeV energy points, which is fitted by
  the predicted spectrum (red line) with components from the $e^+e^-$ (purple line), 
  $\pi^+\pi^-$ (blue line), $\mu^+\mu^-$ events (green line), not seen $3\pi$
  (yellow) and cosmic events (cyan). On the bottom: the ratio of the experimental
  angular distribution to the fitted function, where the dashed green lines correspond to
  the variations of the fit by the $\delta (N_{\pi\pi}/N_{ee}) = \pm 0.5\%$.
}
\label{figthetafit} 
\end{figure}

The polar angle distribution of the detected collinear events can be
used for the determination of the $N_{\pi\pi}/N_{ee}$ ratio. It provides the third
particle separation method ($\theta$-based separation) as a cross check of the particle identification (PID) based 
on either momentum information in the tracker or energy deposition in the calorimeter.
For this purpose, the exact angular distributions for specific final states,
taken from the MC generators, are convolved with all detector effects like
efficiencies, angular resolution and the polar angle corrections obtained in each energy point.
The average corrections per each data taking season are shown in Fig.~\ref{figeffpi}.
The angular spectrum for the cosmic events is taken from the data
itself by clean selection on the event time, and the $3\pi$ contribution is
taken from the full detector MC simulation.
The experimental $dN/d\theta$ distribution summed over the energy
points within the range $\sqrt{s}=0.7\hyph0.82$ GeV and the corresponding predicted
spectra of all components are shown in Fig.~\ref{figthetafit}. The predicted angular
distribution for each final state is averaged over the different energy
points with the weights equal to the numbers of events selected in each c.m.~energy point.
The experimental distribution is fitted in the used angular range $1<\theta<\pi-1$~rad with
free parameters: $N_{ee}$ -- number of
$e^+e^-$ events, the $N_{\pi\pi}/N_{ee}$ ratio and the additional $\delta A_\pi$ asymmetry
correction for the $\pi^+\pi^-$ events. The ratio of the number of muons to the $N_{ee}$ is fixed
($N_{\mu\mu}/N_{ee} \simeq 8.5\%$) according to the QED prediction. 
The number of $3\pi$ events [about 0.2\% of the total number of events ($N_{total}$)] and
the number of cosmic events (0.04\% of $N_{total}$) are fixed at the values
obtained in the momentum-based particle separation.
The ratio of the experimental angular spectrum to the
fitted function is also shown on the right plot in Fig.~\ref{figthetafit}.
It should be pointed out that there is no visible issue in the accounted efficiency
(i.e., no additional sizeable systematics)
at the edges of the good polar angle range, although the notable correction up to 
level 2-5\% at $\theta=1.05-1$~rad is applied as seen in Fig.~\ref{figeffpi}.
The obtained ratio $N_{\pi\pi}/N_{ee} = 1.01727 \pm 0.00127$ from the angular distribution should be compared
with the $N_{\pi\pi}/N_{ee} = 1.01929 \pm 0.00030$ from the momentum-based separation or
with the $N_{\pi\pi}/N_{ee} = 1.01838 \pm 0.00033$ from the energy deposition-based separation method (see Sec.~\ref{separationsection})
as summed within the c.m.~energy range $\sqrt{s}=0.7\hyph0.82$~GeV.
The comparison of $N_{\pi\pi}/N_{ee}$ numbers as relative to the
momentum-based separation
gives deviations $(-0.20\pm0.12)\%$ for $\theta$-separation and $(-0.089\pm0.024)\%$ for the energy deposition-based separation.
If the asymmetry correction parameter is fixed at $\delta A_\pi =0 $ according to the GVMD model
prediction~\cite{Ignatov:2022iou} then the obtained relative $N_{\pi\pi}/N_{ee}$ ratio
difference between the $\theta$-based and momentum-based separations is $(+0.21\pm 0.07)\%$. 
This demonstrates the compatibility of
all three independent particle separation methods and ensures 
that the separation method related systematic uncertainty of the pion form factor measurement
is at level below 0.2\%.

%


\begin{figure}[t]
  \centering
\includegraphics[width=0.55\linewidth]{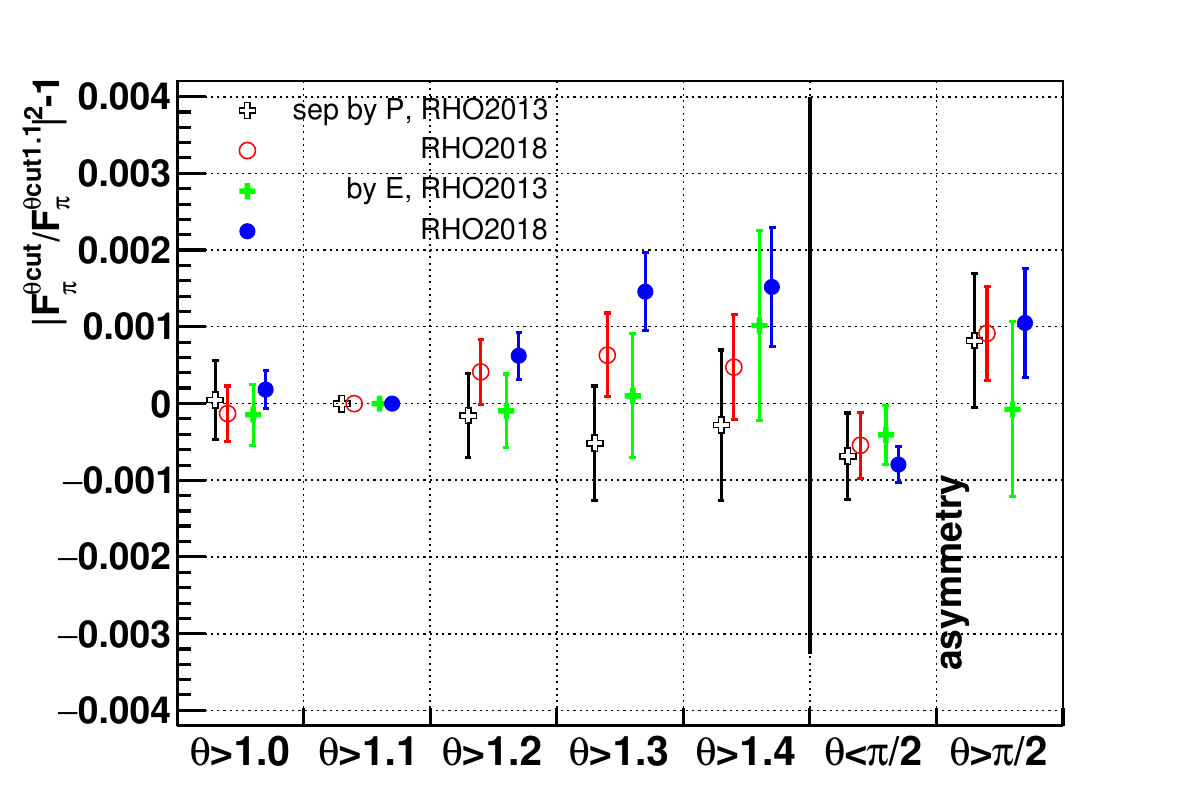}
\caption{The relative differences between the pion form factors in the cases of the various
         $\theta^{cut}$ cut values and that for the $\theta^{cut}=1.1$~rad cut calculated as 
         average for the c.m.~energies $\sqrt{s}=(0.7\hyph0.82)$~GeV.
         Empty markers correspond to the results using the momentum-based separation method,
         filled markers -- the energy deposition-based method. 
         Different data taking seasons are shown by crosses (RHO2013) and circles (RHO2018).
}
\label{fpivsthcut} 
\end{figure}

An additional cross-check of the pion form factor measurement was performed
by changing the selection criterion $\theta^{cut}<\theta^{event}<\pi-\theta^{cut}$
on the average polar angle with the cut value $\theta^{cut}$ varied from 1~rad to
1.4~rad and also separately in both halves of the detection volume $1<\theta^{event}<\pi/2$~rad
 or $\pi/2<\theta^{event}<\pi-1$~rad.
The relative differences between the pion form factors in the cases of the various 
$\theta^{cut}$ cut values and that for the $\theta^{cut}=1.1$~rad cut are shown in 
Fig.~\ref{fpivsthcut} for the momentum-based and energy deposition-based separation methods 
for both RHO2013 and RHO2018 data taking seasons. 
The relative difference is plotted as average for the c.m.~energies $\sqrt{s}=0.7\hyph0.82$~GeV
and the errors of the points are shown as the statistical difference
between the corresponding dataset and the $\theta^{cut}=1.1$~rad case. Additional
correction up to 0.15\% coming from the dependence of the particle
separation systematic bias on the polar angle cut as seen in the simulated datasets was applied specifically
for this plot. No residual effect remains and the measured form factor is
stable on the variation of the polar angle selection criterion at the level below
$0.1\%$ for all cases.
It should be also noted that some of possible 
systematic uncertainties of the $|F_\pi|^{2}$ related to the polar angle measurement
(like the common angle bias or the $z$-length scale
miscalibration) should be 2--3 times smaller for the
$\pi/2<\theta^{event}<1$~rad half of the fiducial volume in comparison with 
the full angle range used in the analysis.
The compatibility observed in Fig.~\ref{fpivsthcut} between this two angle
regions, assuming no other systematic uncertainties arising due to
asymmetrical angle selection, should ensure our $\theta$ angle
related systematic uncertainty estimation with a good safety factor.

\subsection{Asymmetry systematic uncertainty}

The obtained results above and the level of the applied efficiency corrections 
allow to estimate the systematic uncertainty of the charge asymmetry measurement. 
The biggest effect comes from the dependency of the particle
separation systematic bias on datasets within different angles, which
gives effect $\delta A^{PID}_{syst} \simeq -6\cdot 10^{-4}$ for $e^+e^-$. This correction is
not applied to the data, also, the $\delta A^{PID}_{syst}$ value agrees well with
the observed difference between the experimental asymmetry and the prediction based on the
BabaYaga@NLO generator. For the $\pi^+\pi^-$ events 
the particle separation effect is $\delta A^{PID}_{syst}\simeq +3\cdot10^{-4}$.
The efficiency corrections taken into account in the measurement of the charge asymmetry
are shown in Table~\ref{asymeff}, they are averaged in the energy range $\sqrt{s}=0.7-0.82$~GeV.
The main effect comes from the event inefficiency due to the $N_{hits}$ selection, where
the strong degradation is seen at the edges of the used angle range (in the ranges
$1.0\hyph1.05$~rad and $\pi-(1.0\hyph1.05)$~rad), as shown in Figs.~\ref{fignhiteff},~\ref{figeffpi}.
Looking in Fig.~\ref{figthetafit} it can be seen that there are no contributions from some unaccounted
effects at both polar angle edges at least on the $0.2\%$ level. 
Possible additional inefficiency of 0.2\% at one edge of the distribution
is propagated to the related systematic uncertainty of the asymmetry as 
$\delta A^{Nhits}_{syst}= 1.7 \cdot 10^{-4}$ and $0.6 \cdot 10^{-4}$ 
for the $e^+e^-$ and $\pi^+\pi^-$ events, respectively.
The other notable accounted efficiency effects on the asymmetry for the $e^+e^-$ events
come from the $Z$ vertex selection, the angle bias correction and the bremsstrahlung losses.
All these effects and possible systematics on them
are symmetrical over the $\theta=\pi/2$~rad and give negligible correction
$<10^{-4}$ in case of the $\pi^+\pi^-$ events.
Also, it should be noted again, that the
drift chamber was in very different conditions in the RHO2013 and
RHO2018 seasons, it was operated without 4 layers of wires (from 16 in total)
in the middle of the DCH sensitive volume during the RHO2013.
This results in very different effects as on the efficiencies as seen in Fig.~\ref{figeffpi},
and on possible tracking systematics.
This leads to the difference of total
effect from efficiencies on the asymmetry between seasons as
$\delta A^{efficiency}_{RHO2018}-\delta A^{efficiency}_{RHO2013} = (12.0 \pm 1.6) \cdot 10^{-4}$
for the $e^+e^-$ events and only $(-0.9\pm1.3)\cdot 10^{-4}$ for the $\pi^+\pi^-$ events. 
The asymmetry difference for the $e^+e^-$ events comes mainly from the angle bias
correction ($8.9\cdot 10^{-4}$) and from the inefficiency of 
the $N_{hits}$ cut ($5.0\cdot 10^{-4}$).
The experimental differences of the asymmetries of two seasons (after applied corrections) are
$(-1.4\pm 5.6)\cdot 10^{-4}$ and $(-2.7\pm 4.9)\cdot 10^{-4}$ for the $e^+e^-$ 
and $\pi^+\pi^-$ events, respectively, which are consistent with zero.

\begin{table}[t]
\centering
\caption{\label{asymeff}
  Efficiency corrections effects on the asymmetry
  measurement as average over $\sqrt{s}=0.7-0.82$~GeV energy points.
  Total corrections are also shown separately for the RHO2013 and RHO2018 data taking
  seasons. Effects of the calculated radiative
  correction as relative to the lowest order differential cross section
  are given in the last line.
}
\begin{tabular}{lrrr}
efficiency corrections     & $\delta A^{e^+e^-}, 10^{-4}$ &   $\delta A^{\pi^+\pi^-}, 10^{-4}$ \\ \hline      
$N_{hits}$ cut              & $ -4.46  \pm 0.13   $ & $   0.35  \pm 0.14  $ \\
$Z$ vertex selection         & $  3.31  \pm 0.70   $ & $  -0.59  \pm 0.60  $ \\
angle bias correction      & $  3.10  \pm 0.03   $ & $  -0.01  \pm 0.03  $ \\
particle specific loss     & $ -1.73  \pm 0.20   $ & $   0.11  \pm 0.21  $ \\
$\theta,\Delta\theta$ cuts & $  0.95  \pm 0.15   $ & $   0.09  \pm 0.10  $ \\
base efficiency            & $ -0.85  \pm 0.03   $ & $   0.01  \pm 0.05   $ \\
trigger                    & $ -0.013 \pm 0.002  $ & $  -0.008 \pm 0.020 $ \\ \hline
total correction           & $  0.16  \pm 0.84   $ & $  -0.06  \pm 0.69  $ \\
sum for RHO2013            & $ -8.55  \pm 1.19   $ & $   0.70  \pm 0.96  $ \\
sum for RHO2018            & $  3.50  \pm 1.07   $ & $  -0.21  \pm 0.88  $ \\ \hline
radiative corrections      & $ -52.36 \pm 0.38   $ & $ -24.93  \pm 0.20  $ \\
\end{tabular}
\end{table}

Having very asymmetric over $\theta=\pi/2$~rad angular distribution for
the $e^+e^-$ events and mostly symmetric in case of the $\pi^+\pi^-$ events should give
more pronounced any possible systematic effect on the asymmetry for the $e^+e^-$ events 
and suppressed by a factor 2--3 in case of the $\pi^+\pi^-$ events. 
The systematic effect for the $\pi^+\pi^-$ asymmetry becomes even negligible
if a systematic effect has almost symmetric behavior over the $\pi/2$ polar angle 
(like in the particle specific losses because of the nuclear interaction or pion decay). 
The consistency of the experimental asymmetry
with the prediction for the Bhabha events as shown in Fig.~\ref{asymeemm}, 
and assuming the correctness of this prediction,
it can give a level of possible systematic effects for the two pion events.

Taking into account the discussion above, the total systematic uncertainty of the measured
asymmetry for the $\pi^+\pi^-$ process is estimated as $\delta A^{syst} = 0.0005$.

\section{Systematic uncertainties}
\label{systematicsection}

\begin{table*}[t]
\centering
\caption{\label{systfpi} Contributions to the
  systematic error of $|F_{\pi}|^2$. The total systematic uncertainty
  includes scaling of relevant contributions by the factor 
  $(1+a\cdot N_{\mu\mu}/N_{\pi\pi})$ coming from fixing the
  $N_{\mu^+\mu^-}/N_{e^+e^-}$ ratio by the QED prediction. The total
  systematic uncertainty is split between RHO2018, LOW2020 data
  taking seasons and RHO2013.}
\small
\vskip 0.1 in
\begin{tabular}{lc} 
  Source & Contribution \\ \hline
Radiative corrections             & 0.2\% ($\pi^+\pi^-$) $\oplus$ 0.2\%
($F_\pi$, $\sqrt{s}>0.74$~GeV) $\oplus$ 0.1\% ($e^+e^-$)\\       
$e/\mu/\pi$ separation            & 0.5\% ($\sqrt{s}<0.381$~GeV) -- 0.2\% ($\rho$) -- 0.6\% ($\sqrt{s}>1$~GeV) \\        
Fiducial volume                   & 0.5\% / 0.8\% (RHO2013) \\
Trigger                           & $0.05\% $ ($\rho$) -- 0.3\% ($\sqrt{s}>$1~GeV)  \\
Correlation of tracking inefficiencies     & $0.1\%$ ($\rho$) -- 0.15\% ($\sqrt{s}>$1~GeV)  \\
Beam Energy (by Compton)  & 0.1\% , 0.5\% (at $\omega,\phi$-peaks) \\        
Background events                 & 0 -- 0.15\% ($\sqrt{s}=0.9\hyph1.2$~GeV), \\
                                  & 0.05\% (at $\omega$-peak), 0.2\% (at $\phi$-peak)  \\       
Bremsstrahlung loss               & 0.05\% \\
Pion specific loss                & 0.2\% - nuclear interaction \\       
                                  & 0.2\% (low) -- 0.1\% ($\rho$) - pion decay  \\\hline
Total Systematics                 & \multicolumn{1}{l}{\hspace{1cm} 0.8\% (low) -- 0.7\% ($\rho$) -- 1.6\% ($\phi$)} \\
                                  & \multicolumn{1}{l}{\hspace{1cm} 1.1\% (low) -- 0.9\% ($\rho$) -- 2.0\% ($\phi$) (RHO2013) }
\end{tabular}
\end{table*}

Summary of the contributions to the systematic uncertainty of the pion form
factor measurement is shown in Table~\ref{systfpi}. 
The exact dependence of the total systematic uncertainty of the $|F_{\pi}|^2$ on the c.m.~energy is given in Table~\ref{table:fpi}.

One of the important sources of the systematics is a theoretical precision
of the radiative corrections~\cite{WorkingGrouponRadiativeCorrections:2010bjp},
which was discussed in detail in
Sec.~\ref{sectionradcor}. Two contributions should be distinguished:
the one is related to the accuracy of the calculation of the integrated cross sections, and another is related to 
the prediction of the differential cross sections. In first case, the most precise 
lepton pair production generators are well consistent at the level of about 0.1\%.
There is some discrepancy at the threshold region for $\mu^+\mu^-$ process, but its source is known. 
For the $\pi^+\pi^-$ process there is the only MCGPJ generator which declares 0.2\% precision for the total cross section.
In case of the differential cross section, which is important for the momentum-based event
separation method and for the charge asymmetry study, a more precise $e^+e^- \to
e^+e^-(\gamma\gamma)$ generator is very desirable at the exact NNLO with proper matching
to the next orders resummation of logarithmically enhanced corrections. 

For the further reduction of theoretical systematics it is advisable to develop another precise $e^+e^- \to \pi^+\pi^-$
generator based on the theoretical framework beyond the scalar QED approach, as the point-like pion
approximation is already not sufficient.

The other important part of the analysis is the $e/\mu/\pi$
separation. Three methods were developed based on completely independent information: 
the momenta of particles measured in the tracking system, the
energy depositions in the calorimeter and the polar angle distribution (as average at $\rho$ energy region). 
All methods are highly consistent as was described above, which ensures the
systematic uncertainty from the particle separation to be below~0.2\%.

The most important source of the systematics comes from the fiducial volume
determination. In the CMD-3 detector, polar angle of tracks is measured by
the drift chamber with help of the charge division method, providing the $z$-coordinate
resolution of about 2 mm~\cite{Grancagnolo:2010zz}. This measurement is unstable  by itself 
as it depends on a calibration and thermal stability of the parameters of electronics.
For example, fast day-night oscillations are
seen on a reconstructed $z$-coordinate at the level of $\delta z/z \sim 6\times10^{-3}$, 
which requires to perform a temperature dependent calibration of the response of each DCH electronic board.
To determine the $z$-axis scale, an independent calibration is necessary to apply as relative to an external system.
The LXe calorimeter coordinate system was used for this purpose using 
cosmic muon events and with additional regular corrections per a single run.
It was also possible to use the ZC chamber for this purpose.
The ZC chamber was a 2-layer multiwire chamber installed at the outer
radius of the DCH, with the middle radius between two layers at 31.5~cm~\cite{Anashkin:2006zz}.
It had a strip readout along $z$-coordinate, where the strip
width is 6 mm and the $z$-coordinate resolution is about 0.7 mm for
the tracks with 1 radian inclination. The ZC chamber had been working for
25 years (initially at the CMD-2 detector) until the summer of 2017.
Also, the CMD-3 detector has the unique LXe calorimeter where the
ionization is collected in 7 layers with a cathode strip readout~\cite{Anisenkov:2014xpa}.
The combined strip size is 10--15~mm and the coordinate resolution is
about 2~mm. The first layer is located at the radius of 37.9~cm. 
Both subsystems had the design systematic precision for the coordinate strip position 
better than $100\,\rm \mu m$, which is required
to keep the systematic uncertainty of the integrated luminosity at the level 0.1\%. This precision can be also
affected by a noise presence, aggravated by wide strip dimensions.

When the ZC chamber was in operation together with the LXe calorimeter, it was possible to cross check
each other. The consistency during that time was obtained at level up to $\delta z/z <1.8\times10^{-3}$ 
for the $z$-length scale averaged over the angles. The considerable part
of this inconsistency comes from the ZC chamber, 
as seen by performing comparison between own two layers.
Also systematical variations of the DCH track impact point at the LXe inner layer versus
the measured position in LXe is observed up to 0.5~mm at the different coordinate regions and
track inclination angles at the LXe inner layer. This corresponds to the 
$\delta z/z <2.1\times10^{-3}$ at the $\theta=1$~rad for a detected track and $r\simeq 40$~cm radial position.
Both $z$-scale uncertainties at the DCH outer radius correspond to the 
$\sim 0.25 \oplus 0.3 = 0.4\%$ 
systematic uncertainty of the Bhabha cross section determination 
at the $1<\theta < \pi-1$~rad fiducial volume. 

In the reconstruction of the polar angle, the inner layers of the DCH wires 
close to the interaction point are used.
These layers of wires were operated at a reduced high voltage, which results in
much lower measured amplitudes used in the $z$-coordinate reconstruction. 
Together with higher level of beam background in this region and
beam induced correlated noises, the inner layers are highly vulnerable to 
possible coordinate reconstruction systematic effects. 
The polar angle definition of an event as $\theta^{event} = (\theta^+ + \pi -\theta^-)/2$ in the
selection criteria Eq.~(\ref{eq:cutsth}) reduces the angle systematics
coming from the common shift of the $z$-coordinates of the two tracks (or with $Z^{average}_{vtx}=<(Z^++Z^-)/2>$)
in the region close to the interaction point.
Meanwhile, the systematic bias in the mean value of the $\Delta Z^{average}=<(Z^+-Z^-)>$
will introduce a systematic shift in the $\theta^{event}$, and the strong
systematic effect was observed here, which changes number of selected
electrons by up to $\delta N_{ee}/N_{ee} < 0.7\%$. To cope with it, the
vertex constrained fit of both tracks was performed and the modified 
polar angles were used further in the selection criteria. 
Also, the corresponding momenta with improved
resolutions were used further in the event separation. All efficiencies
described before were calculated within these angle and momentum definitions, where it is relevant.
Also, an additional angle correction per each energy point
is introduced to control and account systematic shifts for each definition of the polar angle. 
This correction is obtained from the observed discrepancy between both track's
impact points at the inner wall of the LXe calorimeter and the measured coordinates by
the LXe itself.
The exactly same collinear events sample were used as for the
particle separation itself, where Bhabha and
``MIP'' event types were considered according to the energy deposition.
The correction in case of the polar angle after the vertex constrained fit
is below $0.5\hyph0.7$~mrad at  the $\theta^{event}=1$~rad (and
of the opposite sign over $\theta=\pi/2$), which results in the 
$\delta N_{ee}/N_{ee} \leq 0.15\hyph0.2\%$ for the selected $e^+e^-$ events.
Part of this uncertainty at the level of $\sim0.10\hyph0.15\%$ comes from the 
insufficient knowledge of the strip plane transparency factor for an induced charge 
on the UV coordinates sides of the plane in the LXe ionization layers, which is required to reconstruct properly
the coordinates for the different types of particles and their inclination angles.
Nevertheless, the corresponding contribution from the 
event's polar angle reconstruction to the systematic uncertainty of the pion form factor
measurement is rather conservatively estimated as the difference between results based
on the vertex constrained fit and unconstrained uncorrected angle definitions cases, which are 0.7\%
for the RHO2013 season and 0.3\% for the others. As it was mentioned
before, the tracking performance was much worse in the RHO2013 as the DCH was
operated without HV on the four layers of wires in the middle of the volume, which
leads to the higher weights of the inner layers in the track angle determination.

The overall systematic uncertainty related to the fiducial volume determination is estimated
as $\sim 0.25 \oplus 0.3 \oplus 0.7 = 0.8\%$ for the RHO2013 energy points
and $\sim 0.25 \oplus 0.3 \oplus 0.3 = 0.5\%$ for the others.


Beam parameters were permanently monitored by the method based on Compton backscattering of the
laser photons on the electron beams~\cite{Abakumova:2012pn,Abakumova:2013fsa}.
The statistical uncertainty of the measured beam energy ($\sim 10$~keV) is propagated into the statistical
uncertainty of the measured pion form factor via the energy derivative of
the fitted visible ratio $N_{\pi\pi}/N_{ee}$ from Eq.~(\ref{formdef}).
The systematic precision of the Compton backscattering method is estimated
as 40~keV per beam according to the paper~\cite{Abakumova:2013fsa} and
from a stability of the analysis of the scattered gamma energy distribution. 
This helps to keep the related contribution to the systematic uncertainty 
of the pion form factor measurement below 0.1\% except the c.m.~energies around
$\omega$ and $\phi$-meson resonances, 
where this systematic uncertainty increases up to $0.5\%$ due to the fast 
change of the cross section in the region of the narrow resonances. 
In principal, the systematic bias of the measured beam energy can be different
for the different data taking seasons, as different lasers and analysis
procedures were used for this monitoring. 
A beam energy spread is also provided by the
Compton backscattering method, which gives in average as
$\sigma_{E_{beam}}\sim 200$~keV per beam at the $\omega$-meson resonance and
$\sigma_{E_{beam}}\sim260$~keV at the $\phi$-meson resonance c.m.~energies and 
the corresponding cross section corrections are 
up to $\sim0.08\%(\omega)$ and $\sim0.35\%(\phi)$ at near resonances.

An additional systematic uncertainty comes from the fixing the
$N_{\mu\mu}/N_{ee}$ ratio in the particle separation. The only
relevant contributions for this from Table~\ref{systfpi} are part of
the radiative corrections with $0.1\%(\mu^+\mu^-)\oplus 0.1\%(e^+e^-)$,
bremsstrahlung loss, fiducial volume. The trigger related uncertainty and
the correlated inefficiency are smaller in case of the $\mu^+\mu^-$, they are
 $0.05\% \oplus 0.07\%$ at the $\rho$-meson and  $0.1\% \oplus 0.1\%$ at the $\sqrt{s}>1$~GeV.
Also, the effect of the 40~keV systematic beam energy shift is 0.08\% at the $\omega$-meson
and 0.5\% at the $\phi$-meson resonance c.m.~energies.
All these contributions are fully correlated between $\pi^+\pi^-$ and $\mu^+\mu^-$ events.
In case of the energy deposition-based particle separation an excess of
the $\mu^+\mu^-$ events results in a deficiency of almost the same number of the $\pi^+\pi^-$ events
as they are located in the same ``MIP'' energy deposition region as seen in Fig.~\ref{figmomen2}.
It gives an additional scale factor $(1+a\cdot N_{\mu\mu}/N_{\pi\pi})$ for
the systematic uncertainties of the $|F_\pi|^2$ measurement from
the listed above corresponding sources, with the obtained coefficient $a=0.975\hyph0.9$
at the $\sqrt{s}=0.6\hyph1.2$~GeV. This $a$ coefficient is smaller than 1 because of 
the two-particle correlation in the description of the hadronic interaction tails 
in PDF for the $\pi^+\pi^-$ events.

\begin{figure*}[t]
  \centering
\includegraphics[width=.495\linewidth]{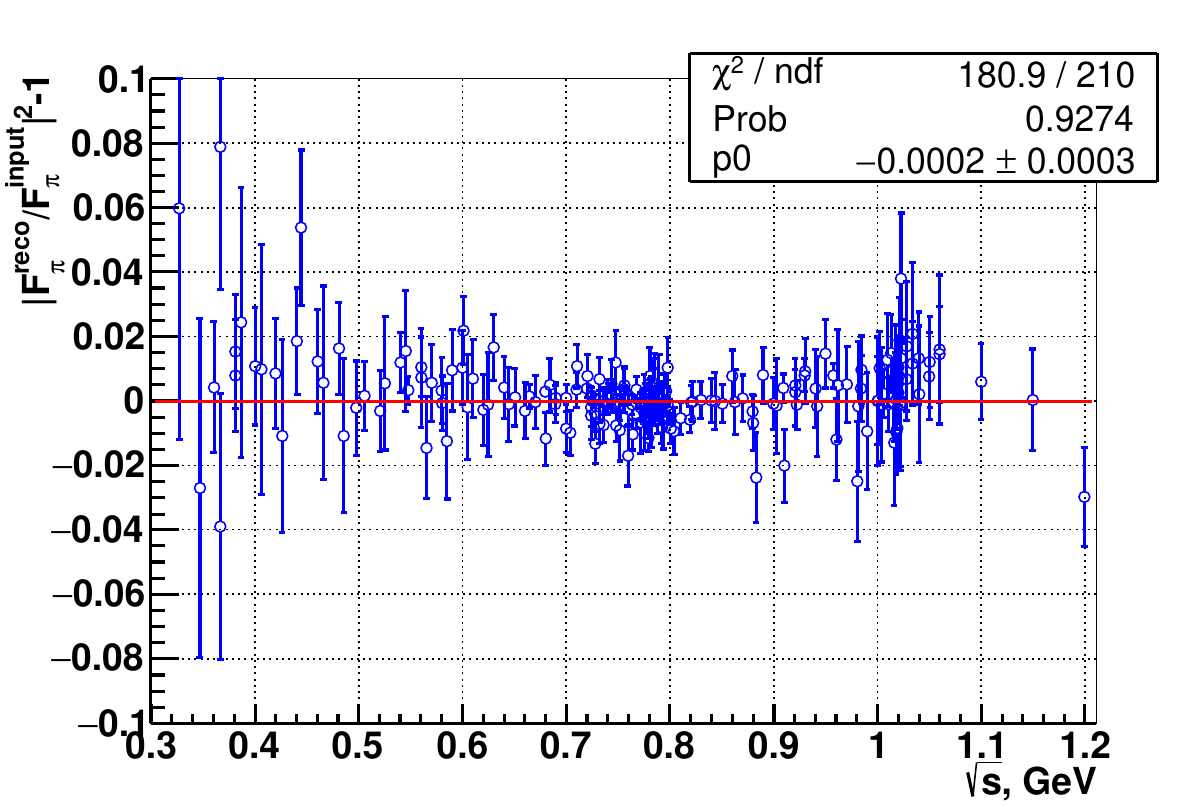}\hfill
\includegraphics[width=.495\linewidth]{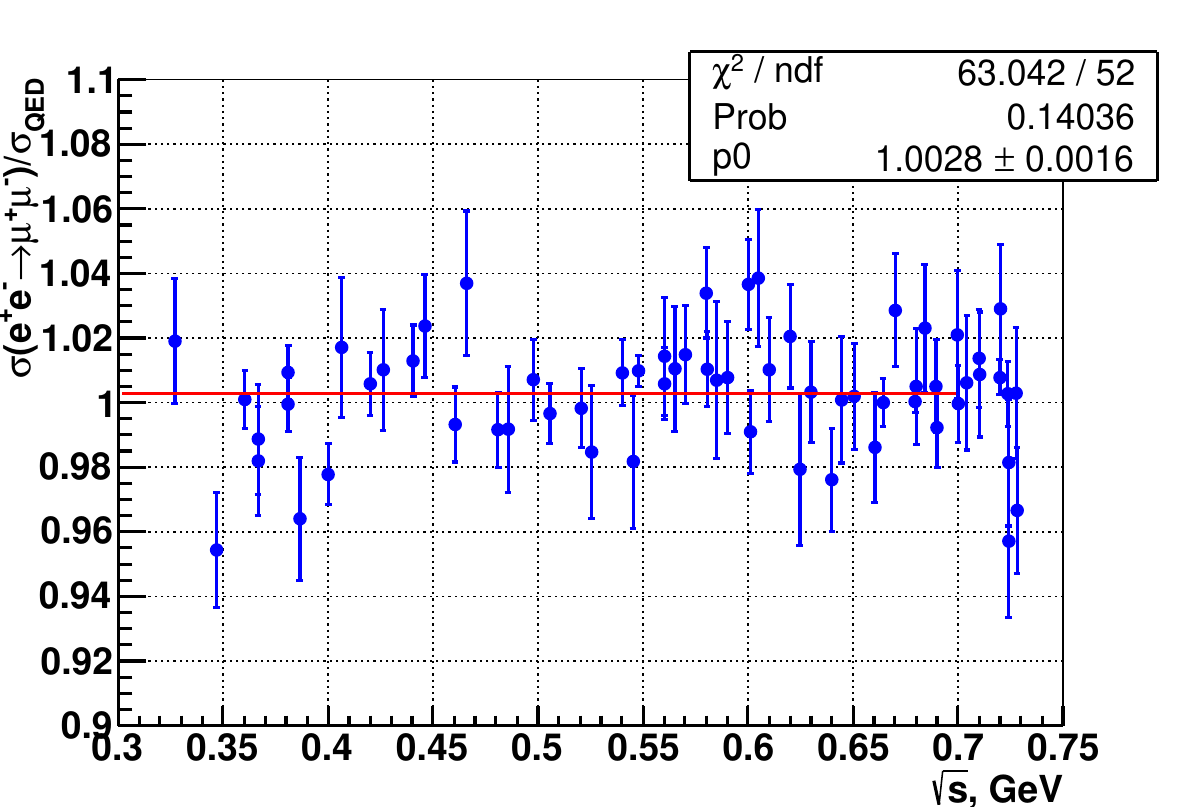}
\caption{Result of the full workflow procedure test on mixed simulated datasets
 for the reconstructed pion form factor (left) and the $e^+e^-\to\mu^+\mu^-$ cross section (right).
 The relative differences of the obtained and implemented in the generator quantities are shown.
}
\label{figfullmcprocess} 
\end{figure*}

In case of the momentum-based particle separation, the
extracted $N_{\pi^+\pi^-}/N_{e^+e^-}$ result is insensitive at
$<0.05\%$ level to fixed/''fixed with 1\% bias'' at the $\sqrt{s}<0.8$~GeV or
fixed/unfixed at the $\sqrt{s}<0.74$~GeV cases for number of $\mu^+\mu^-$ events in
the minimization, with the correlation coefficient $|a|<0.1$ at the $\sqrt{s}<0.6$~GeV,
$a=-0.5$ at the $\sqrt{s}=0.74$~GeV and $a=+0.5$ at the $\sqrt{s}>0.83$~GeV.
The correlation coefficient changes sign at the different energy
ranges as at the lower energies an inaccuracy of muons determination contributes via
the radiative tail of electron while at the higher energies the muons and
pions momenta become closer.

So, fixing the number of the $\mu^+\mu^-$ events to the number of the $e^+e^-$ events according
to the QED prediction and the detector efficiencies results in the increase of the total pion form factor
systematic uncertainty from 1.05\%/ 1.2\%(RHO2013) to 1.6\%/ 2.0\%(RHO2013) at the $\phi$-meson resonance energies with
the $N_{\mu\mu}/N_{\pi\pi}\sim 1$, and from 1.05\% to 1.95\% at
the $\sqrt{s}=1.2$~GeV with the $N_{\mu\mu}/N_{\pi\pi}\sim 2.4$.

The total systematic uncertainty of the $|F_\pi|^{2}$ measurement
from all sources mentioned above is estimated as 0.9\% for
the RHO2013 data taking season and 0.7\% for the other data at the central
$\rho$-meson resonance energies. It is slightly higher at the $\pi^+\pi^-$ production threshold and
grows up to 2\% at above 1~GeV as shown in Table~\ref{systfpi}.

The systematic uncertainty in case of the
$\sigma_{e^+e^-\to\mu^+\mu^-}$ cross section measurement does not have the pion
specific contributions and if not to include the difference between
the MCGPJ and BabaYaga@NLO generators as in Fig.~\ref{fignmmnppnee_radcor}, 
coming from the knowledge of the momentum differential cross section,
then the total systematic uncertainty at the lower energies can be estimated as $0.25\% \oplus
(0.5\%/0.8\%\text{(RHO2013)}) = 0.6\%/0.8\%\text{(RHO2013)}$, where the $0.5\%/0.8\%\text{(RHO2013)}$
corresponds to the fiducial volume uncertainty and the 0.25\%
is a sum of the others.

Further, it is discussed the crucial sources of the systematic uncertainty of the 
pion form factor for purpose of the risk assessment on a possible underestimation
of their contributions.

The estimated systematic uncertainty related to the fiducial volume
determination is taken as the conservative value. It would be difficult to
come up with some tricky model which gives a possible angle bias above
already considered, but at the same time this model should provide
the very remarkable agreement in the
forward-backward charge asymmetry at the level $<0.1\%$ between data and
predictions for both $e^+e^-$
and $\pi^+\pi^-$ processes simultaneously, as described in Sec.~\ref{secasym}.
It will be very helpful to go forward in a precision with available
$e^+e^-\to e^+e^-(\gamma)$ and $\pi^+\pi^-$ generators.
For example, it will be interesting to understand the difference in the C-odd radiative correction
between obtained in the dispersive formalism and the GVMD model
predictions, while the last one reproduces well the obtained experimental result.


The trigger performance is determined by the high track finder efficiency
$>99.9\%$, which is less sensitive to the difference between electron and
pion tracks. The notable variations are present in the cluster
finder, but to affect the overall trigger efficiency it will require some tricky influence from it.
For this purpose, additional hidden correlations between the charged and 
neutral triggers were tried to be scrutinized as it was mentioned above. 

From the analysis point of view, one of the tests was performed on the used workflow in the analysis.
The properly mixed data samples after the full MC simulation,
which take into account the detail detector conditions over time, were
prepared corresponding to the same accumulated luminosities as in the data.
After that the full analysis as on experimental datasets was
performed with the evaluation of efficiencies, particle separations, etc.
Mostly the same procedure, scripts and intermediate files as for the
experimental data were used.
It helps also to ensure that, for example, some inefficiency components
in the described efficiency reconstruction procedure are not double counted.  
The relative difference of the obtained and implemented in the $\pi^+\pi^-$
generator pion form factor is shown in Fig.~\ref{figfullmcprocess}. The average
deviation from the input form factor is $(-0.02\pm 0.03)\%$.
For the three c.m.~energy ranges $0.3\hyph0.6$ / $0.6\hyph0.9$ / $0.9\hyph1.1$~GeV 
the deviations are $(+0.62\pm0.22)\%$ / $(-0.06\pm0.03)\%$ / $(0.49\pm0.13)\%$.
These numbers include the systematic effect from the separation +0.2\%
and +0.6\% at the lowest energies and for the $\sqrt{s}>1$~GeV, respectively.
Separate looks on underneath components were also performed to
understand each single contribution to the final result with better
precision (with the purely statistical Poisson error subtracted or with the larger MC samples).
No notable anomalies in the analysis procedure have been found.
The same compatibility test with MC events is also shown 
for the measurement of the $e^+e^-\to\mu^+\mu^-$ cross section
on the right side in Fig.~\ref{figfullmcprocess}.

\section{Results}
\label{resultsection}

\begin{figure}
\begin{minipage}[t]{.475\textwidth}
\centering
\includegraphics[width=1.05\linewidth]{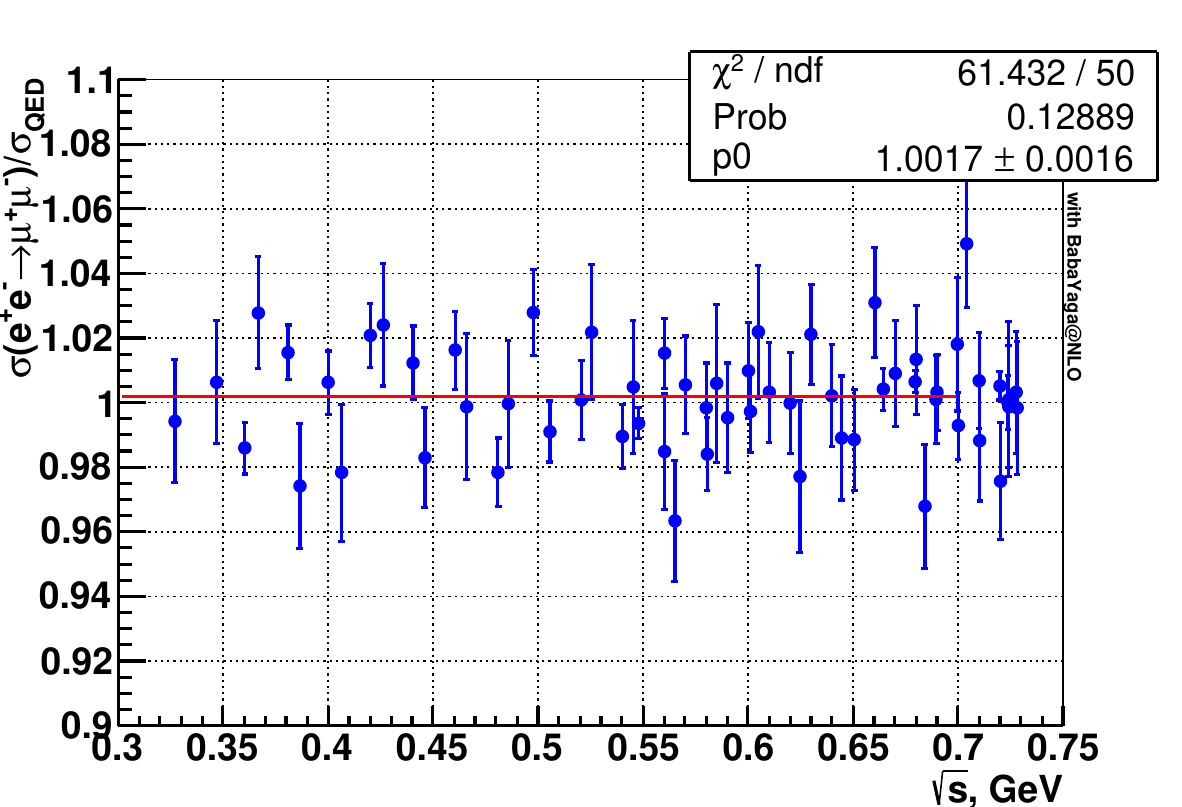}
\caption{The measured cross section of the $\mu^+\mu^-$ production normalized to the QED prediction.
  \label{nmmnee}
} 
\end{minipage}\hfill
\begin{minipage}[t]{.475\textwidth}
\centering
\includegraphics[width=1.05\linewidth]{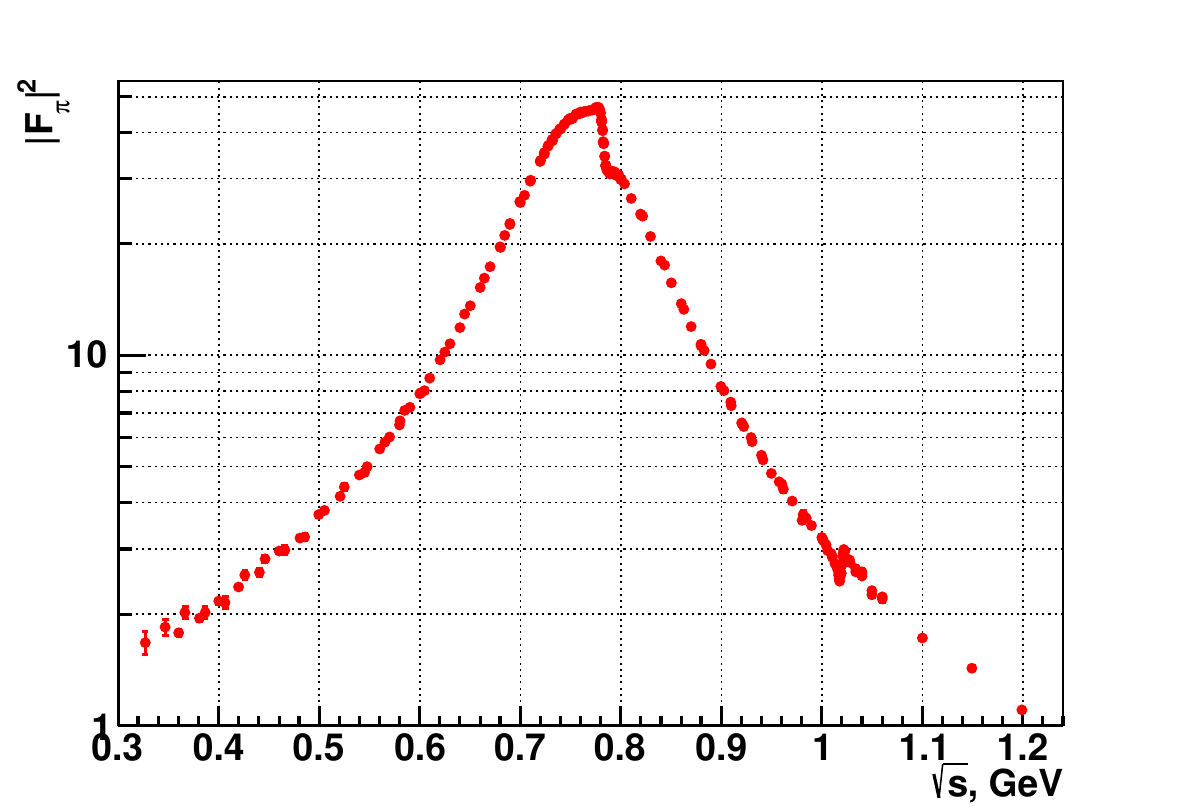}
\caption{The measured CMD-3 pion form factor.}
\label{fpiAbs}
\end{minipage}
\end{figure}

One of the tests in this analysis is the measurement of the 
$e^+e^- \to \mu^+\mu^-$ cross section at low energies, where the
particle separation was performed using momentum information and the number of muons
were extracted separately. The momenta of $\mu^+\mu^-$ particles are
separated from the others up to $\sqrt{s}\lesssim0.7\,\rm GeV$, as, for
example, shown in Fig.~\ref{figmomen}.
The measured cross section is well consistent with the QED prediction with an overall statistical precision of 0.16\%
as shown in Fig.~\ref{nmmnee}. The obtained ratio is
$\sigma_{e^+e^-\to \mu^+\mu^-} / \sigma_{QED} = 1.0017\pm 0.0016$ as
the average at the $\sqrt{s}\leq 0.7$~GeV,
and the separate ratios obtained in the different data taking seasons are also
statistically compatible with each other.
It is very important to take correct differential cross sections as an input of the separation procedure.
The results shown above have been obtained with the BabaYaga@NLO generator based differential
cross sections, while if the MCGPJ generator momentum spectra are used, the relative difference between
the measured and predicted $e^+e^- \to \mu^+\mu^-$ cross sections is about
1.3\% (see the difference from the MCGPJ and BabaYaga@NLO generators usage on the left plot in Fig.~\ref{fignmmnppnee_radcor}).
Similar effect is observed for the $N_{\pi\pi}/N_{ee}$ ratio as shown on the right side in Fig.~\ref{fignmmnppnee_radcor}.
If to compare the cases of the event separations based on energy deposition information 
with one based on momentum information, then 
the inconsistency will be seen at the $\sqrt{s}\sim0.9$~GeV in Fig.~\ref{figPvsE} after applying the difference 
between the MCGPJ and BabaYaga@NLO generators shown in Fig.~\ref{fignmmnppnee_radcor}.

The measured pion form factor values together with the systematic uncertainties are
given in Table~\ref{table:fpi} and shown in Fig.~\ref{fpiAbs}.
This table together with the calculated radiative corrections
$1+\delta_{rad}$ in Eq.~(\ref{formdef}) are provided as Supplemental Material~\cite{supplement}.

The experimental data are fitted using the vector meson dominance (VMD) model:

\begin{multline}
\label{func:GS}
|F_\pi(s)|^2=
\biggr|\Bigr(\mathrm{BW}^{\mathrm{GS}}_{\rho}(s)\cdot
\displaystyle \bigr(\mathstrut 1+\delta_{\omega}  \frac{s}{m_\omega^2}\,
\mathrm{BW}_{\omega}(s)+\delta_{\phi}  \frac{s}{m_\phi^2}\, \mathrm{BW}_{\phi}(s)\bigl)\\+ 
 a_{\rho'} \, \mathrm{BW}^{\mathrm{GS}}_{\rho'}(s)  + a_{\rho''} 
\, \mathrm{BW}^{\mathrm{GS}}_{\rho''}(s)  + a_{cont} \Bigl)/(1+a_{\rho'}+a_{\rho''}+a_{cont})\biggl|^2,
\end{multline}
where the wide $\rho,\rho'$ and $\rho''$ resonances are described by the
Gounaris-Sakurai parametrization (GS)~\cite{Gounaris:1968mw}, and the 
relativistic Breit-Wigner with the constant width,
$\mathrm{BW}_\omega (s) = m^{2}/(m^{2}-s-im\Gamma)$,
is taken for the narrow $\omega$ and $\phi$ meson resonances.
The complex parameters $\delta_{\omega}$, $\delta_{\phi}$, $a_{\rho'}$, $a_{\rho''}$, $a_{cont}$ 
are used to describe the corresponding resonance contributions.
The parameters $\delta_{\omega}$ and $\delta_{\phi}$ are expressed via
the branching fraction $\mathcal{B}_{V\to\pi^+\pi^-}$ values,
using the VMD relation:
\begin{equation}
\sigma_{e^+e^-\to V\to f}(m_V) = \frac{12\pi}{m^2_V} \mathcal{B}_{e^+e^-\to V} \mathcal{B}_{V\to f},
\end{equation}
and together with Eqs.~(\ref{fpitosigma}),~(\ref{func:GS}), it gives the following expression:
\begin{equation}
  \mathcal{B}_{e^+e^-\to V} \mathcal{B}_{V\to\pi^+\pi^-} = 
  \frac{\alpha^{2}\beta_\pi^3(m_V)}{36}
  \bigr| \mathrm{BW}^{\mathrm{GS}}_{\rho}(m_V^2) \frac{m_V}{\Gamma_V}
  \frac{\delta_V}{1+a_{\rho'}+a_{\rho''}+a_{cont}} \bigl|^{2}.
\end{equation}
The phase of the $\delta_\phi$ is introduced as a relative to the phase of the $F_\pi$
without the $\phi$ resonance contribution,
representing the local $\phi$-meson resonance behavior as $F_\pi \sim A(s)\cdot(1+\tilde{\delta}_\phi\mathrm{BW}_{\phi}(s))$.
The parameter $\tilde{\delta}_\phi$ is expressed via the ${\delta}_\phi$ by the following equation:
\begin{equation}
  \mathrm{arg}(\tilde{\delta}_\phi) = \mathrm{arg}(\delta_\phi) 
  - \mathrm{arg}\bigr(F_\pi^{\delta_\phi=0}(m_V^2)\bigl)
  +
  \mathrm{arg}\bigr(\mathrm{BW}^{\mathrm{GS}}_{\rho}(m_\phi^2)/(1+a_{\rho'}+a_{\rho''}+a_{cont})\bigl).
\end{equation}
Such definition of the $\mathrm{arg}(\tilde{\delta}_\phi)$ provides more consistent result
for possible different $\rho'$ and $\rho''$ parametrizations.

\begin{table*}[ht]
\begin{center}
  \caption{\label{fitpar}
    Result of the fit of the CMD-3 data using the pion form factor parametrization from Eq.~(\ref{func:GS}).
    All uncertainties here are only statistical of the fit.
    The second errors come from the propagation of statistical uncertainties of 
    the $\rho'$ and $\rho''$ parameters fitted with the help of external
    data in the extended energy range.
    The second column corresponds to the fit
    where $m$ and $\Gamma$ of the $\phi$ and $\omega$ meson resonances are
    constrained by the values and their errors from the PDG.
    The result of the combined fit of the CMD-3, CMD-2 (at the $\sqrt{s}>1.1$~GeV) and DM2 data to
    determine $\rho'$ and $\rho''$ parameters is shown in the bottom part.
  }
  \small
  \begin{tabular}{lccc}
    Parameter & value & $M_{\phi,\omega}, \Gamma_{\phi,\omega}$ constrained& PDG(2022)~\cite{Workman:2022ynf} \\
                   &  & by PDG's values &  \\ \hline
$m_\rho$, MeV                       & $ 775.41 \pm 0.08 \pm 0.07 $ & $ 775.4 \pm 0.07 \pm 0.07 $ & $ 775.26 \pm 0.23 $ \\ 
$\Gamma_\rho$, MeV                  & $ 148.8 \pm 0.16 \pm 0.05 $ & $ 148.76 \pm 0.16 \pm 0.06 $ & $ 147.4 \pm 0.8 $ \\
$m_\omega$, MeV                     & $ 782.43 \pm 0.03 \pm 0.01 $ & $ 782.44 \pm 0.03 \pm 0.01 $ & $ 782.66 \pm 0.13 $ \\ 
$\Gamma_\omega$, MeV                & $ 8.57 \pm 0.06 \pm 0.01 $ & $ 8.59 \pm 0.06 \pm 0.01 $ & $ 8.68 \pm 0.13 $ \\
$\mathcal{B}_{\omega\to\pi^+\pi^-}\mathcal{B}_{\omega\to e^+e^-}, 10^{-6}$ & $ 1.204 \pm 0.009 \pm 0.003 $ & $ 1.204 \pm 0.009 \pm 0.004 $ & $ 1.28 \pm 0.05 $ \\ 
$\mathrm{arg}(\delta_\omega)$, rad  & $ 0.167 \pm 0.008 \pm 0.01 $ & $ 0.169 \pm 0.008 \pm 0.012 $ &  \\ 
$m_\phi$, MeV                       & $ 1019.761 \pm 0.128 \pm 0.022 $ & $ 1019.465 \pm 0.016 \pm 0 $ & $ 1019.461 \pm 0.016 $ \\ 
$\Gamma_\phi$, MeV                  & $ 4.681 \pm 0.271 \pm 0.058 $ & $ 4.25 \pm 0.013 \pm 0 $ & $ 4.249 \pm 0.013 $ \\ 
$\mathcal{B}_{\phi\to\pi^+\pi^-}\mathcal{B}_{\phi\to e^+e^-}, 10^{-8}$   & $ 3.65 \pm 0.24 \pm 0.02 $ & $ 3.51 \pm 0.22 \pm 0.03 $ & $ 2.2 \pm 0.4 $ \\ 
$\mathrm{arg}(\tilde{\delta}_\phi)$, rad & $ 2.883 \pm 0.052 \pm 0.011 $ & $ 2.77 \pm 0.023 \pm 0.006 $ & \\ 
$|a_{cont}|$                        & $ 0.0975 \pm 0.0011 \pm 0.0096 $ & $ 0.0971 \pm 0.001 \pm 0.0106 $ & \\ 
$\mathrm{arg}(a_{cont})$, rad       & $ 2.337 \pm 0.021 \pm 0.286 $ & $ 2.344 \pm 0.02 \pm 0.309 $ & \\ \hline
$\chi^{2}/ndf$                      &  212.53  /  195  &  223.42  /  199  & \\ 
$m_\rho'$, MeV                      & \multicolumn{2}{c}{ $ 1226.22 \pm 24.76 $ } & $ 1465 \pm 25 $ \\ 
$\Gamma_\rho'$, MeV                 & \multicolumn{2}{c}{ $ 272.97 \pm 45.53 $ } & $ 400. \pm 60 $ \\ 
$m_\rho''$, MeV                     & \multicolumn{2}{c}{ $ 1604.66 \pm 30.8 $ } & $ 1720 \pm 20 $ \\ 
$\Gamma_\rho''$, MeV                & \multicolumn{2}{c}{ $ 249.39 \pm 52.24 $ } & $ 250. \pm 100 $ \\ 
$|a_\rho'|$                         & \multicolumn{2}{c}{ $ 0.3589 \pm 0.0693 $ } & \\ 
$|a_\rho''|$                        & \multicolumn{2}{c}{ $ 0.1042 \pm 0.031 $ } & \\ 
$\mathrm{arg}(a_\rho')$, rad        & \multicolumn{2}{c}{ $ -1.831 \pm 0.07 $ } & \\ 
$\mathrm{arg}(a_\rho'')$, rad       & \multicolumn{2}{c}{ $ 3.384 \pm 0.234 $ } & \\ \hline

$\chi^2/ndf$  & \multicolumn{2}{c}{288.87/240} \\
{\tiny CMD3+CMD2+DM2 } 
& \multicolumn{3}{l}{ \small $\chi^2=$ 220.08(CMD3)+25.30(CMD2)+40.10(DM2)+3.39(PDG)} \\
& \multicolumn{3}{l}{\small ndf= 207+29+20+4 - 12($\rho,\omega,\phi,cont$) - 8($\rho',\rho''$)} \\
\end{tabular}
\end{center}
\end{table*}

Parameters of $\rho'$ and $\rho''$ resonances are obtained from the combined fit of
the CMD-3 pion form factor at the $\sqrt{s}\leq 1.1$~GeV together with the
CMD-2 points at the $\sqrt{s}\geq 1.1$~GeV~\cite{CMD-2:2005mvb} and
DM2 result at the $\sqrt{s}\geq 1.35$~GeV~\cite{DM2:1988xqd}.
The obtained $\rho'$ and $\rho''$ parameters are relevant here
only for the functional form factor behavior, and they should not be directly compared
to the PDG's values, since for the latter more dedicated description of these resonances
have to be taken into account.
The fitted $a_{cont}$ constant represents a continuum
contribution, which also could partially absorb the part of the $\rho'$ and $\rho''$
resonance contributions and can account for left tails of the
excitations $\rho'''$ and other higher resonances.
The obtained parameters of the $\rho'$ and $\rho''$ are strongly depend on
freedom of the $a_{cont}$ contribution and the model used for these
resonances description, systematic uncertainty analysis of which is
out of the scope of this paper since they rely on the external data in the extended energy range.
The result of the fit of the solely CMD-3 points at the c.m.~energies
$\sqrt{s}\leq 1.1$~GeV is given in the Table~\ref{fitpar},
were the second errors includes the additional propagation of
statistical uncertainties of the fixed $\rho'$ and $\rho''$ parameters
obtained in the extended energy range.
The masses of the $\omega$ and $\phi$ mesons, $M_\omega$ and $M_\phi$, are slightly shifted from their
world average PDG's values~\cite{Workman:2022ynf}.
This was also observed in more dedicated analysis in the framework of
dispersion relations of the previously measured $e^+e^-\to\pi^+\pi^-$
experimental data~\cite{Colangelo:2018mtw,Colangelo:2022prz} in
comparison with the $\pi^0\gamma, 3\pi, KK$ channels~\cite{Hoid:2020xjs,Hoferichter:2019mqg,Stamen:2022uqh}.
The shift is not originated from the beam energy miscalibration since
the $\omega$-meson cross section peak in the $3\pi$ channel (Fig.~\ref{fig3pi} and similar result at the $\phi$-meson resonance energies) 
and CMD-3 analysis of the $K\bar{K}$ production 
on and around $\phi$-meson resonance~\cite{CMD-3:2016nhy,Kozyrev:2017agm} show mass values consistent with
those in PDG. 
The origin of the obtained mass shift in the $\pi^+\pi^-$
channel can be the $\phi$-meson parametrization together with $\rho$ interference. 
To account for the last case, a constrained fit was performed with the $M_{\omega,\phi}, \Gamma_{\omega,\phi}$ values
and their corresponding errors taken from the PDG(2022)~\cite{Workman:2022ynf}.
Parameters obtained in this fit are presented in the second column
of the Table~\ref{fitpar} and considered as the baseline result, while
the differences of the two sets of parameters are assigned as additional systematic uncertainties
for some of the parameters discussed below.

\begin{figure}[t]
  \centering
\includegraphics[width=0.85\linewidth]{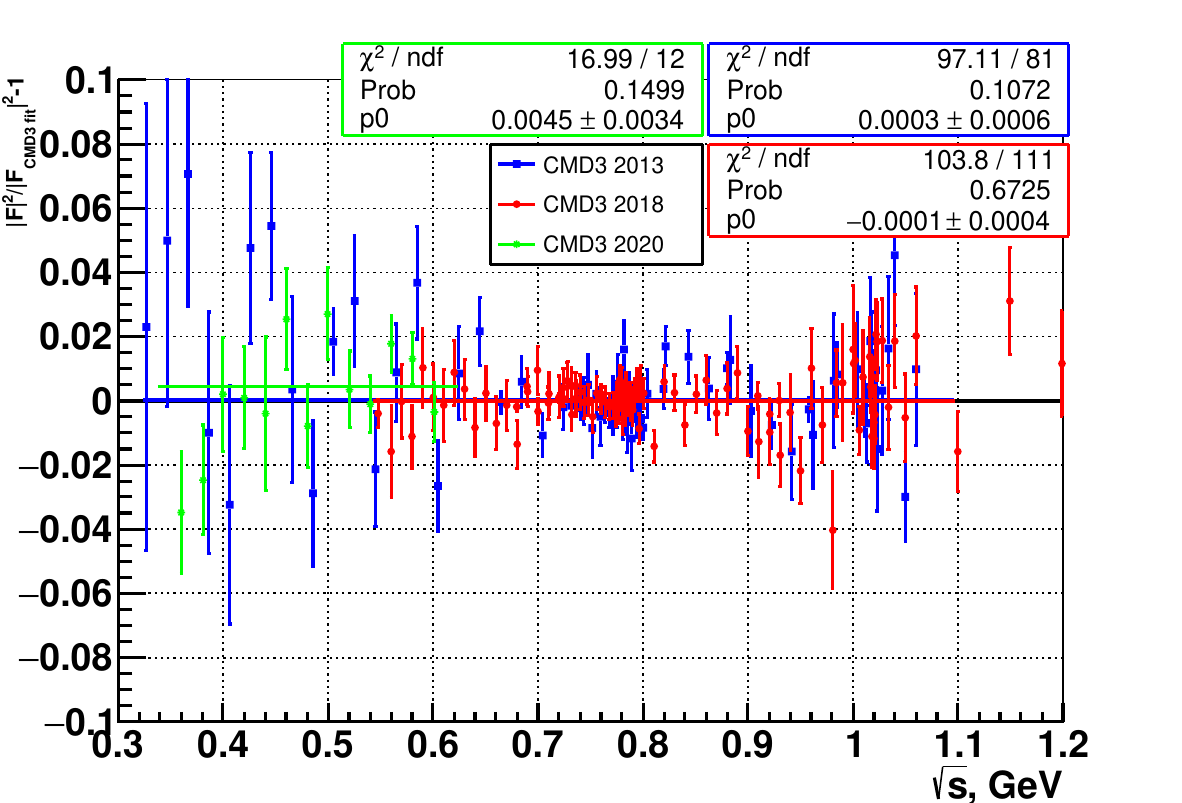}
\caption{The relative deviation of the measured pion form factor from the
  fitted function. Different types of markers correspond to different
  datasets of RHO2013, RHO2018, and LOW2020 data taking seasons.
  The fit parameters show the average difference between the data and the fitted function.
}
\label{fpifitseasons} 
\end{figure}

The data samples used in the analysis were accumulated during three seasons: RHO2013, RHO2018 and LOW2020. 
The relative deviations of the measured form factors from the fitted curve are shown in
Fig.~\ref{fpifitseasons}. 
Data of the different datasets are shown by different colors.
The datasets show good compatibility with the average differences:
$\Delta(RHO2018-RHO2013) = (-0.04 \pm 0.07)\%$ and
$\Delta(LOW2020-RHO2013) = (-0.5 \pm 0.6)\%$ at the $\sqrt{s}\leq 0.6$~GeV.
In spite of the different efficiencies and the polar angle related systematics 
in the drift chamber during RHO2013 season, the results from all seasons are very similar.

\begin{figure}
\begin{minipage}[t]{.475\textwidth}
\centering
\includegraphics[width=1.1\linewidth]{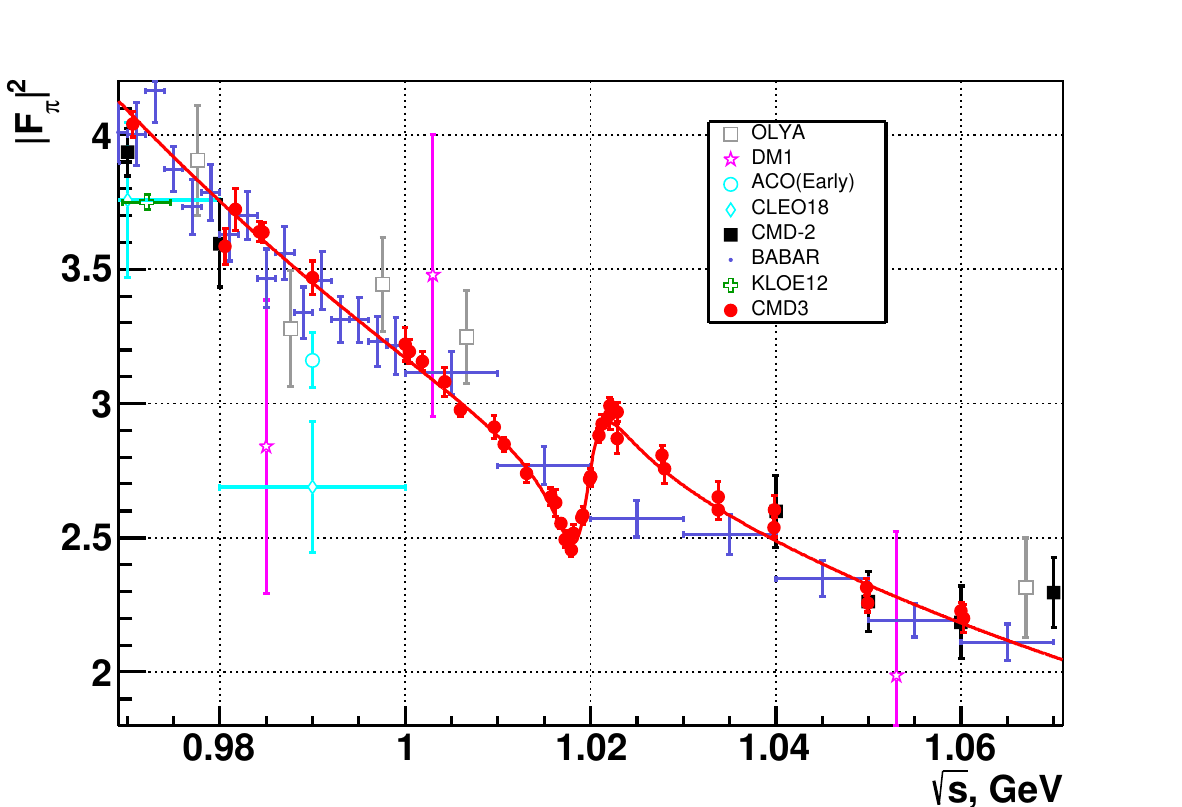}
\caption{The pion form factor measured by CMD-3 and the other experiments near the $\phi$-meson resonance.}
\label{fpiphi}
\end{minipage}
\hfill 
\begin{minipage}[t]{.475\textwidth}
\centering
\includegraphics[width=1.1\linewidth]{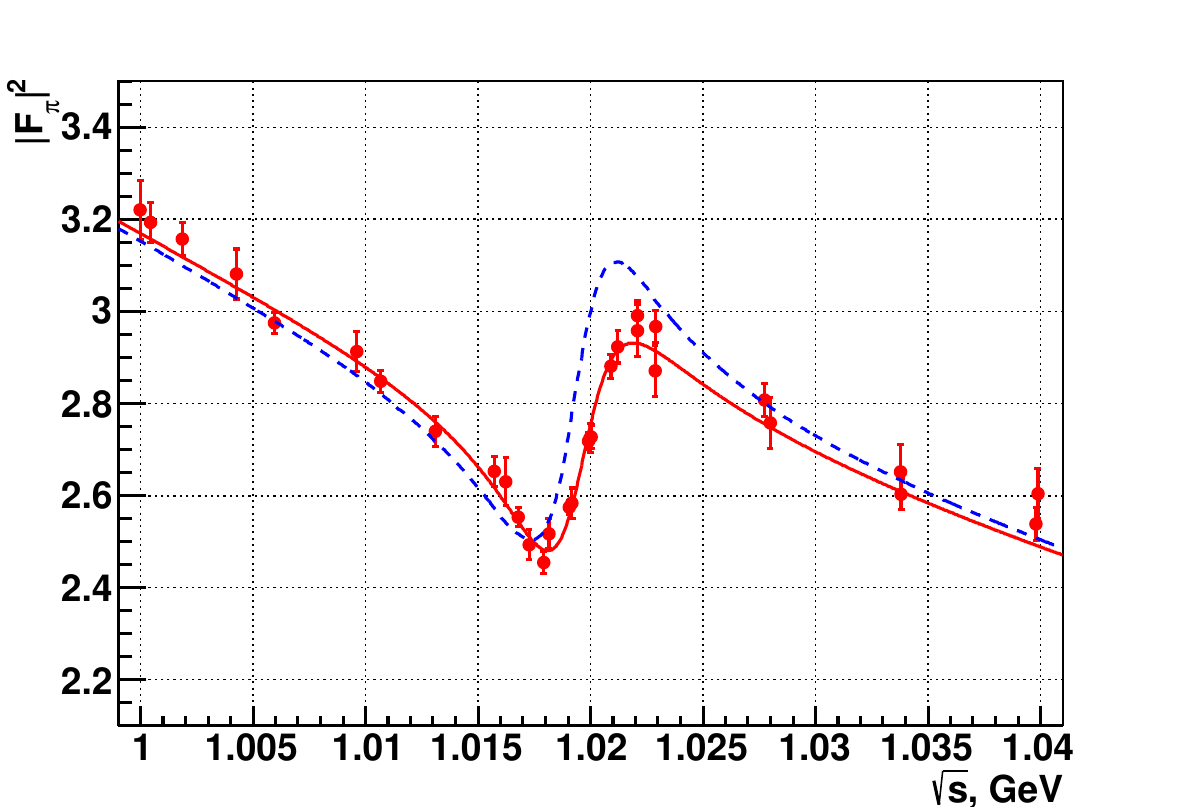}
\caption{The comparison of the measured form factor
  with the $\phi$-interference coming from the vacuum polarization term,
  shown by the blue-dashed line and calculated as $|F_\pi^{\delta_\phi=0}(s)\frac{1- {\cal P}_{not-\phi-res}(m_\phi^2)}{1- {\cal P}(s)}|^{2}$}
\label{fpiphiVP} 
\end{minipage}
\end{figure}

This analysis provides the measurement of the pion form factor at the $\phi$-meson
resonance for the first time. 
Figure~\ref{fpiphi} shows the form factor around $\phi$-meson resonance 
together with the constrained fit function and the measurements of the other experiments
at the off-peak energies: OLYA~\cite{Barkov:1985ac},
DM1~\cite{Quenzer:1978qt}, ACO~\cite{Cosme:1976ft},
CLEO~\cite{Xiao:2017dqv}, CMD-2~\cite{CMD-2:2005mvb},
$BABAR$~\cite{BaBar:2012bdw}, KLOE~\cite{KLOE:2012anl}.
The obtained parameters of the $\phi\to\pi^+\pi^-$ interference together
with the systematics uncertainties are the following:
\begin{align}
  \label{phi2pi}
|F_\pi^{\delta_\phi=0}(m_\phi^2)|^{2} &= 2.808 \pm 0.008 \pm 0.042 \nonumber \\
\psi_{\pi}=\mathrm{arg}(\tilde{\delta}_\phi)-\pi &= -(21.3\pm2.0\pm 2.6 \pm 9.7)\degree \\
\mathcal{B}_{\phi\to\pi^+\pi^-}\mathcal{B}_{\phi\to e^+e^-} &=
(3.51\pm0.33 \pm 0.10 \pm 0.22)\times 10^{-8} \nonumber
\end{align}
Where the first and second errors are statistical and systematic parts, and the third error corresponds to
the difference between PDG constrained/unconstrained fits.

The systematic errors of the $\phi\to\pi^+\pi^-$ decay parameters include the
variations of the cross section within the total systematic uncertainty in the Table~\ref{systfpi}, 
variations of the $\phi$-meson resonance backgrounds,
different fit function parametrizations of the $\phi$, $\rho'$, $\rho''$ in the global fit.
The significant contribution comes from the uncertainty of the radiative
correction calculation.
The initial systematic or statistical shift of the measured branching ratio $\mathcal{B}_{\phi\to\pi^+\pi^-}$ or phase $\psi_\pi$ 
after the iterative procedure to calculate the radiative correction results in the corresponding bias in the 
radiative correction. If to repeat the fit of the data with the biased radiative correction and extract the 
$\mathcal{B}_{\phi\to\pi^+\pi^-}$ and $\psi_\pi$ values again, 
their shift will be inflated by an additional factor in comparison with the initial one. 
By repeating this iteratively, the final inflation factor will be $1.45-1.50$.
This is an effect of the inflation of the uncertainties 
of $\mathcal{B}_{\phi\to\pi^+\pi^-}$ and $\psi_\pi$ due to the radiative corrections.
The listed errors of the $\mathcal{B}_{\phi\to\pi^+\pi^-}\mathcal{B}_{\phi\to e^+e^-}$ and
$\psi_{\pi}$ include this scale factor, both for statistical and systematic uncertainties.









The obtained phase $\psi_{\pi}$ of $\rho-\phi$ mixing is in the
agreement with the theoretical prediction $-(11 \hyph 28)\degree$ in the
papers~\cite{Achasov:1989kz,Achasov:1990at}(Eqs. (3.20),(3.21) from ~\cite{Achasov:1990at}).

The $\phi$ interference was previously studied using only the detected numbers of
events by the OLYA~\cite{Vasserman:1981ax} experiment or by using visible cross section
measured by ND~\cite{Golubev:1986vv} and SND~\cite{Achasov:2000pu} at
the VEPP-2M $e^+e^-$ collider.
In the SND case, a simultaneous fit together with possible background
contributions in the detected cross section was performed. SND result
for the $\phi\to\pi^+\pi^-$ interference is the following:
$|F_\pi^{\phi}|^{2} = 2.98 \pm 0.02 \pm 0.16$,
$\psi_{\pi}=-(34\pm 5)\degree$ 
and $\mathcal{B}_{\phi\to\pi^+\pi^-}\mathcal{B}_{\phi\to e^+e^-} = (2.1\pm 0.4)\times 10^{-8}$
(according to Eq.(14) from the SND paper~\cite{Achasov:2000pu}). 
Their parameters show $2.5$ standard deviation difference from the CMD-3 result.
This deviation can be partially
explained by the overestimated pion form factor value (and possibly a
resonance background underestimation) and by an uncertainty of the applied
radiative corrections due to the input resonance parameters, 
which together gives back compatibility to $1.5\hyph2 \sigma$.

It is interesting to note that the obtained branching fraction is smaller
than that expected from the vacuum polarization term, in the assumption that there is no direct
$\phi\to\pi^+\pi^-$ transition:
\begin{equation}
(\mathcal{B}_{\phi\to\pi^+\pi^-}\mathcal{B}_{\phi\to e^+e^-})^{VP} = 
\frac{\beta_\pi^3(m_\phi^2)}{4}|F_\pi^{\delta_\phi=0}(m_\phi^2)|^{2}
\mathcal{B}_{\phi\to e^+e^-}^2|1- {\cal P}_{not-\phi-res}(m_\phi^2)|^2 
\sim 5.3\times 10^{-8},
\end{equation}
where $|1- {\cal P}_{not-\phi-res}(m_\phi)|^2 \sim 0.9613 $ -- non
$\phi$-resonant part of the vacuum polarization~\cite{Ignatov:2008bfz,WorkingGrouponRadiativeCorrections:2010bjp}.
Figure ~\ref{fpiphiVP} shows the comparison of the measured form factor with the expectation coming only from
the vacuum polarization described by the
$|F_\pi^{\delta_\phi=0}(s)\frac{1- {\cal P}_{not-\phi-res}(m_\phi^2)}{1- {\cal P}(s)}|^{2}$ function.

The obtained result for the $\omega\to\pi^+\pi^-$ decay is the following:
\begin{align}
  \label{omega2pi}
\mathcal{B}_{\omega\to\pi^+\pi^-}\mathcal{B}_{\omega\to e^+e^-} &=
(1.204\pm0.013 \pm 0.023 )\times 10^{-6} ,
\end{align}
where the first and second errors correspond to the statistical and
systematic uncertainties. Both errors include the $\sim1.3$ scale factor related to
the uncertainty of the radiative correction. It is obtained in the same way as
discussed above for the $\phi$-meson case.

\begin{figure}
\begin{minipage}[t]{.475\textwidth}
\centering
\includegraphics[width=1.1\linewidth]{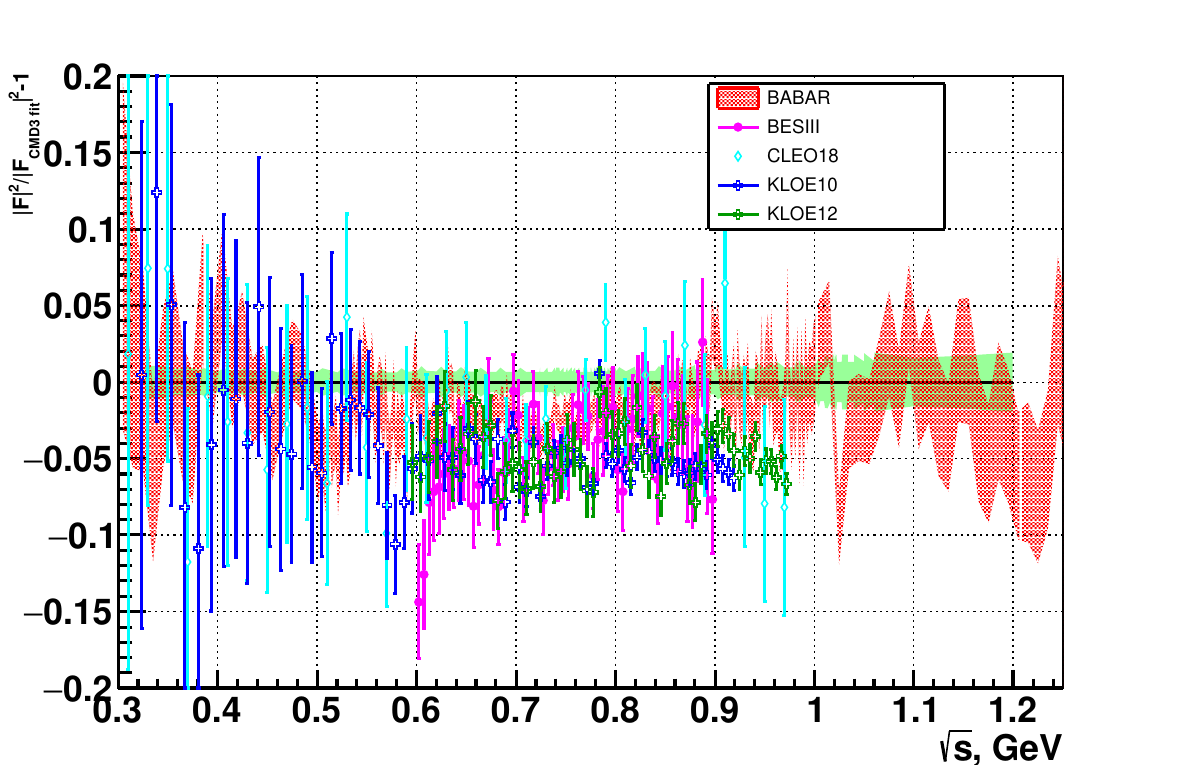}
\caption{The relative differences of the pion form factors obtained 
in the ISR experiments ($BABAR$, BESIII, CLEO, KLOE) and the CMD-3 fit
result. The green band corresponds to the systematic uncertainty of the 
CMD-3 measurement.}
\label{fpiisrvscmd3}
\end{minipage}
\hspace{.03\textwidth}
\begin{minipage}[t]{.475\textwidth}
\centering
\includegraphics[width=1.1\linewidth]{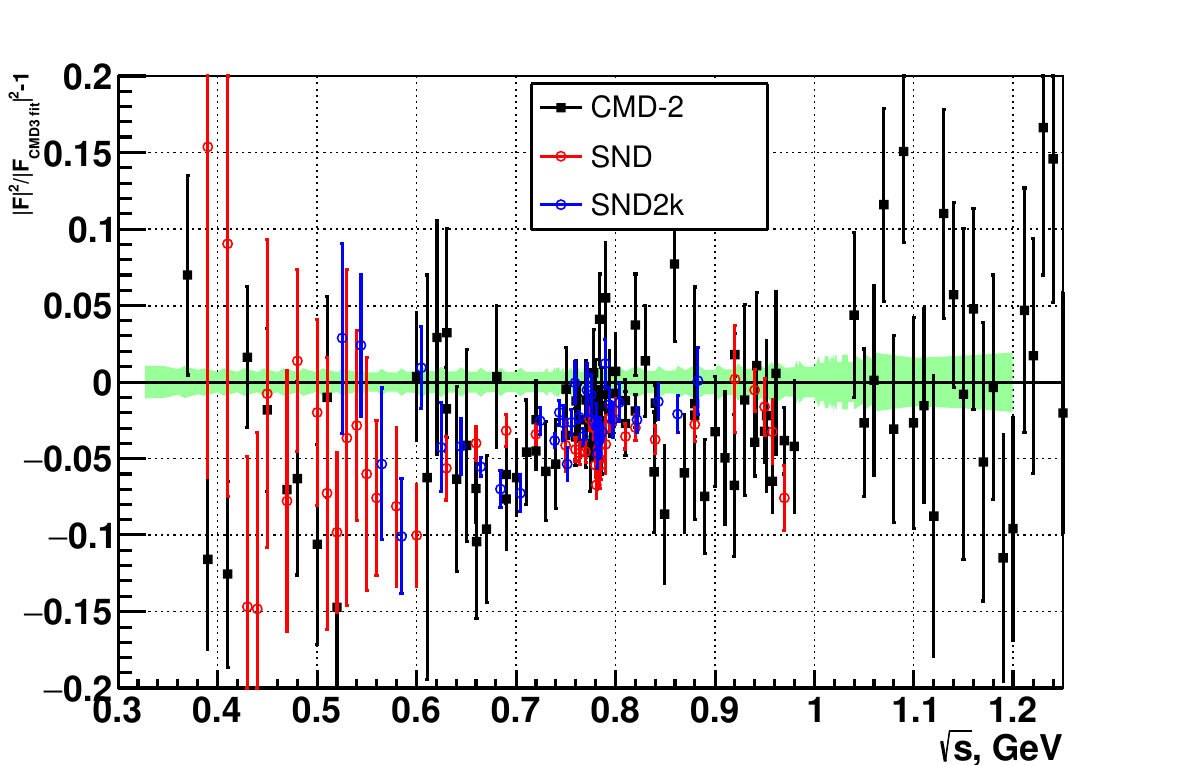}
\caption{The relative differences of the pion form factors obtained 
in the previous energy scan measurements (CMD-2, SND, SND2k) and the
CMD-3 fit result. The green band corresponds to the systematic uncertainty of the 
CMD-3 measurement.}
\label{fpicmd2sndvscmd3} 
\end{minipage}
\end{figure}

\begin{table*}[t]
  \begin{minipage}[b]{.515\textwidth}
    \centering
    \includegraphics[width=1.1\linewidth]{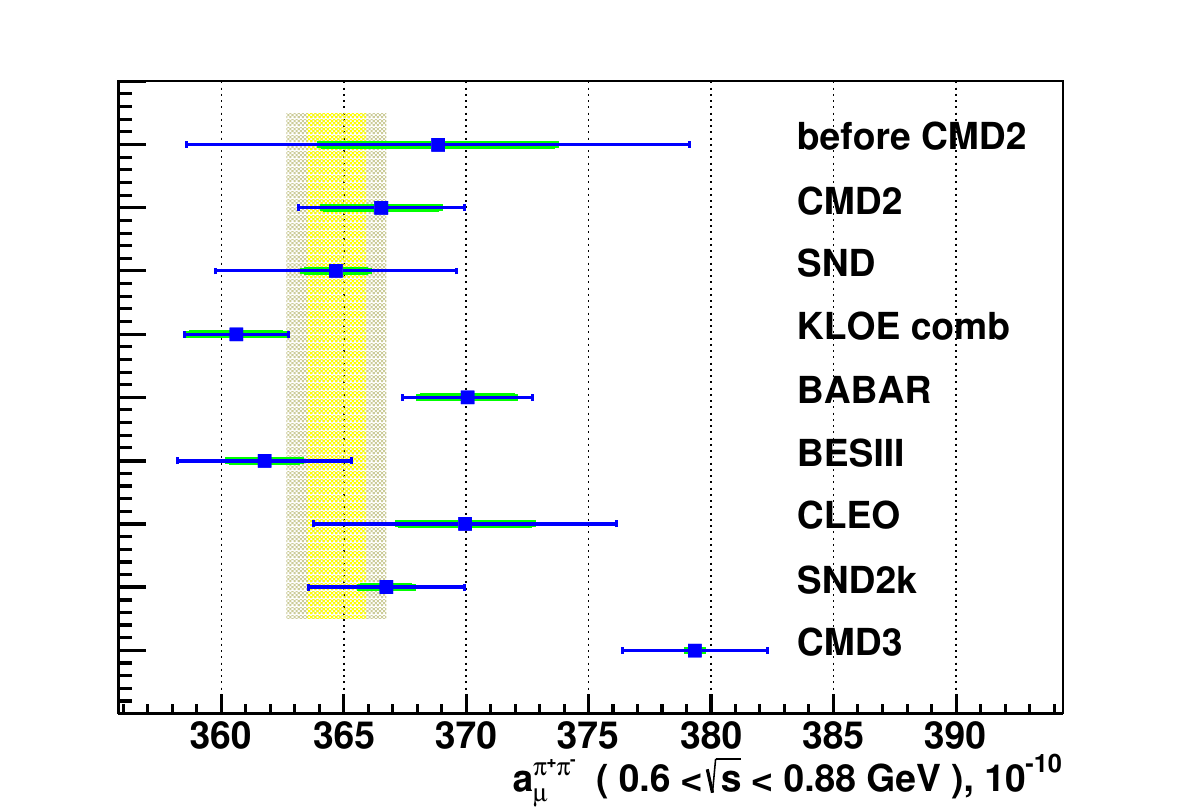}
    \captionof{figure}{The $\pi^+\pi^-(\gamma)$ contribution to the $a_\mu^{had,LO}$
      from the energy range $0.6<\sqrt{s}<0.88$~GeV obtained from the CMD-3 data and 
      the results of the other experiments.
 }
    \label{amurho}
  \end{minipage}\hfill
  \begin{minipage}[b]{.435\textwidth}
    \vskip -0.25in
    \centering
    \small
    \begin{tabular}{lc} 
      Experiment & $a_\mu^{\pi^+\pi^-, LO}, 10^{-10}$ \\ \hline
      before CMD2	 & $368.8 \pm 10.3$ \\ 
      CMD2	         & $366.5 \pm 3.4$ \\ 
      SND              	 & $364.7 \pm 4.9$ \\ 
      KLOE 		 & $360.6 \pm 2.1$ \\ 
      BABAR		 & $370.1 \pm 2.7$ \\ 
      BESIII		 & $361.8 \pm 3.6$ \\ 
      CLEO		 & $370.0 \pm 6.2$ \\ 
      SND2k		 & $366.7 \pm 3.2$ \\ 
      CMD3		 & $379.3 \pm 3.0$ \\ \hline
    \end{tabular}

   \vskip 0.25in
    \caption{\label{amufpinumbers} The $\pi^+\pi^-(\gamma)$ contribution to the $a_\mu^{had,LO}$
      from the energy range $0.6<\sqrt{s}<0.88$~GeV obtained from the CMD-3 data and 
      the results of the other experiments.}
  \end{minipage}
\end{table*}

The comparison of the pion form factor measured in this work with the results obtained in 
the most recent ISR experiments
($BABAR$~\cite{BaBar:2012bdw}, KLOE~\cite{KLOE:2010qei,KLOE:2012anl}, BESIII~\cite{BESIII:2015equ}) is shown in 
Fig.~\ref{fpiisrvscmd3}. 
The comparison with the most precise previous energy scan experiments 
(CMD-2~\cite{CMD-2:2003gqi,CMD-2:2005mvb,Aulchenko:2006dxz,CMD-2:2006gxt},
SND~\cite{Achasov:2006vp} at the VEPP-2M and SND~\cite{SND:2020nwa} at the VEPP-2000) 
is shown in Fig.~\ref{fpicmd2sndvscmd3}.
The $e^+e^-\to\pi^+\pi^-$ cross section obtained by CMD-3
significantly deviates from the results of previous measurements,
including the one performed by the CMD-2 experiment, the predecessor
of CMD-3. It should be noted that the same scale discrepancies have
already been observed in previous measurements, e.g., between KLOE and
$BABAR$ as seen in Fig.~\ref{fpiisrvscmd3}. The reason for these discrepancies is unknown at the
moment. CMD-3 and CMD-2, as well as SND, are the same-type
experiments, out of which CMD-3 is the next-generation one,
considering improvements in detector performance, much more
sophisticated data analysis and a comprehensive study of systematic
effects based on more than an order of magnitude larger statistics.
The limitation in available statistics before may have led to some
systematic contributions being missed from consideration
as an effect may be hidden under statistical precision.
The CMD-3 and CMD-2 share only one detector subsystem, the
$Z$-chamber. Therefore, CMD-3 and CMD-2 should be considered as
independent experiments in series of $e^+e^-\to\pi^+\pi^-$
cross-section measurements. Further studies based on data from
existing experiments or confirmations from new measurements are
required to understand the observed discrepancies.

One of the main applications of the measured $e^+e^-\to hadrons$
cross sections is the evaluation of the hadronic part of
the anomalous magnetic moment of the muon $a_\mu=(g_\mu-2)/2$, which is calculated
via the dispersive integral~\cite{Gourdin:1969dm}:
\begin{equation}
  a^{had,LO}_{\mu} =
  \frac{m^{2}_{\mu}}{12\pi^3}\int_{4m_\pi^2}^{\infty}
  \frac{\sigma_{e^{+}e^{-}\to\gamma^*\to hadrons}(s) K(s)}{s}ds,
\end{equation}
where $K(s)$ -- is known analytical kernel function. 
The dispersive integral requires the bare cross section, 
which can be obtained from the measured one $\sigma_{\pi^+\pi^-} (s)$:
\begin{equation}
  \sigma_{e^{+}e^{-}\to\gamma^*\to\pi^+\pi^-(\gamma)}^{bare} (s)= 
  \sigma_{e^{+}e^{-}\to\pi^+\pi^-} (s) \cdot
|1- {\cal P}(s)|^2 \cdot (1+\frac{\alpha}{\pi}\Lambda(s)),
\end{equation}
by subtracting the vacuum polarization ${\cal P}(s)$ of the intermediate photon and adding the
final state radiation in the pointlike approximation with the $\Lambda(s)$ term~\cite{Bystritskiy:2005ib,Hoefer:2001mx}. 
The most recent the muon (g-2)/2 evaluations can be found in the
references~\cite{Keshavarzi:2019abf,Davier:2019can,Aoyama:2020ynm},
and the $\pi^+\pi^-$ hadronic channel gives the dominant contribution to the hadronic
part and determines the overall precision of the full $a_\mu$ value.

In order to evaluate the impact of our result it is calculated the contribution to the $a_\mu^{had,LO}$ 
from the $e^+e^-\to \pi^+\pi^-$ process using data from the various experiments
in the common energy range $0.6<\sqrt{s}<0.88$~GeV. 
This particular energy range was chosen for two reasons: 
it is covered by many high-precision experiments and 
it gives more than 50\% of the full $a_\mu^{had,LO}$ integral.
The dispersive integral calculation is performed by using linear
interpolation between the experimental points with the proper account of
the statistical and systematic errors.
The results of the $a_\mu^{had,LO}$ calculations are shown in Fig.~\ref{amurho} and
given in Table~\ref{amufpinumbers}. The yellow band corresponds to
the average of all experiments before CMD-3, where the gray band
includes additional uncertainty inflation due to the KLOE/$BABAR$
inconsistency. The first line in 
Table~\ref{amufpinumbers} corresponds to the combined result of all
measurements before CMD-2 experiment. 

The pion form factor measurements
from the RHO2013 and RHO2018 seasons of the CMD-3 experiment give the statistically
consistent results for the $a_\mu^{had,LO}$ integral as: \\
\begin{align}
  a_\mu^{\pi\pi,LO}(\mathrm{RHO2013}) &= \quad (380.06 \pm 0.61 \pm 3.64)\times 10^{-10} \nonumber \\
  a_\mu^{\pi\pi,LO}(\mathrm{RHO2018}) &= \quad (379.30 \pm 0.33 \pm 2.62)\times 10^{-10} \nonumber \\
  a_\mu^{\pi\pi,LO}(\mathrm{average}) &= \quad (379.35 \pm 0.30 \pm 2.95)\times 10^{-10} 
\end{align}
Two CMD-3 values are in very good agreement 
in spite of a very different data taking conditions (as was discussed earlier). 
The combined CMD-3 result is obtained in a very conservative assumption of 100\% 
correlation between systematic errors of the two datasets. 
The CMD-3 result on the $a_\mu^{had,LO}$ is significantly higher in comparison with 
the results of the other $e^+e^-$ energy scan and ISR experiments. 
Although this evaluation is done in the limited 
energy range and the full evaluation of the $a_\mu^{had,LO}$ is yet to be done, 
it is clear that our measurement reduces the tension between 
the experimental value of the anomalous magnetic moment of muon and its Standard Model prediction. 
The correlated impact on the $a_\mu^{had,LO}$, $\alpha_{QED}(M_{Z})$, $<r^{2}_\pi>$ from different pion form factor behaviors and possible
increase in the hadronic cross section at different energy ranges have
been discussed in a number of papers~\cite{Passera:2008jk,Keshavarzi:2020bfy,Malaescu:2020zuc,Colangelo:2020lcg}.

\begin{table*}[t]
\centering
\caption{\label{table:fpi} CMD-3 pion form factor $|F_{\pi}|^2$.
The first error is statistical, and the second error is systematic. 
The sources of the $|F_{\pi}|^2$ systematic uncertainty see in Table~\ref{systfpi}.
}
\begin{footnotesize}
\tabcolsep=0.1cm
\renewcommand{\arraystretch}{1.15}
\scalebox{0.77}{
 \begin{tabular}{|r|r|r|r|r|r|r|r|}
\hline
$\sqrt{s}, MeV$ &  \multicolumn{1}{|c|}{${|F_{\pi}|}^{2}$} & $\sqrt{s}, MeV$ &  \multicolumn{1}{|c|}{${|F_{\pi}|}^{2}$} & $\sqrt{s}, MeV$ &  \multicolumn{1}{|c|}{${|F_{\pi}|}^{2}$} & $\sqrt{s}, MeV$ &  \multicolumn{1}{|c|}{${|F_{\pi}|}^{2}$} \\ \hline

\multicolumn{2}{|l|}{RHO 2013 season} 	 &   819.674 & 24.091 $\pm$ 0.146 $\pm$ 0.239 	 &   723.766 & 34.985 $\pm$ 0.092 $\pm$ 0.235 	 &   889.769 & 9.469 $\pm$ 0.075 $\pm$ 0.075  \\
326.980 & 1.677 $\pm$ 0.114 $\pm$ 0.018 	 &   821.643 & 23.779 $\pm$ 0.141 $\pm$ 0.237 	 &   724.208 & 35.281 $\pm$ 0.208 $\pm$ 0.237 	 &   899.809 & 8.250 $\pm$ 0.063 $\pm$ 0.067  \\
346.820 & 1.844 $\pm$ 0.091 $\pm$ 0.019 	 &   843.428 & 17.527 $\pm$ 0.143 $\pm$ 0.180 	 &   727.684 & 36.723 $\pm$ 0.210 $\pm$ 0.247 	 &   909.413 & 7.475 $\pm$ 0.031 $\pm$ 0.061  \\
366.660 & 2.024 $\pm$ 0.078 $\pm$ 0.021 	 &   862.676 & 13.327 $\pm$ 0.127 $\pm$ 0.142 	 &   728.170 & 36.908 $\pm$ 0.227 $\pm$ 0.249 	 &   910.113 & 7.310 $\pm$ 0.082 $\pm$ 0.060  \\
386.520 & 2.024 $\pm$ 0.077 $\pm$ 0.018 	 &   879.938 & 10.710 $\pm$ 0.054 $\pm$ 0.117 	 &   731.973 & 38.303 $\pm$ 0.226 $\pm$ 0.258 	 &   920.556 & 6.579 $\pm$ 0.032 $\pm$ 0.055  \\
406.360 & 2.152 $\pm$ 0.082 $\pm$ 0.019 	 &   883.188 & 10.309 $\pm$ 0.137 $\pm$ 0.114 	 &   732.038 & 37.993 $\pm$ 0.183 $\pm$ 0.256 	 &   920.648 & 6.535 $\pm$ 0.066 $\pm$ 0.055  \\
426.200 & 2.550 $\pm$ 0.073 $\pm$ 0.023 	 &   902.620 & 8.037 $\pm$ 0.114 $\pm$ 0.093 	 &   735.576 & 39.634 $\pm$ 0.156 $\pm$ 0.268 	 &   929.645 & 6.006 $\pm$ 0.030 $\pm$ 0.051  \\
446.060 & 2.827 $\pm$ 0.061 $\pm$ 0.025 	 &   922.517 & 6.420 $\pm$ 0.047 $\pm$ 0.078 	 &   736.114 & 39.662 $\pm$ 0.220 $\pm$ 0.268 	 &   930.447 & 5.855 $\pm$ 0.058 $\pm$ 0.050  \\
465.900 & 2.985 $\pm$ 0.086 $\pm$ 0.027 	 &   941.793 & 5.231 $\pm$ 0.079 $\pm$ 0.067 	 &   739.946 & 41.031 $\pm$ 0.219 $\pm$ 0.278 	 &   940.281 & 5.375 $\pm$ 0.063 $\pm$ 0.047  \\
485.760 & 3.232 $\pm$ 0.076 $\pm$ 0.029 	 &   957.787 & 4.555 $\pm$ 0.017 $\pm$ 0.060 	 &   744.065 & 42.147 $\pm$ 0.213 $\pm$ 0.285 	 &   949.992 & 4.805 $\pm$ 0.051 $\pm$ 0.043  \\
505.202 & 3.816 $\pm$ 0.038 $\pm$ 0.034 	 &   961.925 & 4.352 $\pm$ 0.073 $\pm$ 0.059 	 &   748.035 & 43.218 $\pm$ 0.209 $\pm$ 0.292 	 &   960.354 & 4.506 $\pm$ 0.056 $\pm$ 0.041  \\
525.162 & 4.409 $\pm$ 0.086 $\pm$ 0.040 	 &   981.691 & 3.723 $\pm$ 0.077 $\pm$ 0.053 	 &   751.984 & 43.850 $\pm$ 0.218 $\pm$ 0.296 	 &   970.587 & 4.041 $\pm$ 0.047 $\pm$ 0.038  \\
545.093 & 4.829 $\pm$ 0.088 $\pm$ 0.042 	 &   984.654 & 3.638 $\pm$ 0.037 $\pm$ 0.052 	 &   750.003 & 43.655 $\pm$ 0.037 $\pm$ 0.295 	 &   980.546 & 3.585 $\pm$ 0.068 $\pm$ 0.035  \\
565.149 & 5.824 $\pm$ 0.087 $\pm$ 0.051 	 &   1004.253 & 3.082 $\pm$ 0.054 $\pm$ 0.051 	 &   756.081 & 44.870 $\pm$ 0.230 $\pm$ 0.303 	 &   984.251 & 3.641 $\pm$ 0.037 $\pm$ 0.036  \\
585.012 & 7.096 $\pm$ 0.120 $\pm$ 0.064 	 &   1010.669 & 2.848 $\pm$ 0.024 $\pm$ 0.048 	 &   760.023 & 45.277 $\pm$ 0.208 $\pm$ 0.306 	 &   990.000 & 3.469 $\pm$ 0.063 $\pm$ 0.034  \\
604.872 & 8.027 $\pm$ 0.117 $\pm$ 0.084 	 &   1013.113 & 2.740 $\pm$ 0.033 $\pm$ 0.047 	 &   760.477 & 45.390 $\pm$ 0.096 $\pm$ 0.306 	 &   999.985 & 3.221 $\pm$ 0.063 $\pm$ 0.033  \\
624.779 & 10.210 $\pm$ 0.147 $\pm$ 0.103 	 &   1016.231 & 2.630 $\pm$ 0.052 $\pm$ 0.046 	 &   763.916 & 45.568 $\pm$ 0.216 $\pm$ 0.308 	 &   1000.450 & 3.194 $\pm$ 0.043 $\pm$ 0.039  \\
644.624 & 12.939 $\pm$ 0.134 $\pm$ 0.126 	 &   1017.269 & 2.494 $\pm$ 0.032 $\pm$ 0.044 	 &   767.793 & 45.769 $\pm$ 0.182 $\pm$ 0.309 	 &   1001.857 & 3.158 $\pm$ 0.035 $\pm$ 0.039  \\
664.549 & 16.172 $\pm$ 0.055 $\pm$ 0.154 	 &   1018.142 & 2.517 $\pm$ 0.032 $\pm$ 0.044 	 &   771.827 & 46.100 $\pm$ 0.194 $\pm$ 0.313 	 &   1005.961 & 2.976 $\pm$ 0.023 $\pm$ 0.037  \\
684.425 & 21.127 $\pm$ 0.165 $\pm$ 0.198 	 &   1019.174 & 2.584 $\pm$ 0.033 $\pm$ 0.048 	 &   773.790 & 46.238 $\pm$ 0.227 $\pm$ 0.314 	 &   1009.606 & 2.913 $\pm$ 0.042 $\pm$ 0.036  \\
704.207 & 27.099 $\pm$ 0.182 $\pm$ 0.252 	 &   1019.987 & 2.725 $\pm$ 0.032 $\pm$ 0.054 	 &   775.033 & 46.516 $\pm$ 0.089 $\pm$ 0.316 	 &   1015.730 & 2.652 $\pm$ 0.032 $\pm$ 0.035  \\
724.120 & 35.028 $\pm$ 0.243 $\pm$ 0.325 	 &   1021.217 & 2.924 $\pm$ 0.036 $\pm$ 0.053 	 &   775.089 & 46.534 $\pm$ 0.244 $\pm$ 0.316 	 &   1016.793 & 2.553 $\pm$ 0.020 $\pm$ 0.034  \\
731.860 & 38.033 $\pm$ 0.241 $\pm$ 0.353 	 &   1022.089 & 2.959 $\pm$ 0.056 $\pm$ 0.052 	 &   776.027 & 46.676 $\pm$ 0.213 $\pm$ 0.317 	 &   1017.915 & 2.455 $\pm$ 0.024 $\pm$ 0.032  \\
739.149 & 40.685 $\pm$ 0.133 $\pm$ 0.378 	 &   1022.874 & 2.871 $\pm$ 0.056 $\pm$ 0.051 	 &   776.942 & 46.530 $\pm$ 0.247 $\pm$ 0.316 	 &   1019.056 & 2.574 $\pm$ 0.015 $\pm$ 0.036  \\
740.515 & 41.005 $\pm$ 0.147 $\pm$ 0.381 	 &   1027.981 & 2.757 $\pm$ 0.055 $\pm$ 0.049 	 &   777.927 & 46.521 $\pm$ 0.080 $\pm$ 0.316 	 &   1019.916 & 2.718 $\pm$ 0.018 $\pm$ 0.043  \\
743.785 & 42.088 $\pm$ 0.312 $\pm$ 0.391 	 &   1033.769 & 2.652 $\pm$ 0.058 $\pm$ 0.047 	 &   778.023 & 46.759 $\pm$ 0.197 $\pm$ 0.318 	 &   1020.028 & 2.728 $\pm$ 0.026 $\pm$ 0.043  \\
747.749 & 43.356 $\pm$ 0.413 $\pm$ 0.402 	 &   1039.883 & 2.604 $\pm$ 0.054 $\pm$ 0.047 	 &   779.029 & 46.128 $\pm$ 0.211 $\pm$ 0.316 	 &   1020.904 & 2.881 $\pm$ 0.026 $\pm$ 0.040  \\
751.715 & 43.641 $\pm$ 0.407 $\pm$ 0.405 	 &   1049.905 & 2.256 $\pm$ 0.033 $\pm$ 0.043 	 &   779.976 & 45.171 $\pm$ 0.216 $\pm$ 0.318 	 &   1022.077 & 2.991 $\pm$ 0.032 $\pm$ 0.039  \\
755.702 & 44.806 $\pm$ 0.310 $\pm$ 0.416 	 &   1060.255 & 2.200 $\pm$ 0.052 $\pm$ 0.042 	 &   780.943 & 42.837 $\pm$ 0.210 $\pm$ 0.322 	 &   1022.897 & 2.968 $\pm$ 0.035 $\pm$ 0.039  \\
756.059 & 44.794 $\pm$ 0.122 $\pm$ 0.416 	 &   \multicolumn{2}{|l|}{} 	 &   780.989 & 42.858 $\pm$ 0.057 $\pm$ 0.323 	 &   1027.733 & 2.807 $\pm$ 0.036 $\pm$ 0.036  \\
759.566 & 44.978 $\pm$ 0.410 $\pm$ 0.418 	 &   \multicolumn{2}{|l|}{RHO 2018 season} 	 &   781.886 & 40.449 $\pm$ 0.204 $\pm$ 0.329 	 &   1033.810 & 2.603 $\pm$ 0.034 $\pm$ 0.034  \\
763.598 & 45.392 $\pm$ 0.135 $\pm$ 0.422 	 &   547.784 & 5.015 $\pm$ 0.021 $\pm$ 0.032 	 &   782.727 & 37.800 $\pm$ 0.059 $\pm$ 0.318 	 &   1039.804 & 2.539 $\pm$ 0.036 $\pm$ 0.034  \\
767.976 & 45.690 $\pm$ 0.206 $\pm$ 0.425 	 &   560.253 & 5.461 $\pm$ 0.080 $\pm$ 0.034 	 &   782.799 & 37.493 $\pm$ 0.171 $\pm$ 0.316 	 &   1049.794 & 2.315 $\pm$ 0.032 $\pm$ 0.032  \\
771.573 & 46.072 $\pm$ 0.421 $\pm$ 0.430 	 &   570.210 & 6.019 $\pm$ 0.069 $\pm$ 0.038 	 &   783.978 & 34.514 $\pm$ 0.185 $\pm$ 0.275 	 &   1059.990 & 2.227 $\pm$ 0.033 $\pm$ 0.031  \\
775.119 & 46.278 $\pm$ 0.137 $\pm$ 0.432 	 &   580.133 & 6.481 $\pm$ 0.064 $\pm$ 0.041 	 &   784.985 & 32.619 $\pm$ 0.042 $\pm$ 0.240 	 &   1099.963 & 1.723 $\pm$ 0.022 $\pm$ 0.027  \\
775.729 & 46.823 $\pm$ 0.423 $\pm$ 0.437 	 &   590.167 & 7.245 $\pm$ 0.087 $\pm$ 0.047 	 &   785.018 & 32.613 $\pm$ 0.157 $\pm$ 0.239 	 &   1149.610 & 1.430 $\pm$ 0.023 $\pm$ 0.025  \\
778.561 & 46.321 $\pm$ 0.151 $\pm$ 0.433 	 &   600.144 & 7.884 $\pm$ 0.082 $\pm$ 0.058 	 &   786.086 & 31.727 $\pm$ 0.145 $\pm$ 0.222 	 &   1199.168 & 1.105 $\pm$ 0.018 $\pm$ 0.021  \\
779.141 & 45.928 $\pm$ 0.192 $\pm$ 0.431 	 &   610.064 & 8.671 $\pm$ 0.098 $\pm$ 0.063 	 &   787.783 & 31.394 $\pm$ 0.070 $\pm$ 0.216 	 &   \multicolumn{2}{|l|}{}  \\
780.738 & 43.394 $\pm$ 0.377 $\pm$ 0.424 	 &   620.169 & 9.722 $\pm$ 0.097 $\pm$ 0.069 	 &   788.007 & 31.228 $\pm$ 0.146 $\pm$ 0.215 	 &   \multicolumn{2}{|l|}{LOW 2020 season}  \\
781.141 & 42.692 $\pm$ 0.071 $\pm$ 0.427 	 &   630.083 & 10.765 $\pm$ 0.099 $\pm$ 0.075 	 &   790.108 & 31.219 $\pm$ 0.150 $\pm$ 0.214 	 &   360.352 & 1.781 $\pm$ 0.035 $\pm$ 0.015  \\
782.031 & 40.671 $\pm$ 0.363 $\pm$ 0.422 	 &   639.997 & 11.898 $\pm$ 0.108 $\pm$ 0.082 	 &   791.835 & 31.402 $\pm$ 0.087 $\pm$ 0.216 	 &   380.968 & 1.950 $\pm$ 0.034 $\pm$ 0.016  \\
782.894 & 37.457 $\pm$ 0.122 $\pm$ 0.397 	 &   650.596 & 13.632 $\pm$ 0.114 $\pm$ 0.093 	 &   792.083 & 31.369 $\pm$ 0.152 $\pm$ 0.216 	 &   400.059 & 2.168 $\pm$ 0.039 $\pm$ 0.014  \\
782.910 & 37.316 $\pm$ 0.073 $\pm$ 0.396 	 &   660.375 & 15.238 $\pm$ 0.126 $\pm$ 0.104 	 &   794.070 & 31.307 $\pm$ 0.152 $\pm$ 0.215 	 &   420.041 & 2.367 $\pm$ 0.037 $\pm$ 0.015  \\
783.880 & 34.563 $\pm$ 0.316 $\pm$ 0.356 	 &   670.204 & 17.377 $\pm$ 0.135 $\pm$ 0.117 	 &   796.050 & 30.605 $\pm$ 0.144 $\pm$ 0.210 	 &   440.460 & 2.596 $\pm$ 0.062 $\pm$ 0.017  \\
784.880 & 32.572 $\pm$ 0.315 $\pm$ 0.321 	 &   679.684 & 19.675 $\pm$ 0.029 $\pm$ 0.132 	 &   797.942 & 30.680 $\pm$ 0.168 $\pm$ 0.211 	 &   460.467 & 2.962 $\pm$ 0.045 $\pm$ 0.019  \\
785.031 & 32.569 $\pm$ 0.056 $\pm$ 0.320 	 &   680.181 & 19.576 $\pm$ 0.148 $\pm$ 0.131 	 &   800.654 & 29.948 $\pm$ 0.142 $\pm$ 0.207 	 &   480.895 & 3.210 $\pm$ 0.042 $\pm$ 0.021  \\
786.703 & 31.595 $\pm$ 0.128 $\pm$ 0.299 	 &   689.578 & 22.610 $\pm$ 0.127 $\pm$ 0.151 	 &   810.560 & 26.552 $\pm$ 0.130 $\pm$ 0.188 	 &   499.440 & 3.712 $\pm$ 0.052 $\pm$ 0.024  \\
788.574 & 31.213 $\pm$ 0.108 $\pm$ 0.295 	 &   690.000 & 22.705 $\pm$ 0.107 $\pm$ 0.152 	 &   820.258 & 23.960 $\pm$ 0.121 $\pm$ 0.172 	 &   520.620 & 4.160 $\pm$ 0.049 $\pm$ 0.027  \\
789.450 & 30.924 $\pm$ 0.317 $\pm$ 0.293 	 &   699.705 & 26.044 $\pm$ 0.190 $\pm$ 0.174 	 &   829.849 & 20.952 $\pm$ 0.117 $\pm$ 0.152 	 &   540.167 & 4.752 $\pm$ 0.042 $\pm$ 0.030  \\
792.469 & 31.184 $\pm$ 0.106 $\pm$ 0.296 	 &   700.044 & 25.833 $\pm$ 0.103 $\pm$ 0.173 	 &   839.997 & 18.004 $\pm$ 0.117 $\pm$ 0.132 	 &   560.250 & 5.647 $\pm$ 0.050 $\pm$ 0.036  \\
794.466 & 31.053 $\pm$ 0.307 $\pm$ 0.295 	 &   709.962 & 29.512 $\pm$ 0.152 $\pm$ 0.198 	 &   850.530 & 15.698 $\pm$ 0.088 $\pm$ 0.117 	 &   580.504 & 6.662 $\pm$ 0.054 $\pm$ 0.042  \\
797.656 & 30.521 $\pm$ 0.281 $\pm$ 0.290 	 &   710.357 & 29.753 $\pm$ 0.190 $\pm$ 0.199 	 &   860.216 & 13.811 $\pm$ 0.105 $\pm$ 0.104 	 &   601.222 & 7.930 $\pm$ 0.074 $\pm$ 0.058  \\ 
799.987 & 29.797 $\pm$ 0.112 $\pm$ 0.284 	 &   719.917 & 33.457 $\pm$ 0.048 $\pm$ 0.225 	 &   870.262 & 11.962 $\pm$ 0.083 $\pm$ 0.092 &&	 \\
803.980 & 29.031 $\pm$ 0.212 $\pm$ 0.280 	 &   720.301 & 33.621 $\pm$ 0.218 $\pm$ 0.226 	 &   880.145 & 10.615 $\pm$ 0.084 $\pm$ 0.083 &&	 \\ \hline
\end{tabular}
}
\end{footnotesize}
\end{table*}

\section{Conclusions}

The measurement of the $e^+e^-\to\pi^+\pi^-$ cross section was performed by 
the CMD-3 experiment at the VEPP-2000 collider in the energy range
$\sqrt{s} = 0.32 \hyph 1.2$~GeV in 209 energy points.
The analysis was based on the biggest ever used 
collected statistics at the $\rho$ resonance region with $34\times10^{6}$
$\pi^+\pi^-$ events at $\sqrt{s}<1$~GeV.
The large statistics
allows to study various systematic effects in detail.
The total systematic uncertainty of the pion form factor 
in the central $\rho$-meson energy region is estimated as 0.7/0.9\%,
with the two numbers reflecting the difference in the detector performance in
the different data taking seasons.
The new result generally shows larger
$e^+e^-\to\pi^+\pi^-$ cross section compared to the previous measurements 
in the whole energy range under discussion.

The $\rho-\omega$ interference observed in the data allowed to measure the
branching ratio of the $\omega\to\pi^+\pi^-$ decay as given in
Eq.~(\ref{omega2pi}).
As a by-product of the background analysis, the branching ratio
of the $\omega\to\pi^+\pi^-\pi^0$ decay (given in Sec.~\ref{omega3pi}) is
measured using the background $3\pi$ events which contaminate the signal 
sample of $2\pi$ events within used collinear samples.
The obtained result is in good agreement with the other measurements of
the $e^+e^-\to\pi^+\pi^-\pi^0$ cross section, 
which indicates the proper handling of the $3\pi$ background in the $2\pi$
sample. More precise study of the $e^+e^-\to\pi^+\pi^-\pi^0$ process
will be done in the ongoing dedicated analysis of the full $3\pi$ events sample.

At the energies around $\phi$-meson resonance the first measurement of
the $e^+e^-\to\pi^+\pi^-$ cross section is performed with high energy resolution.
The corresponding $\phi\to\pi^+\pi^-$ decay parameters are given in Eq.~(\ref{phi2pi}).

As one of the consistency checks, the $e^+e^-\to\mu^+\mu^-$ cross
section was measured at the lowest energies $\sqrt{s}<0.7$~GeV, where
the muon pairs can be cleanly separated from the others. The experimental values 
are well consistent with the QED prediction with the average ratio as:
$$\sigma_{e^+e^-\to \mu^+\mu^-} / \sigma_{QED} = 1.0017\pm 0.0016 \pm 0.0025 \pm 0.0056,$$
where the first error is statistical and the others are estimated systematic uncertainties, 
the last term corresponds to the uncertainty of the fiducial volume determination.

The charge asymmetry in the $\pi^+\pi^-$ final state was extracted
using the forward and backward parts of the measured cross sections.
The strong deviation was observed from the prediction based on the conventional
scalar QED approach for the calculation of the radiative corrections. The improved
GVMD model was proposed in the paper~\cite{Ignatov:2022iou}, which gives the remarkable agreement with
the experimental data. The significant corrections beyond the scalar QED was also confirmed by the calculation in 
the dispersive formalism in the paper~\cite{Colangelo:2022lzg}.
It will be still interesting to understand the difference between the C-odd radiative correction
obtained in the dispersive formalism and that estimated in the GVMD model,
which is sensed by the experimental statistical precision.
The obtained result shows the importance of the appropriate choice of
the model for the calculation of the radiative corrections for the
$\pi^+\pi^-$ channel. It is important to revise possible effect of the scalar QED limitations 
for other calculations including two photon exchange processes.
The observed differences between the measured and predicted charge asymmetries 
for the $\pi^+\pi^-$ and $e^+e^-$ events averaged in the $\sqrt{s}=0.7\hyph0.82$~GeV energy range
are $\delta A^{\pi^+\pi^-} = -0.00029 \pm 0.00023 $ and $\delta A^{e^+e^-} = -0.00060 \pm 0.00026$.
This consistency better than 0.1\% should additionally
ensure our polar angle related systematic uncertainty estimation in the measurement of 
the pion form factor.

The measured $e^+e^-\to\pi^+\pi^-$ cross section was used to evaluate $2\pi$ contribution
to the hadronic part of the anomalous magnetic moment of the muon $a_\mu^{had,LO}$ in the energy
range $0.6<\sqrt{s}<0.88$~GeV. The value based on the CMD-3 data is notably larger than 
the evaluations, based on the results of the previous measurements. 
The CMD-3 result reduces the tension between the
experimental value~\cite{Muong-2:2021ojo} of the $a_\mu$ and its Standard Model prediction.

\subsection{Acknowledgments}

The authors are grateful to R.~Lee
for his help with the theoretical development of the GVMD model.
We thank the VEPP-2000 team for the excellent machine operation. 

\bibliography{mybibfile}
\end{document}